\documentclass[prx, twocolumn, superscriptaddress]{revtex4-2}
\usepackage{bm, amsmath, amsfonts, amssymb, braket}
\usepackage{multirow}
\usepackage[pdftex]{graphicx}
\usepackage{float}
\usepackage{color}
\usepackage{comment}

\usepackage{mathtools}
\usepackage{amsmath,amssymb, amsfonts}
\usepackage{amsthm}
\usepackage{comment}
\usepackage{mathdots}

\usepackage[pdftex]{hyperref}
\let\Re\relax
\DeclareMathOperator{\Re}{Re}
\let\Im\relax
\DeclareMathOperator{\Im}{Im}

\begin{document}

\newcommand{\ka}{\kappa}
\newcommand{\norm}[1]{\left\lVert#1\right\rVert}
\newcommand{\abs}[1]{\left\lvert#1\right\rvert}
\renewcommand{\thesection}{\Roman{section}}
\renewcommand{\thesubsection}{\Alph{subsection}}

\newcommand{\EOBC}[1]{\mathcal{E}_{\mathrm{OBC}}^{(#1)}}
\newcommand{\EPBC}{\mathcal{E}_{\mathrm{PBC}}}
\newcommand{\AIIdag}[1]{H_{\mathrm{#1}}^{\mathrm{AII}^\dag}}
\newcommand{\UAIIdag}[1]{U_{\mathrm{#1}}^{\mathrm{AII}^\dag}}
\newcommand{\HN}[1]{H_{\mathrm{#1}}^{\mathrm{HN}}}
\newcommand{\kket}[1]{|#1\rangle\!\rangle}

\title{Topological enhancement of non-normality in non-Hermitian skin effects}
\author{Yusuke O. Nakai}
    \email{yusuke.nakai@yukawa.kyoto-u.ac.jp}
	\affiliation{Center for Gravitational Physics and Quantum Information, Yukawa Institute for Theoretical Physics, Kyoto University, Kyoto 606-8502, Japan}
\author{Nobuyuki Okuma}
    \email{okuma@hosi.phys.s.u-tokyo.ac.jp}
	\affiliation{Graduate School of Engineering, Kyushu Institute of Technology, Kitakyushu 804-8550, Japan}
\author{Daichi Nakamura}
    \email{daichi.nakamura@yukawa.kyoto-u.ac.jp}
	\affiliation{Center for Gravitational Physics and Quantum Information, Yukawa Institute for Theoretical Physics, Kyoto University, Kyoto 606-8502, Japan}
\author{Kenji Shimomura}
    \email{kenji.shimomura@yukawa.kyoto-u.ac.jp}
	\affiliation{Center for Gravitational Physics and Quantum Information, Yukawa Institute for Theoretical Physics, Kyoto University, Kyoto 606-8502, Japan}
\author{Masatoshi Sato}
    \email{msato@yukawa.kyoto-u.ac.jp}
	\affiliation{Center for Gravitational Physics and Quantum Information, Yukawa Institute for Theoretical Physics, Kyoto University, Kyoto 606-8502, Japan}
\date{\today}

\begin{abstract}
The non-Hermitian skin effects are representative phenomena intrinsic to non-Hermitian systems: the energy spectra and eigenstates under the open boundary condition (OBC) drastically differ from those under the periodic boundary condition (PBC). Whereas a non-trivial topology under the PBC characterizes the non-Hermitian skin effects, their proper measure under the OBC has not been clarified yet. This paper reveals that topological enhancement of non-normality under the OBC accurately quantifies the non-Hermitian skin effects. Correspondingly to spectrum and state changes of the skin effects, we introduce two scalar measures of non-normality and argue that the non-Hermitian skin effects enhance both macroscopically under the OBC. 
We also show that the enhanced non-normality correctly describes phase transitions causing the non-Hermitian skin effects and reveals the absence of non-Hermitian skin effects protected by average symmetry.
The topological enhancement of non-normality governs the perturbation sensitivity of the OBC spectra and the anomalous time-evolution dynamics through the Bauer-Fike theorem.
\end{abstract}

\maketitle

\section{introduction}
Non-Hermitian systems, whose dynamics are effectively described by non-Hermitian Hamiltonians, have been extensively studied recently \cite{r6.1-nonhermitian-review,r6-kunst2021,r6.2okuma-sato-review}. In condensed matter physics, non-Hermicity emerges in various situations. For example, gain or loss of particles or energy causes non-Hermicity in open quantum or classical systems \cite{r9.5-song-yao-wang-openskin-2019,r14.5-longhi-2020,r19-yi-2020,r22-haga-liouvilian-2021,r24.11-zhou-2021,r25-wu-2014,r26-hu-2011,r27-malzard-2015,r28-lee-2014,r29-xu-2017,r30-nakagawa-2018,r31-kawabata-parity-2018,r32-wu-2019,r33-yamamoto-2019,r34-xiao-2019,r62-lieu-tenfold-2020}. One can also obtain non-Hermitian systems in terms of the one-particle Green's functions, where the non-Hermicity originates in the self-energy from the correlations or disorders \cite{r22.5-okuma-correlated-2021,r35-fu-quasi-2017,r36-philip-2018,r37-chen-2018,r38-zyuzin-2018,r39-yoshida-2018,r40-yoshida-2019,r41-bergholtz-2019,r42-moors-2019,r43-michishita-exceptional-2020,r44-michishita-equivalence-2020,r45-rausch-2021,r46-yoshida-exceptional-2020,r47-michishita-renormalization}. 

Unlike Hermitian Hamiltonians, the eigenvalues of non-Hermitian Hamiltonians can be complex. In addition, the conjugate of a ket (right) eigenvector is not always a bra (left) eigenvector. Such mathematical properties lead to unique phenomena in non-Hermitian systems \cite{r1-Gongashida,r2-ksus}. In particular, the non-Hermitian skin effects occur, which are extreme sensitivities of the complex spectra to the boundary condition \cite{r6.5-yao-wang-edge-2018,r7-yao-song-wang-2018,r8-kunst-biorthogonal-2018,r9-yokomizo-murakami-nonbloch-2019,r9.5-song-yao-wang-openskin-2019,r10-song-yao-wang-toporeal-2019,r11-okss,r13-zhang-yang-fang-2020,r12-yoshida-mirror-2020,
r14-kss-hons-2020,r14.5-longhi-2020,r15-okugawa-takahashi-yokomizo-2020,r16-martinez-2018,r17-longhi-2019,r18-lee-2019,r18.5-okuma-2019,r20-li-critical-2020,r21-okuma-anomaly-2021,r22-haga-liouvilian-2021,r22.5-okuma-correlated-2021,r23-taylor-2021,r24-zhang-kai-2021,r24.11-zhou-2021,Robert}.

Recent progress on non-Hermitian topological phases \cite{r3-esakisato2011,r4-zhen-fu2018,r5-kawabata-higashikawa-2019} has revealed that a non-trivial topology under the periodic boundary condition (PBC) is the origin of the non-Hermitian skin effects \cite{r11-okss,r13-zhang-yang-fang-2020}.
By mapping the non-Hermitian system to a Hermitian one, we can show that the non-Hermitian skin modes coincide with the topological boundary modes of the Hermitian system.
However, their proper characterization under the open boundary condition (OBC) 
has not been clarified yet.  
Whereas the boundary modes in the Hermitian system have their own topological structure, the consequence of the topology on the skin modes has rarely been discussed. 
In this paper, we introduce scalar measures $\kappa_{\rm R}$ and ${\rm dep}_{\rm D}$ characterizing the non-Hermitian skin effects under the OBC. 
In particular, we show that the Hermitian topology leads to macroscopic enhancement of non-normality under the OBC,
and the non-Hermitian skin effect occurs when $\kappa_{\rm R}=O(e^{cL})$  
and ${\rm dep}_{\rm D}=O(L^{1/2})$, where $L$ is the system size and $c$ is a positive constant. 
%In particular, we show that the main characteristic features of the non-Hermitian skin effects, {\it i.e.} the extreme sensitivity of spectra to the boundary condition and the non-conjugate nature of right and left eigenstates, lead to {\it topological enhancement of non-normality under the OBC}. 
Furthermore, the enhanced non-normality under the OBC well-describes phase transitions of the non-Hermitian skin effects and governs the perturbation sensitivity of the OBC spectra and the anomalous time-evolution dynamics.

To examine the usefulness of the new scalar measures, we consider two different non-Hermitian systems. The first one is the Hatano-Nelson model, a disordered one-dimensional non-Hermitian tight-binding model \cite{r50-hatano-1996}.
%which belongs to class A in the Altland-Zirnbauer classification \cite{r56-altland-1997}. 
Our analytical and numerical results confirm that the non-Hermitian skin effect in the Hatano-Nelson model exhibits a macroscopically enhanced non-normality under the OBC. 
Moreover, we show that this characterization captures the disorder-induced phase transition where the non-Hermitian skin effect vanishes. 

Second, we consider a one-dimensional model with the symmetry-protected skin effect \cite{r11-okss}.
This model has the non-Hermitian version of time-reversal symmetry,
$
    TH^{\mathrm{T}}T^{-1}=H,
$
where $H$ is the Hamiltonian and $T$ is a unitary matrix that satisfies $TT^*=-1$ \cite{r2-ksus}.
%The corresponding symmetry class is class $\text{AII}^{\dag}$\cite{r2-ksus}.
Whereas the topological number characterizing the conventional skin effect becomes trivial owing to the symmetry,
one can instead introduce a $\mathbb{Z}_2$ topological number that ensures the symmetry-protected skin effect \cite{r11-okss}.
Our numerical and analytical results confirm again that the symmetry-protected skin effect exhibits an enhanced non-normality under the OBC. 
The enhanced non-normality also describes the disorder-induced phase transition for the symmetry-protected skin effect.

The rest of this paper is organized as follows. In Sec.~\ref{sec:skinskin}, we summarize the basic properties of non-Hermitian skin effects. 
In particular, we explain how non-trivial topology under the PBC leads to the non-Hermitian skin effects.
In Sec.\ref{enhance}, we introduce scalar measures of non-normality 
and show that the non-Hermitian skin effects macroscopically enhance them under the OBC.  
We also demonstrate that the enhanced non-normality correctly describes topological phase transitions for the non-Hermitian skin effects analytically in Sec.\ref{sec:examples} and numerically in Sec.\ref{sec:ditpt}. 
In Sec.\ref{sec:ditpt}, we also clarify that average time-reversal symmetry fails to protect the symmetry-protected skin effect.
We discuss the implications of our results in Sec.\ref{sec:Implications} and give a conclusion.in Sec.\ref{conclusion}

\section{Non-Hermitian skin effect}
\label{sec:skinskin}
We first summarize the basic properties of the non-Hermitian skin effects. 

\subsection{Hatano-Nelson model without disorder}
\label{sec:skin}

The non-Hermitian skin effect is a phenomenon in which the bulk eigenspectrum strongly depends on the boundary conditions. 
The simplest model showing the non-Hermitian skin effect is the Hatano-Nelson model without the disorder \cite{r50-hatano-1996,r51-hatano-1997,r52-hatano-1998}
\begin{equation}
  \label{2-2-3-1-}
  \hat{H}=\sum_{j=1}^{L}\left[(t+g)\hat{c}_{j+1}^{\dagger}\hat{c}_{j}+(t-g)\hat{c}^{\dagger}_{j}\hat{c}_{j+1}\right],
\end{equation}
where ($\hat{c}_{j}$, $\hat{c}^{\dagger}_{j}$) are annihilation and creation operators at site $j$, $L$ is the number of sites, $t\in \mathbb{R}$ is a Hermitian symmetric hopping, and $g\in \mathbb{R}$ is a non-Hermitian asymmetric hopping. Below we assume $t> g\ge 0$ for simplicity.

The one-particle spectrum of the above model is given by the eigenvalues of the matrix Hamiltonian $H$ whose $(ij)$ component $H_{ij}$ is given by $\hat{H}=\sum_{ij}\hat{c}_i^\dag H_{ij}\hat{c}_j$:
\begin{equation}
\label{a-1}
  H=
  \begin{pmatrix}
    0&t-g &0&\cdots&\\
    t+g&0&t-g&\cdots&\\
    0 &t+g&0 &\cdots &\\
    \vdots&\vdots &\vdots & \ddots&
  \end{pmatrix}.\\
\end{equation}
\begin{figure*}[btp]
 \begin{center}
  \includegraphics[scale=0.15]{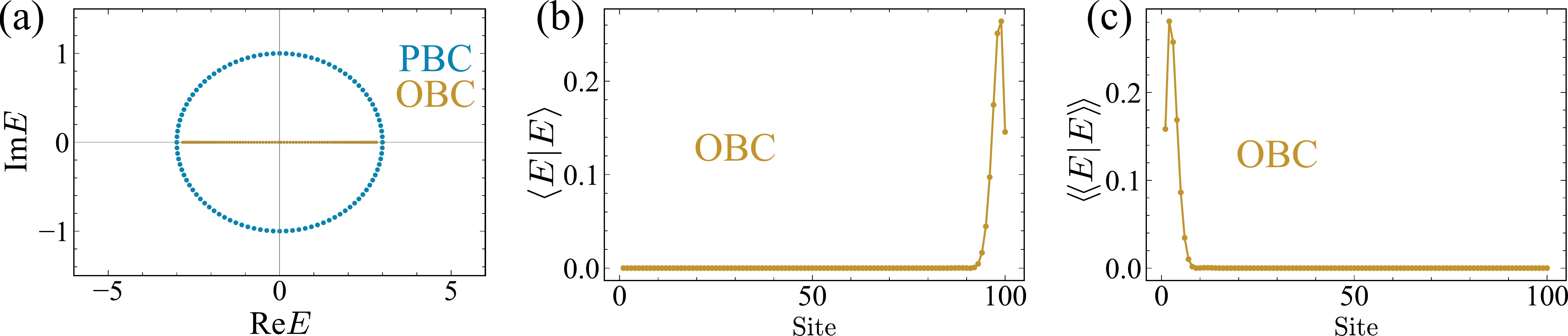}
  \caption{Skin effect. (a) The blue and yellow dots represent the PBC and the OBC spectra of Eq.(\ref{a-1}) with $t=1.5$,\ $g=0.5$,\ $L=100$. (b) Site dependence of the weight function for a right eigenstate of Eq.(\ref{a-1}) under the OBC with $E=-2.664$. (c) Site dependence of the weight function for the left eigenstate corresponding to the right eigenstate in (b).  The left and right eigenstates are localized at opposite boundaries. \label{FIG1}}
  \end{center}
\end{figure*}
Under the PBC, the plane waves 
\begin{align}
    \ket{n}_{\rm PBC}=\frac{1}{\sqrt{L}}
    \begin{pmatrix}
    e^{ik_n}\\
    e^{ik_n2}\\
    \vdots\\
    e^{ik_n(L-1)}\\
    e^{ik_nL}
  \end{pmatrix},
\end{align}
with the crystal momentum $k_n$ 
\begin{equation}
    k_n=\frac{2\pi}{L}n\ \ \ (n=0,\ 1,\ \cdots,\ L-1).
\end{equation}
give the eigenstates. The corresponding PBC spectrum is
\begin{equation}
\label{2-2-4-1-}
    E_n=(t+g)e^{-ik_n}+(t-g)e^{ik_n},
\end{equation}
which forms  an ellipse in the complex energy plane if $g\neq 0$, as illustrated in Fig.~\ref{FIG1}(a).
In contrast, under the OBC, the eigenspectrum becomes real (Fig.~\ref{FIG1}(a)):
\begin{equation}
\label{2-2-9-1-}
    E_n=2\sqrt{t^2-g^2}\cos \left(\frac{\pi}{L+1}n\right)~~(n=1, 2, \cdots, L),
\end{equation}
which is drastically different from the PBC one.  
The corresponding OBC eigenstates are not plane waves but localized when $g\neq 0$.
The explicit form of the right eigenstate is 
\begin{align}
    \ket{n}_{\rm OBC}=\sqrt{\frac{2}{L+1}}
    \begin{pmatrix}
    r^1 \sin{\left(\frac{n\pi}{L+1}\right)}\\
    r^2 \sin{\left(\frac{n\pi}{L+1}\times 2\right)}\\
    \vdots\\
    r^{L-1}\sin{\left(\frac{n\pi}{L+1}\times (L-1)\right)}\\
    r^{L}\sin{\left(\frac{n\pi}{L+1}\times L\right)}
  \end{pmatrix},
\label{eq:skinmode}    
\end{align}
and that of the left eigenstate is
\begin{align}
    |n\rangle\!\rangle_{\rm OBC} =\sqrt{\frac{2}{L+1}}
    \begin{pmatrix}
    r^{-1} \sin{\left(\frac{n\pi}{L+1}\right)}\\
    r^{-2} \sin{\left(\frac{n\pi}{L+1}\times 2\right)}\\
    \vdots\\
    r^{-L+1}\sin{\left(\frac{n\pi}{L+1}\times (L-1)\right)}\\
    r^{-L}\sin{\left(\frac{n\pi}{L+1}\times L\right)}
  \end{pmatrix},
\label{eq:skinmode2}    
\end{align}
with $r=\sqrt{(t+g)/(t-g)}$.
Thus, for $g\neq 0$, $O(L)$ bulk modes are localized at the boundary under the OBC.
Such a macroscopic change of the bulk spectrum and bulk states is called the non-Hermitian skin effect. 

We should note here that the localization in the non-Hermitian skin effect is very unique:
The right eigenstate $|n\rangle_{\rm OBC}$ and the left one ${}_{\rm OBC}\langle\!\langle n|$  are localized at opposite ends, as seen in Eqs.(\ref{eq:skinmode}) and (\ref{eq:skinmode2}). (See also Figs.~\ref{FIG1}(b) and (c).)
This type of localization never happens in Hermitian systems since their right eigenstates are identical to the left ones.
In the next subsection, we see that this exceptional localization comes from the topological nature of the non-Hermitian skin effect.

\subsection{Non-Hermitian topology and skin effect}
\label{sec:skintopo}

It has been known that the non-Hermitian skin effect originates from a topological number of the non-Hermitian system under the PBC \cite{r11-okss,r13-zhang-yang-fang-2020}.
As shown below, the topological origin also ensures the particular localization of skin modes mentioned above.

Let us consider a one-dimensional non-Hermitian system under the PBC.
The topological number relevant to the skin effect is the energy winding number \cite{r11-okss,r13-zhang-yang-fang-2020}:
If the system has the lattice translation symmetry with  the corresponding crystal momentum $k$, and 
the system size $L$ is large enough to regard $k$ as a continuous number with $k\in [-\pi,\pi]$,
the winding number around a reference energy $E\in \mathbb{C}$ is given by~\cite{r1-Gongashida}
\begin{align}\label{21-1}
  W(E)= \int_{-\pi}^{\pi}\frac{dk}{2\pi i}\frac{\partial}{\partial k}\log \det (H(k)-E),
\end{align}
where $H(k)$ is the matrix representation of the Hamiltonian in the momentum space.

To see the relation between this topological number and the skin effect, we introduce the following Hermitian Hamiltonian
\begin{align}
\tilde{H}(k)=
\begin{pmatrix}
0 & H(k)-E\\
H^\dagger(k)-E^* & 0
\end{pmatrix},
\label{eq:doubled_Hamiltonian}
\end{align}
which has chiral symmetry, 
\begin{align}
\{\tilde{H}(k), \Sigma\}=0, \quad \Sigma=\sigma_z=
\begin{pmatrix}
1 & 0 \\
0 & -1
\end{pmatrix}.
\label{eq:chiral}
\end{align}
First, we consider the semi-infinite boundary condition where the system has only one boundary. 
Because $W(E)$ in Eq.(\ref{21-1}) is also a topological number of the Hermitian system $\tilde{H}(k)$, we can apply the conventional bulk-boundary correspondence to the Hermitian system. For a nonzero $W(E)$,   
the Hermitian system $\tilde{H}$ support exact zero energy boundary modes with definite eigenvalues of $\Sigma$.  
Importantly, each of the exact zero modes gives a boundary mode with energy $E$ of the original non-Hermitian Hamiltonian:
If the zero mode has positive chirality $\Sigma=1$, it obeys
\begin{align}
\tilde{H}
\begin{pmatrix}
|E\rangle\\
0
\end{pmatrix}
=0,
\quad
\Sigma
\begin{pmatrix}
|E\rangle\\
0
\end{pmatrix}
=\begin{pmatrix}
|E\rangle\\
0
\end{pmatrix},
\label{eq:pc}
\end{align}
which implies that $H|E\rangle=E|E\rangle$.
Also, for the exact zero mode with negative chirality $\Sigma=-1$, we have
\begin{align}
\tilde{H}
\begin{pmatrix}
0\\
|E\rangle\!\rangle
\end{pmatrix}
=0,
\quad
\Sigma
\begin{pmatrix}
0\\
|E\rangle\!\rangle
\end{pmatrix}
=-\begin{pmatrix}
0\\
|E\rangle\!\rangle
\end{pmatrix},
\label{eq:nc}
\end{align}
which gives $H^\dagger|E\rangle\!\rangle=E^*|E\rangle\!\rangle$, 
namely $\langle\!\langle E |H =\langle\!\langle E|E$.

Now we consider the OBC, where the system has two boundaries.
Whereas any $E$ with a nonzero $W(E)$ provides boundary modes of the non-Hermitian system under the semi-infinite boundary condition, the situation becomes slightly different under the OBC.
Under the OBC, most of $E$ with a nonzero $W(E)$ only provides approximately zero modes of the Hermitian system $\tilde{H}$, not exact ones, so can not give energy eigenstates of the non-Hermitian system.
Therefore, we have boundary modes only for a particular subset of $E$ with $W(E)\neq 0$ where the Hermitian Hamiltonian has exact zero modes \cite{r11-okss}.
Moreover, since the zero modes in the Hermitian system always appear in a pair with opposite chiralities at opposite boundaries, they give a pair of right and left eigenstates of $H$ localized at opposite boundaries.
In other words, we have skin modes with the non-conjugate nature of the right and left eigenstates. 
Since the chirality of the zero modes in the Hermitian system is a topological number, this particular localization of the skin modes is topologically protected.

As an example, let us consider the Hatano-Nelson model without disorder in Eq.(\ref{a-1}). 
In the momentum space, the Hamiltonian is given by  $H(k)=(t+g)e^{-ik}+(t-g)e^{ik}$, and thus if $g\neq 0$ the winding number is $W(E)=-1$ for $E$ inside the region enclosed by the PBC spectrum in Fig.\ref{FIG1} (a).
This non-trivial topological number explains why this model supports the skin modes in Eqs.(\ref{eq:skinmode}) and (\ref{eq:skinmode2}).

\subsection{Symmetry protected skin effect}
\label{sec:spse}

\begin{figure*}[tb]
 \begin{center}
  \includegraphics[scale=0.1]{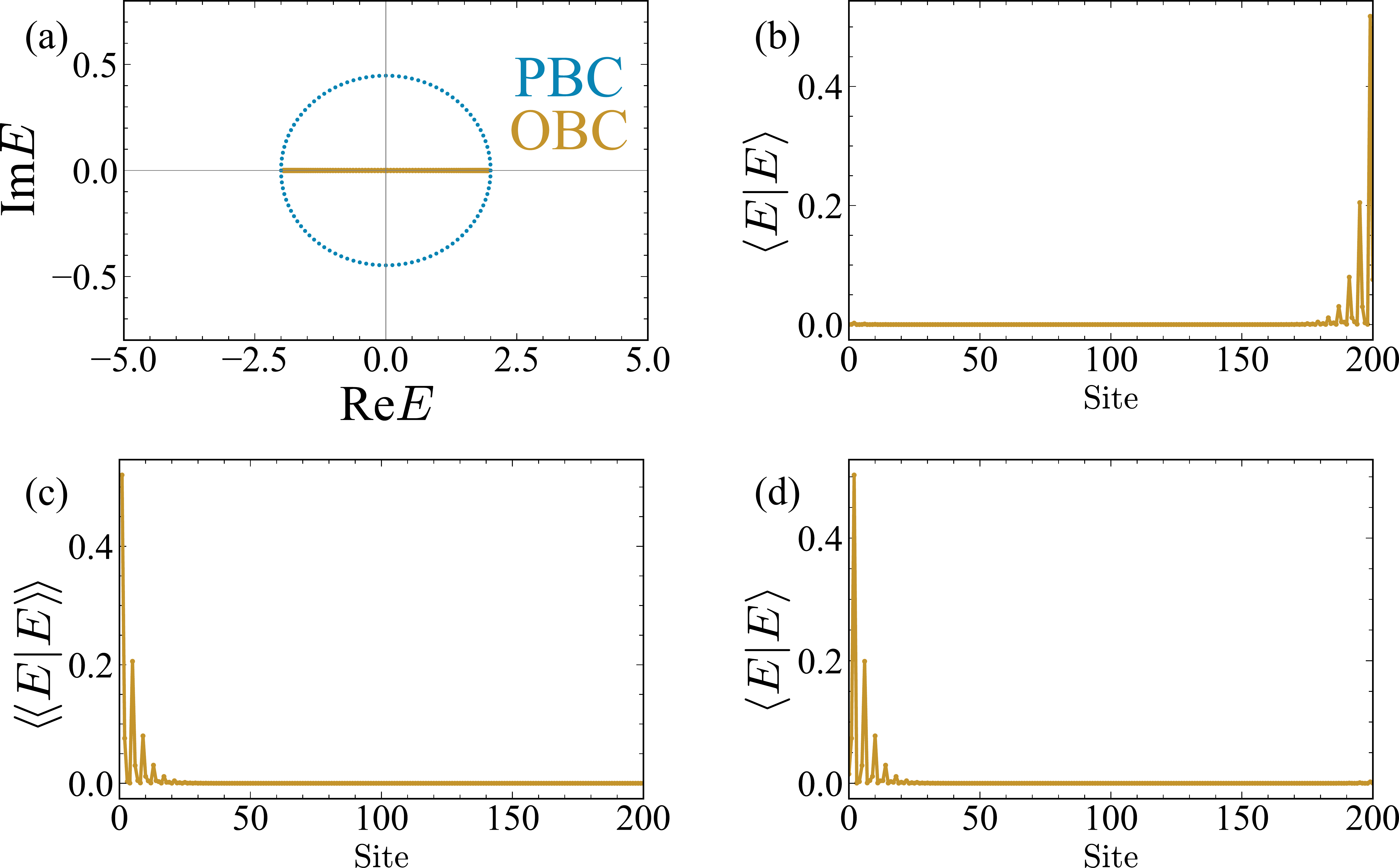}
 \caption{Symmetry protected skin effect. (a) The blue and yellow dots represent the PBC and the OBC spectra of Eq.(\ref{41-1}) with $t=1$, $g=0.3$, $\Delta=0.2$, $L=100$.  (b) Site dependence of the weight function for a right eigenstate of Eq.(\ref{41-1}) under the OBC with $E=-9.091\times 10^{-2}$. (c) Site dependence of the weight function for the left eigenstate corresponding to the right eigenstate in (b). The left and right eigenstates are localized at different ends.  (d) A right eigenstate corresponding to the Kramers partner of the right eigenstate in (b).}\label{FIG2} 
  \end{center}
\end{figure*}

So far, we do not assume any symmetry. 
Here, we consider a one-dimensional non-Hermitian system with the transpose version of time-reversal symmetry, 
\begin{align}
    TH^{T}(k)T^{-1}=H(-k), \quad TT^*=-1,
    \label{trsdagger}
\end{align}
where $T$ is a unitary matrix.
(In the classification scheme in Ref.\cite{r2-ksus}, the transpose version of time-reversal symmetry is dubbed as TRS$^\dagger$ and a system with Eq.(\ref{trsdagger}) is called class AII$^\dagger$.)
This symmetry makes $W(E)$ identically zero but
enables a different type of non-Hermitian skin effect, which we call symmetry-protected skin effect \cite{r11-okss}.
The symmetry-protected skin effect originates from the one-dimensional $\mathbb{Z}_2$ topological invariant $\nu(E)\in\{0,1\}$,
\begin{align}
  \label{10-13-0-}
  &(-1)^{\nu(E)}\notag\\
  &=\mathrm{sgn}\left[\frac{\mathrm{Pf}[(H(\pi)-E)T]}{\mathrm{Pf}[(H(0)-E)T]}\right.\notag\\
  &\times\exp\left.\left[-\frac{1}{2}\int_{k=0}^{k=\pi}dk\frac{\partial}{\partial k}\log \det[(H(k)-E)T]\right] \right],
\end{align}
where $E$ is a reference energy \cite{r2-ksus}. 
Note that the symmetry in Eq.(\ref{trsdagger})
is necessary to have the $\mathbb{Z}_2$ topological invariant:
For the Pfaffian in Eq.(\ref{10-13-0-}) to be well-defined, $(H(k_0)-E)T$ with $k_0=0,\pi$ must be anti-symmetric, which is derived from Eq.(\ref{trsdagger}).

In a manner similar to the ordinary skin effect, to see the relation between the $\mathbb{Z}_2$ invariant and the symmetry-protected skin effect, we introduce the Hermitian Hamiltonian in Eq.(\ref{eq:doubled_Hamiltonian}).
In the present case, the Hermitian Hamiltonian has time-reversal symmetry,
\begin{align}
\tilde{T}\tilde{H}^*(k)\tilde{T}^{-1}=\tilde{H}(-k),    
\quad
\tilde{T}=
\begin{pmatrix}
0& T\\
T & 0
\end{pmatrix},
\end{align}
in addition to chiral symmetry in Eq.(\ref{eq:chiral}).
Therefore, the Hermitian Hamiltonian belongs to class DIII in terms of the Altland-Zirnbauder classification\cite{r56-altland-1997}, which describes a one-dimensional time-reversal superconductor. As $(-1)^{\nu(E)}$ in Eq.(\ref{10-13-0-}) coincides with the $\mathbb{Z}_2$ invariant of the Hermitian system, the one-dimensional superconductor hosts a Kramers pair of Majorana zero-energy boundary modes if the $\mathbb{Z}_2$ invariant is non-trivial.
When the Kramers pair are exact zero modes, it provides skin modes of the original non-Hermitian system.
However, different from conventional ones, the resultant skin modes are localized at both boundaries of the OBC.
To see this property, let us consider a skin mode of which right (left) eigenstate $|E\rangle$ ($|E\rangle\!\rangle$) 
is localized at, say, the right (left) end of the system.
Then, using time-reversal symmetry in Eq.(\ref{trsdagger}),  we have a skin mode localized at the left (right) end, of which right (left) eigenstate is given by $T|E\rangle\!\rangle^*$ ($T|E\rangle^*$). 
These two skin modes $|E\rangle$ and $T|E\rangle\!\rangle^*$ (or their left eigenstates $|E\rangle\!\rangle$ and $T|E\rangle^*$) form a (biorthogonal) Kramers pair \cite{r2-ksus} because we have  $\langle\!\langle E| (T|E\rangle\!\rangle^*)=\langle E| (T|E\rangle^*)=0$.

%The $\mathbb{Z}_2$ non-Hermitian skin effect protected by transpose-type time-reversal symmetry occurs under the OBC if $\nu(E)$ is nonzero for a point $E$ enclosed by the PBC curve 

To illustrate the symmetry-protected skin effect, we consider a time-reversal invariant version of the Hatano-Nelson model without disorder \cite{r11-okss}: 
\begin{align}
  \hat{H}&=\sum_{j=1}^{L}((t+g)\hat{c}_{j+1,\uparrow}^{\dagger}\hat{c}_{j,\uparrow}
  +(t-g)\hat{c}_{j,\uparrow}^{\dagger}\hat{c}_{j+1,\uparrow}))\notag\\
  &+\sum_{j=1}^{L}((t+g)\hat{c}_{j,\downarrow}^{\dagger}\hat{c}_{j+1,\downarrow}+(t-g)\hat{c}_{j+1,\downarrow}^{\dagger}\hat{c}_{j,\downarrow}))\notag\\
  &-i\Delta\sum_{j=1}^{L}(\hat{c}_{j+1,\uparrow}^{\dagger}\hat{c}_{j,\downarrow}-\hat{c}_{j,\uparrow}^{\dagger}\hat{c}_{j+1,\downarrow})\notag\\
  &-i\Delta\sum_{j=1}^{L}(\hat{c}_{j+1,\downarrow}^{\dagger}\hat{c}_{j,\uparrow}-\hat{c}_{j,\downarrow}^{\dagger}\hat{c}_{j+1,\uparrow})
\label{a2d}
\end{align}
where $j$ is the site index, $L$ is the number of sites, $\uparrow$ and $\downarrow$ represent the spin degrees of freedom. The first and the second lines of the above Hamiltonian describe the Hatano-Nelson model without disorder, and its time-reversal partner, respectively, and $\Delta\in \mathbb{R}$ is a time-reversal invariant coupling between them.

In the momentum space, the matrix representation of the Hamiltonian reads
\begin{align}
H(k)
% &=
%  \begin{pmatrix}
%    (t+g)e^{-ik}+(t-g)e^{ik}&2\Delta \sin k &\\
%    2\Delta \sin k&(t+g)e^{ik}+(t-g)e^{-ik}&
%  \end{pmatrix}\notag\\
%  
=2t\cos k+2\Delta (\sin k)\sigma_x+2ig(\sin k)\sigma_z ,
 \label{41-1}
 \end{align}
where $\sigma_{i}$s' are the Pauli matrices in spin space.   
The transpose version of time-reversal symmetry is given by 
 \begin{align}
 \label{eq:TRSdag}
 TH^T(k)T^{-1}=H(-k), \quad T=i\sigma_y.
\end{align}
%If the internal degree of freedom of the Hamiltonian represents the spin degree of freedom, the off-diagonal terms can be interpreted as spin-orbit interactions.
The energy eigenvalues of Eq.(\ref{41-1}) give the PBC spectrum,
\begin{equation}
    E_{\pm}(k)=2t\cos k\pm 2i\sqrt{g^2-\Delta^2}\sin k.\label{a2ene}
\end{equation}
For $|g|>|\Delta|$, the PBC spectrum forms an ellipse on the complex energy plane. Under the same condition, the model has a non-trivial $\mathbb{Z}_2$ number for $E$ inside the ellipse.
Therefore, we can expect the symmetry-protected skin effect.
Actually, as shown in Fig.~\ref{FIG2}(a), 
the OBC spectrum forms a line on the real axis of the complex plane, being completely different from the PBC spectrum. 
Moreover, for each $E$ of the OBC spectrum, we have a pair of the right eigenstates, one is localized at the right end (Fig.~\ref{FIG2}(b)), and the other is localized at the left end (Fig.~\ref{FIG2}(d)).
As expected from the argument above,
the corresponding left eigenstates are localized at opposite ends (Fig.~\ref{FIG2}(c)).
 
\subsection{Anderson localization versus skin effects}
\label{subsec:Anderson}

\begin{figure*}[tbp]
 \begin{center}
  \includegraphics[scale=0.13]{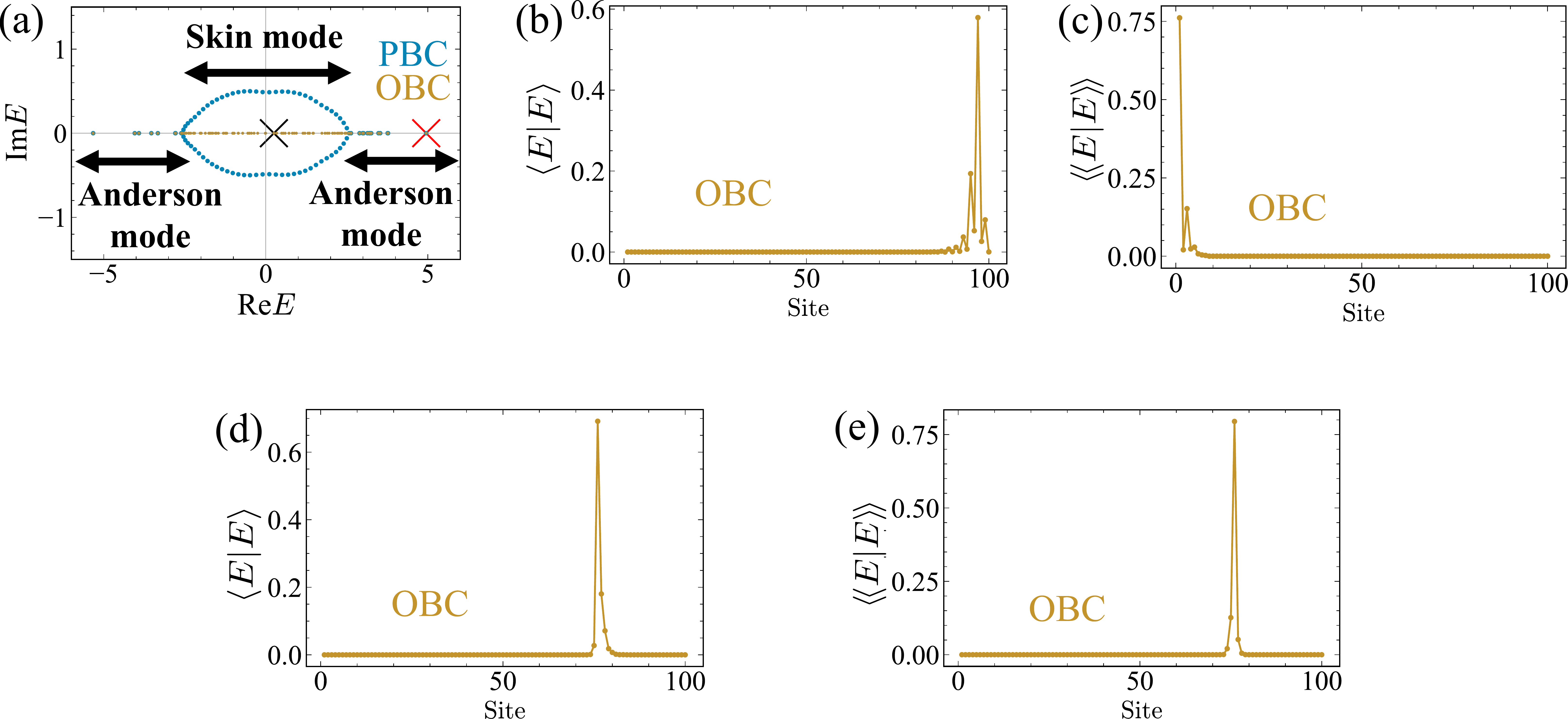}
  \caption{(a) The blue and yellow dots represent the PBC and the OBC spectra of Eq.(\ref{61-1}) with $t=1.5$,~$g=0.5$,~$L=100$,~$\gamma=0.5$. Eigenstates with the OBC energy inside the PBC spectrum are skin modes, and eigenstates with the OBC energy outside the PBC spectrum are Anderson modes. (b) Site dependence of weight function for a right eigenstate of Eq.(\ref{61-1}) under the OBC with eigenenergy at the black cross in (a). (c) Site dependence of weight function for the left eigenstate corresponding to the right eigenstate in (c). The right eigenstate is localized on one side while the left counterpart is localized on the other side. (d) Site dependence of weight function for a right eigenstate of Eq.(\ref{61-1}) under the OBC with eigenenergy at the red cross in (a). (e) Site dependence of weight function for the left eigenstate corresponding to the right eigenstate in (d). The right and left eigenstates are localized at the same position. \label{FIG3}}
  \end{center}
\end{figure*}

Disorders induce another localization, {\it i.e.} the Anderson localization\cite{r114-anderson,r112-Scaling,r113-Anderson-review}. 
Here we compare the non-Hermitian skin effects with the Anderson localization and show that the former keeps the non-conjugation nature of the left and right eigenstates even in the presence of disorder, whereas the latter does not. 

We first consider the Hatano-Nelson model \cite{r50-hatano-1996,r51-hatano-1997,r52-hatano-1998}:
\begin{equation}
  \label{61-1}
  \hat{H}=\sum_{j=1}^{L}\left[(t+g)\hat{c}_{j+1}^{\dagger}\hat{c}_{j}+(t-g)\hat{c}^{\dagger}_{j}\hat{c}_{j+1}+w_j\hat{c_j}^\dagger\hat{c_j}\right],
\end{equation}
where $w_j\in\mathbb{R}\ (j=1,2,\cdots, L)$ is the strength of the on-site disorder. 
For later convenience, we choose the random potential with the Cauchy distribution, of which the probability density function is given by
\begin{align}
    P(w_j)=\frac{\gamma}{\pi}\frac{1}{w_j^2+\gamma^2}.
\label{cauchy}
\end{align}
%Below we refer to $\gamma$ as the disorder strength.
We numerically investigate the model (\ref{61-1}) with $t=1.5,~g=0.5,~L=100$, and $\gamma=0.5$. 
%Here $\gamma (\geq0)$ represents the half-width of the Cauchy distribution. 
The obtained PBC and OBC energy spectra are shown in Fig.~\ref{FIG3}(a).
We observe two different behaviors in the spectra:
In the central region of the complex energy plane, the PBC spectrum and the OBC one are completely different, and thus the non-Hermitian skin effect occurs.
As is illustrated in Figs.~\ref{FIG3}(b) and (c), the right and left eigenstates of the corresponding skin modes are localized at different ends, so they are non-conjugate.
In contrast, in the outer region of the complex energy plane, the PBC spectrum collapses and does not show a clear distinction from the OBC spectrum.
This collapse originates from the Anderson localization \cite{r50-hatano-1996,r51-hatano-1997,r52-hatano-1998}.
As shown in Figs.~\ref{FIG3}(d) and (e), the right and left eigenstates under the OBC
are almost the same.

\begin{figure*}[tbp]
 \begin{center}
  \includegraphics[scale=0.15]{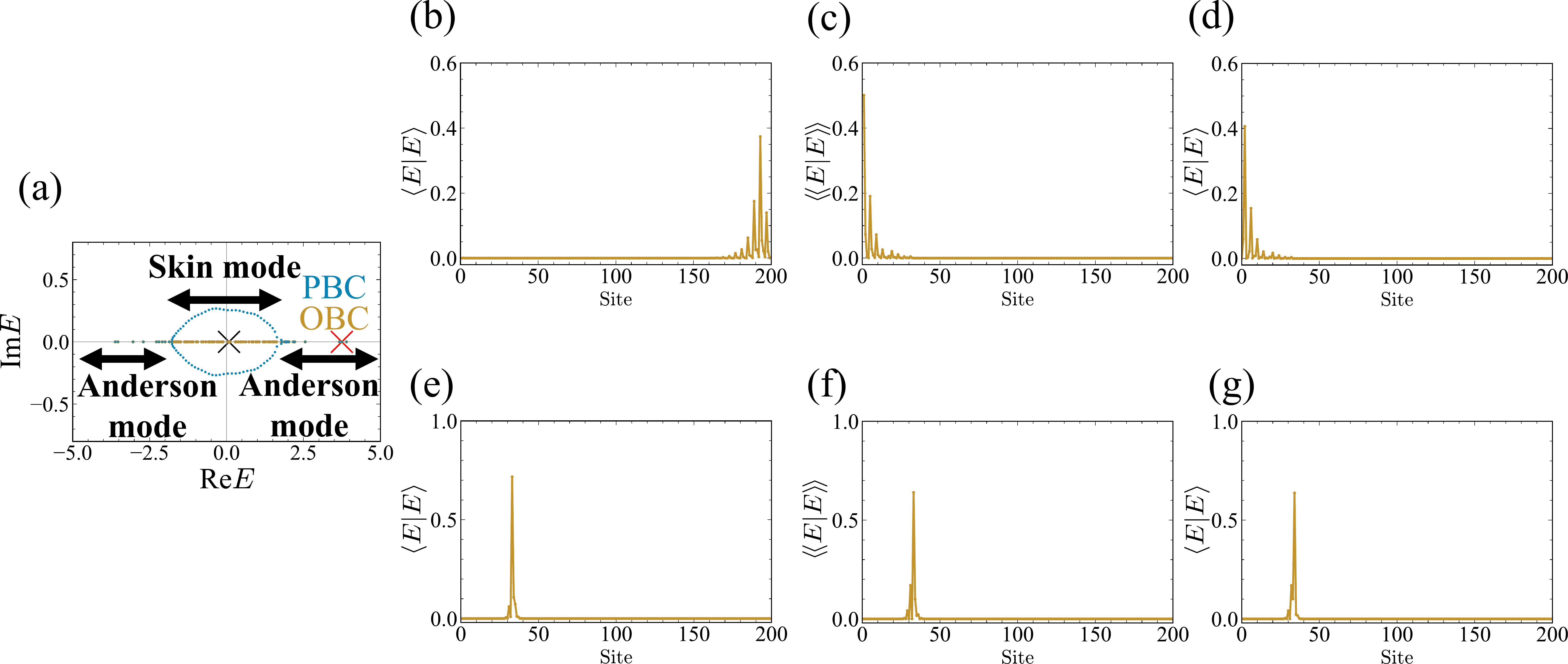}
  \caption{
  (a) The blue and yellow dots represent the PBC and the OBC spectra of Eq.(\ref{a4}) with ~$t=1$,~$g=0.3$,~$\Delta=0.2$, $N=100$,~$\gamma=0.2$. Eigenstates with the OBC energy inside the PBC spectrum are skin modes, and eigenstates with the OBC energy outside the PBC spectrum are Anderson modes. (b) Site dependence of weight function for a right eigenstate of Eq.(\ref{a4}) under the OBC with eigenenergy at the black cross in (a). (c) Site dependence of weight function for the left eigenstate corresponding to the right eigenstate shown in (b). (d) Site dependence of weight function for the Kramers partner corresponding to the right eigenstate shown in (b). The right eigenstate is localized at one end while the left counterpart and the Kramers partner are localized at the other end.  (e) Site dependence of weight function for a right eigenstate of Eq.(\ref{a4}) under the OBC with eigenenergy at the red cross in (a). (f) Site dependence of weight function for the left eigenstate corresponding to the right eigenstate shown in (e). (g) Site dependence of weight function for the Kramers partner corresponding to the right eigenstate shown in (e). The right, left eigenstates and the Kramers partner are localized at the same position. \label{FIG7}}
  \end{center}
\end{figure*}

We also consider the model in Eq.(\ref{a2d}) with disorder,
\begin{align}
  \hat{H}&=\sum_{j=1}^{L}\left[
  (t+g)\hat{c}_{j+1,\uparrow}^{\dagger}\hat{c}_{j,\uparrow}+(t-g)\hat{c}_{j,\uparrow}^{\dagger}\hat{c}_{j+1,\uparrow}
  +w_j\hat{c}_{j,\uparrow}^{\dagger}\hat{c}_{j,\uparrow}\right]
  \notag\\
  &+\sum_{j=1}^{L}\left[
  (t+g)\hat{c}_{j,\downarrow}^{\dagger}\hat{c}_{j+1,\downarrow}
  +(t-g)\hat{c}_{j+1,\downarrow}^{\dagger}\hat{c}_{j,\downarrow}
  +w_j\hat{c}_{j,\downarrow}^{\dagger}\hat{c}_{j,\downarrow}\right]\notag\\
  &-i\Delta\sum_{j=1}^{L}(\hat{c}_{j+1,\uparrow}^{\dagger}\hat{c}_{j,\downarrow}-\hat{c}_{j\uparrow}^{\dagger}\hat{c}_{j+1\downarrow})\notag\\
  &-i\Delta\sum_{j=1}^{L}(\hat{c}_{j+1,\downarrow}^{\dagger}\hat{c}_{j,\uparrow}-\hat{c}_{j,\downarrow}^{\dagger}\hat{c}_{j+1,\uparrow})
\label{a2}
\end{align}
where $w_j$ is the onsite disorder with the Cauchy distribution in Eq.(\ref{cauchy}).
The corresponding matrix Hamiltonian reads
\begin{widetext}
\begin{align}
  \label{a4}
  H=
  \begin{pmatrix}
    w_{1}&0 &t-g & i\Delta&\cdots&0&0&0&0&\\
    0&w_{1}&i\Delta &t+g &\cdots&0&0&0&0&\\
    t+g &-i\Delta &w_{2} & 0&\cdots &0&0&0&0&\\
    -i\Delta&t-g &0 & w_{2}&\cdots &0&0&0&0&\\
    \vdots&\vdots&\vdots & \vdots&\ddots &\vdots&\vdots&\vdots&\vdots&\\
    0&0&0 & 0&\cdots &w_{L-1}&0&t-g&i\Delta&\\
    0&0&0 & 0&\cdots &0&w_{L-1}&i\Delta&t+g&\\
    0&0&0 & 0&\cdots &t+g&-i\Delta&w_{L}&0&\\
    0&0&0 & 0&\cdots &-i\Delta&t-g&0&w_{L}
  \end{pmatrix},
\end{align}
\end{widetext}
under the OBC. The matrix Hamiltonian has additional entries in the upper right and lower left corners under the PBC. 
The disordered matrix Hamiltonian preserves the transpose-type time-reversal symmetry:
\begin{align}
    T H^{T}T^{-1}=H,
    \label{trsd}
\end{align}
where $T=i\sigma_{y}\otimes {\bm 1}_{L\times L}$ satisfies $T T^{\ast}=-1$. 
We show the energy spectrum of the system with $t=1$, $g=0.3$, $\Delta=0.2$, $L=100$ and $\gamma=0.2$ in Fig.~\ref{FIG7}(a). 
We find again that the PBC spectrum and the OBC one are completely different in the central region of the complex energy plane, whereas they are almost identical in the outer region.
In the former region, the right and left eigenstates under the OBC are localized at opposite ends,  
as shown in Figs.~\ref{FIG7}(b) and (c). 
It also has a Kramers partner localized at the opposite end [Fig.~\ref{FIG7}(d)] because of the transpose-type time-reversal symmetry in Eq.(\ref{trsd}).
Therefore, they are symmetry-protected skin modes.
In contrast, the right and left eigenstates in the outer region are localized at almost the same position, as shown in Figs.~\ref{FIG7}(e) and (f). Moreover, the Kramers partner occupies the same position as shown in Fig.~\ref{FIG7}(g). This result indicates that they are not skin modes but modes with Anderson localization.

\section{Topological enhancement of non-normality}
\label{enhance}

As was seen in previous sections, the non-Hermitian skin effects exhibit the following characteristic features: (i) 
The OBC spectrum is completely different from the PBC spectrum:
When the PBC spectrum forms a loop in the complex energy plane, the OBC spectrum becomes a line (or an arc), which is inside the PBC spectrum.
%In particular, the spectrum of skin modes collapses into arcs, which are inside the PBC spectrum.
(ii) The skin modes show a peculiar localization specific to non-Hermitian systems.
In particular, the right and left eigenstates of each skin mode are localized at opposite boundaries. 

Remarkably, these features originate from the topological nature of the skin effects.
The OBC spectrum of skin modes must be inside the PBC spectrum since the topological numbers for the skin effects are non-zero only inside the PBC spectrum. Moreover, the chirality of topological boundary modes results in the peculiar localization of the skin modes. 

In this section, we discuss the relation between these topological properties of the skin effects and the non-normality of the system. 
A non-Hermitian Hamiltonian $H$ is said to be non-normal (normal) if it obeys $[H, H^\dagger ]\neq 0$ ($[H,H^\dagger]=0$).
%For $H$ to be non-normal, $H$ must be non-Hermitian.
Below, we introduce scalar measures of non-normality and clarify that the topological properties of the non-Hermitian skin effects macroscopically enhance the non-normality of the system under the OBC.

\subsection{condition number}\label{sec:condition-number}

First, we introduce the condition number as a useful scalar measure of non-normality \cite{r67-trefethen-book-2005,r110-okuma-pseudo-2020}.
For a diagonalizable matrix Hamiltonian $H$, 
the condition number $\kappa(V)$ is defined as follows:
Let $E_\alpha$ ($\alpha=1,\dots, p$) be distinct eigenvalues of $H$, and $|E^a_\alpha\rangle$ ($a=1,\dots,d_\alpha$) be corresponding right eigenstates for $E_\alpha$, 
\begin{align}
H|E^a_\alpha\rangle=E_\alpha|E^a_\alpha\rangle,    
\end{align}
where $a$ labels independent eigenstates with the same eigenvalue $E_\alpha$, and $d_\alpha$ is the number of the independent eigenstates. 
We can diagonalize $H$ as
\begin{align}
V^{-1}HV=
\begin{pmatrix}
E_1{\bm 1}_{d_1\times d_1}& &  \\    
 & E_2{\bm 1}_{d_2\times d_2} & \\
 & &\ddots 
\end{pmatrix}
\equiv \Lambda
\label{eq:vhv}
\end{align}
by the regular matrix 
\begin{align}\label{eq:def-of-V}
V=(|E_1^1\rangle, \dots, |E_1^{d_1}\rangle, |E_2^1\rangle, \dots, |E_2^{d_2}\rangle, \dots).  
\end{align}
Then, the condition number is defined as
 \begin{align}
 \kappa(V)=\| V \|_2 \| V^{-1}\|_2,
 \end{align}
where $\|\cdot\|_2$ is the 2-norm of a matrix
\begin{align}
\|\cdot \|_2={\rm max}_{\bm x}\left[\sqrt{|\cdot{\bm x}|^2}/\sqrt{|{\bm x}|^2}\right].     
\end{align}
Because the 2-norm of a matrix coincides with the largest singular value of the matrix, 
we can evaluate $\kappa(V)$ as
\begin{align}
1\le \kappa(V)=s_{\rm max}(V)/s_{\rm min}(V)<\infty,    
\label{eq:kappa_evaluation}
\end{align}
where $s_{\rm max}(V)$ and $s_{\rm min}(V)$ are the largest and smallest singular values of $V$.
For a non-diagonalizable matrix Hamiltonian $H$, we formally define $\kappa(V)=\infty$.

In general, $\kappa(V)$ is not unique for a given $H$ since
we have a different $V$ by the linear transformation 
$|E^a_\alpha\rangle\rightarrow \sum_{b}|E^b_\alpha\rangle G^\alpha_{ba}$ with ${\rm det} G^\alpha \neq 0$.
Thus, we need to impose an additional constraint on $V$ to obtain a unique $\kappa(V)$.
For a special case when the eigenvalues are distinct, Ref.\cite{r67-trefethen-book-2005} proposed a constraint to obtain a unique $\kappa(V)$.
Here, generalizing this constraint, we adapt the normalization condition 
\begin{align}
\langle E_\alpha^a|E_\alpha^b\rangle=\delta_{ab}.  
\label{eq:normalization}
\end{align}
Still, we can obtain a different $V$ by a unitary transformation $|E_\alpha^a\rangle\rightarrow\sum_{b}|E^b_\alpha\rangle W^\alpha_{ba}$ with a unitary matrix $W^\alpha$, but  
the unitary transformation does not change the 2-norm of $V$.
Hence, we have a unique $\kappa(V)$.

This choice of $\kappa(V)$ has nice properties. First, $\kappa(V)=1$ is possible if and only if $H$ is normal. Actually, if $\kappa(V)=1$, we have $s_{\rm max}(V)=s_{\rm min}(V)$, so all the eigenvalues of the Hermitian matrix $V^\dagger V$ are equal. Therefore, we can diagonalize $V^\dagger V$ as $V^\dagger V=\sqrt{s_{\rm max}(V)} {\bm 1}$, which implies that $V^{-1}=V^\dagger/\sqrt{s_{\rm max}(V)}$, and
$H$ is normal. The inverse is also true because if $H$ is normal, we can diagonalize $H$ by a unitary matrix $V$, which satisfies the above constraint and $\kappa(V)=1$.
Second, as we prove in Appendix \ref{app:uniquely-defined-condition-number}, the resultant $\kappa(V)$ exceeds the minimal value of $\kappa(V)$ by at most a factor of $\sqrt{L}$, where $L$ is the size of $H$ (see Eq.(\ref{eq:ineq-cond-num})). 
Thus, it does not overestimate the non-normality of $H$ more than necessary.

Now we relate $\kappa(V)$ with the non-Hermitian skin effect.
By introducing left eigenstates of $H$, 
\begin{align}
H^\dagger |E_\alpha^a\rangle\!\rangle=E_\alpha^*|E_\alpha^a\rangle\!\rangle,
\end{align}
with the biorthogonal normalization
\begin{align}
\langle\!\langle E_\alpha^a|E_\beta^b\rangle=\delta_{\alpha\beta}\delta_{ab},
\label{eq:biorthogonal}
\end{align}
$V^{-1}$ is explicitly written as
\begin{align}
V^{-1}=(\kket{E_1^1}, \dots, \kket{E_1^{d_1}},\kket{E_2^1},\dots, \kket{E_2^{d_2}}, \dots)^\dag.
\end{align}
Then, from the inequality 
\begin{align}
\sqrt{{\rm tr}A^\dagger A/L}\le \|A\|_2 \le \sqrt{{\rm tr}A^\dagger A}   
\end{align}
for a $L\times L$ square matrix, $\kappa(V)$ satisfies
\begin{align}
\xi(V)
\le \kappa(V)
\le 
\xi(V) L,
\label{eq:bound}
\end{align}
where $L$ is the size of $H$ and $\xi(V)$ is given by
\begin{align}
\xi(V)&=\frac{\sqrt{\sum_{\alpha,a}\langle E_{\alpha}^a|E_{\alpha}^a\rangle\sum_{\beta,b}\langle\!\langle E_{\beta}^b|E_{\beta}^b\rangle\!\rangle}}{L}
\nonumber\\
&=\sqrt{\frac{\sum_{\beta,b}\langle\!\langle E_{\beta}^b|E_{\beta}^b\rangle\!\rangle}{L}}.
\label{eq:xi}
\end{align}
Here we have used the normalization condition $\langle E_\alpha^a|E_\alpha^a\rangle=\delta_{ab}$ in the last equality.
The lower bound $\xi(V)$ of $\kappa(V)$ describes how the right and left eigenstates of $H$ differ from each other: When they rarely overlap with each other, $\xi(V)$ becomes huge because $\langle\!\langle E_\beta^b|E_\beta^b\rangle\!\rangle$ must be extremely large to satisfy Eqs.(\ref{eq:normalization}) and (\ref{eq:biorthogonal}).

As mentioned in Sec.\ref{sec:skintopo}, for topological reasons, the right and left eigenstates of skin modes are spatially well-separated from each other. 
Thus, under the conditions of Eqs.(\ref{eq:normalization}) and (\ref{eq:biorthogonal}), $\langle\!\langle E_\alpha^a|E_\alpha^a\rangle\!\rangle$ becomes exponentially large with respect to the system size $L$,
which gives $\xi(V)\sim e^{cL}/\sqrt{L}$ with a positive constant $c$.
As a result, $\kappa(V)$ is also exponentially large in the presence of the non-Hermitian skin effects.
In general, we may have contributions for $\kappa(V)$ other than the skin modes, but they are insensitive to the boundary conditions.
Therefore, we can extract the contribution from the skin effects by considering the ratio $\kappa_{\rm R}=\kappa_{\rm OBC}(V)/\kappa_{\rm PBC}(V)$ between $\kappa(V)$s' under the OBC and the PBC.
This ratio is exponentially enhanced with the system size $L$ in the presence of the skin effects, 
\begin{align}
\kappa_{\rm R}\sim e^{cL}    
\label{eq:krscale}
\end{align}
while $\kappa_{\rm R}=O(1)$ in the absence of the skin effects.
Thus, we can use $\kappa_{\rm R}$ as an order parameter for the non-Hermitian skin effects
\footnote{
For more detailed arguments on the relation between $\kappa_{\rm R}$ and the non-Hermitian skin effect, see Appendix \ref{app:kappa}}. 

\subsection{departure from normality}

Another useful quantity is Henrici's departure from normality \cite{r67-trefethen-book-2005}. 
As shown below, for relatively simple models like the Hatano-Nelson model, this quantity also relates the non-Hermitian skin effects to the non-normality of the Hamiltonian. 

The central idea of  Henrici's departure from normality is to use the Schur decomposition to characterize non-normality.
For any given Hamiltonian $H$, we can perform the Schur decomposition
\begin{equation}
    H=U(\Lambda+R)U^{\dag},
\end{equation}
where $U$ is a unitary matrix, $\Lambda$ is a diagonal matrix of eigenvalues, and $R$ is a strictly upper triangular matrix.
When $R$ is zero, this is a unitary diagonalization; hence, $H$ must be normal. Thus the norm of $R$ measures the non-normality of $H$. 
Since the Shur decomposition is not unique in general,
Henrici introduced the departure from normality as the minimum value of $\|R\|$ over all possible decompositions, 
\begin{equation}
    \mathrm{dep}(H)=\min_{H=U(\Lambda+R)U^{\dag}}\|R\|.
\end{equation}
Note that ${\rm dep}(H)$ is insensitive to the origin of energy, and thus we can shift the reference energy $E$ arbitrarily,
\begin{align}
{\rm dep}(H)={\rm dep}(H-E).    
\end{align}

We obtain a direct relation between the departure from normality and the spectrum of the system by considering the Frobenius norm $\|R\|_{\rm F}=\sqrt{{\rm tr}(R^\dagger R)}$.
From the equation
\begin{align}
\|H\|_{\rm F}
&=\|\Lambda+R\|^2_{\rm F}\nonumber\\
&=\|\Lambda\|^2_{\rm F}+\|R\|^2_{\rm F},
\end{align}
the Frobebnius norm version of the departure from normality $\mathrm{dep}_{\mathrm{F}}(A)$ does not depend on a particular Schur decomposition and is given by
\begin{align}
    \mathrm{dep}_{\mathrm{F}}(H)&=\sqrt{\|H\|^2_{\mathrm{F}}-\|\Lambda\|^2_{\mathrm{F}}}\notag\\
    &=\sqrt{\sum_i s_i^2-\sum_i|E_i|^2},
\label{depp}
\end{align}
where $s_i$ and $E_i$ are the singular values of $H$ and the energy eigenvalues of $H$, respectively. 
While $s_i$ rarely depends on the boundary condition as it is given by an eigenvalue of the Hermitian matrix $HH^\dagger$, $E_i$ crucially depends on the boundary condition in the presence of non-Hermitian skin effects.
Therefore, we can expect that ${\rm dep}_{\rm F}(H)$ measures the non-normality caused by the non-Hermitian skin effects. 

Actually, we find that 
the non-Hermitian skin effects also enhance ${\rm dep}_{\rm F}(H)$ under the OBC if the PBC spectrum forms a single loop in the complex energy plane.
To see this property, we first note that the average of the energy eigenvalues coincides between the PBC and the OBC since the total of the energy eigenvalues is given by ${\rm tr}(H)$, which is insensitive to the boundary conditions.
Thus, without changing ${\rm dep}_{\rm F}(H)$, we can simultaneously set the average energies both under the PBC and the OBC to zero by shifting the origin of the energy. 
Now $\sum_i|E|_i^2/L$ becomes the variance of the complex energy eigenvalues, which measures the deviation of the eigenvalues from the average.
Then, we can show that the non-Hermitian skin effects make the variance under the OBC smaller than that under the PBC: 
As the average energy is set to zero, the PBC spectrum is a loop whose center coincides with the origin in the complex energy plane.  
Then, because the OBC spectrum with a skin effect is obtained by an adiabatic shrinking of the PBC spectrum, keeping the same average value \cite{r11-okss}, the variance of the OBC spectrum becomes smaller than that of the PBC one.
As a result, the skin effect enhances ${\rm dep}_{\rm F}(H)$ under the OBC.

For the characterization of the non-Hermitian skin effects, it is convenient to consider the difference between  ${\rm dep}_{\rm F}(H)$s under the OBC and the PBC.  Since the non-Hermitian skin effects affect $O(L)$ modes, the difference behaves as
\begin{align}
{\rm dep}_{\rm D}={\rm dep}_{\rm OBC}(H)-{\rm dep}_{\rm PBC}(H)\sim c'\sqrt{L}    
\label{eq:ddscale}
\end{align}
with a positive constant $c'$ in the presence of the skin effects, while we have ${\rm dep}_{\rm D}\sim O(1)$ in the absence of the skin effects.
Thus, we can also use ${\rm dep}_{\rm D}$ as another order parameter for the skin effects.

%\vspace{1ex}
\subsection{Non-quantization property}
Whereas the non-Hermitian skin effect originates from the non-trivial bulk topology, 
the enhanced non-normality in the above does not show the quantization.
We would like to point out that 
this non-quantization property is intrinsic to the non-Hermitian skin effects.
%is because non-Hermitian skin modes have a different property than conventional topological boundary states.
%so their topological characterization is not necessarily quantized. 
For conventional topological boundary states, their number is determined by the bulk topological number. For example, the Chern number in a quantum Hall state determines the number of chiral edge modes. Because of this correspondence, the quantized quantity characterizes topological boundary states. In contrast, this correspondence does not hold in the non-Hermitian skin effects. Whereas the non-Hermitian skin effects occur when the bulk topological number becomes nonzero, the total number of skin modes is insensitive to the nonzero value of the 
topological number, and is determined by the number of bulk degrees of freedom, such as the system size.
In other words, different values of the bulk topological number may give the same number of skin modes. Therefore, their topological characterization is not necessarily quantized.

\section{Examples}
\label{sec:examples}
In this section, we analytically confirm the enhanced non-normality of the non-Hermitian skin effects in concrete models without the disorder.

\subsection{Hatano-Nelson model without disorder}
\label{exhata}

Here we consider the Hatano-Nelson model in Eq.(\ref{2-2-3-1-}). We assume that $t>g\ge 0$ for simplicity. 
As shown in Sec.\ref{sec:skin}, this model shows the non-Hermitian skin effect if $g\neq 0$.

\subsubsection{condition number}\label{sec:cond_HN}

First, we calculate the condition number under the PBC. Under the PBC, the matrix $V$ diagonalizing the Hamiltonian in Eq.(\ref{a-1})
is given by
\begin{widetext}
\begin{equation}
\label{9a}
  V_{\mathrm{PBC}}=\frac{1}{\sqrt{L}}
  \begin{pmatrix}
    \exp\left(i\frac{2\pi}{L}\times0\times1\right)&\exp\left(i\frac{2\pi}{L}\times1\times1\right) &\cdots&\exp(i\frac{2\pi}{L}\times(L-1)\times1)&\\
    \exp(i\frac{2\pi}{L}\times0\times2)&\exp(i\frac{2\pi}{L}\times1\times2)&\cdots&\exp(i\frac{2\pi}{L}\times (L-1)\times2)&\\
    \vdots &\vdots&\ddots &\vdots &\\
    \exp(i\frac{2\pi}{L}\times0\times L)&\exp(i\frac{2\pi}{L}\times1\times L)&\cdots & \exp(i\frac{2\pi}{L}\times(L-1)\times L)&
  \end{pmatrix},
\end{equation}
\end{widetext}
which satisfies the normalization condition in Eq.(\ref{eq:normalization}).
Since this matrix is unitary, 
the condition number of $V_{\mathrm{PBC}}$ becomes 1,
\begin{equation}
    \label{11a}
    \kappa(V_{\rm PBC})=1.
\end{equation}
This result is consistent with the fact that the Hamiltonian in Eq.(\ref{a-1}) is normal under the PBC.

Now we evaluate the condition number under the OBC. 
Under the OBC, we can diagonalize the Hamiltonian in Eq.(\ref{a-1}) by using the right eigenstates $|n\rangle_{\mathrm{OBC}}$ in Eq.(\ref{eq:skinmode}); 
the matrix $V$ diagonalizing the Hamiltonian is given by 
\begin{align}
V_{\rm OBC}=\left(\frac{|1\rangle_{\mathrm{OBC}}}{\||1\rangle_{\mathrm{OBC}}\|_2}, \frac{|2\rangle_{\mathrm{OBC}}}{\||2\rangle_{\mathrm{OBC}}\|_2}, \dots,  \frac{|L\rangle_{\mathrm{OBC}}}{\||L\rangle_{\mathrm{OBC}}\|_2}\right), 
\label{1-1}
\end{align}
where we have imposed the normalization condition in Eq.(\ref{eq:normalization}) on $V_{\rm OBC}$.
To evaluate $\kappa(V_{\rm OBC})$, we decompose $V_{\rm OBC}$ into three matrices $R$, $U$ and $N$,
\begin{align}
V_{\rm OBC}=RUN.    
\end{align}
Here $R$ and $U$ are the diagonal and unitary matrices, respectively, of which $(m,n)$ components are given by
\begin{align}
\label{r}
(R)_{m,n}&=r^{m}\delta_{m,n}
\nonumber\\
    (U)_{m,n}&=\sqrt{\frac{2}{L+1}}\sin\left(\frac{\pi m n}{L+1}\right),
\end{align}
with $r=\sqrt{(t+g)/(t-g)}$.
Note that  $RU$ gives the right eigenstate $|n\rangle_{\rm OBC}$ in Eq.(\ref{eq:skinmode}), 
\begin{align}
    RU
    = \left(|1\rangle_{\mathrm{OBC}}, |2\rangle_{\mathrm{OBC}},\dots, |L\rangle_{\mathrm{OBC}}\right).
\end{align}
Then, the matrix $N$ is the diagonal matrix,
\begin{align}
    (N)_{m,n}=\||n\rangle_{\mathrm{OBC}}\|_2^{-1}\delta_{m,n},
\end{align}
which forces $V_{\rm OBC}$ to obey the normalization condition in Eq.(\ref{eq:normalization}).
As shown in Appendix \ref{48-inequality-d}, using this decomposition, we have the inequality
\begin{align}
\label{48-inequality}
    \left(\frac{2}{L+1}\right)^{3/2}r^{L-1}<\kappa(V_{\rm OBC})\leq \sqrt{L}r^{L-1},
\end{align}
which implies that $\kappa(V_{\rm OBC})$ behaves as 
$\kappa(V_{\rm OBC})\sim e^{L\ln r}$. 

By combining the above result with Eq.(\ref{11a}), the ratio $\kappa_{\rm R}$ becomes
\begin{align}
\kappa_{\rm R}\sim e^{L\ln r}.    
\end{align}
Thus, $\kappa_{\rm R}$ grows exponentially with the system size $L$ when $g\neq 0$ since $r>1$ in this case.
On the other hand, if $g=0$, $\kappa_{\rm R}\sim O(1)$ since $r=1$.  
This behavior agrees with the fact that the non-Hermitian skin effect occurs (does not occur) when $g\neq 0$ ($g=0$).

\subsubsection{departure from normality}
Now we evaluate the departure from normality. 
The energy spectrum of the model in Eq.(\ref{2-2-3-1-}) under the PBC in the infinite-volume limit is given as 
\begin{equation}
    E(k)=2t\cos k- 2ig\sin k,
\end{equation}
where $k$ is the corresponding crystal momentum. Thus, the second term of Eq.(\ref{depp}) is given by 
\begin{align}
    \sum_{j}|E_j|^2
    &\to \frac{L}{2\pi}\int_0^{2\pi}dk\left|2t\cos k-2ig\sin k\right|^2\notag\\
    &=2L(t^2+g^2)\label{q1A}.
\end{align}
%Next, we calculate the first term of Eq.~(\ref{depp}). 
On the other hand, the square of the singular value of the model is given as
\begin{align}
    (s(k))^2&=(2t\cos k-2ig\sin k)(2t\cos k+2ig\sin k)\notag\\
    &=4t^2\cos^2 k+4g^2\sin^2 k.
\end{align}
Note that singular values of normal matrices are the absolute values of the eigenvalues.
Thus, under the PBC, we have
\begin{align}
    \sum_j (s_j)^2
    &\to\frac{L}{2\pi}\int_0^{2\pi}dk (4t^2\cos^2 k+4g^2\sin^2 k)\notag\\
    &=2L(t^2+g^2).\label{q2A}
\end{align}
From Eq.(\ref{q1A}) and Eq.(\ref{q2A}), we find that ${\rm dep}_{\rm F}(H)$ of the model (\ref{2-2-3-1-}) vanishes under the PBC, 
\begin{align}
    {\rm dep}_{\rm PBC}(H)=0.\label{q3A}
\end{align}

Under the OBC, the eigenvalues of the model (\ref{2-2-3-1-}) are given as 
\begin{equation}
    E_n=2\sqrt{t^2-g^2}\cos \left(\frac{\pi}{L+1}n\right)~~(n=1, 2, \cdots, L),
\end{equation}
which leads to
\begin{align}
    \sum_j|E_j|^2
    =\sum_n \left(4(t^2-g^2)\cos^2\left(\frac{\pi}{L+1}n\right)\right)\label{q4A}.\notag\\
\end{align}
Thus, in the infinite-volume limit~($L\to\infty$), we have
\begin{align}
\label{42A}
    \sum_j|E_j|^2
    &\to\frac{L}{\pi}\int_0^{\pi}dk\left(4(t^2-g^2)\cos^2 k\right)\notag\\
    &=2L(t^2-g^2).
\end{align}
Since the square of the singular value of $H$ is independent of the boundary conditions in the
infinite volume limit, Eq.(\ref{q2A}) also gives the first term of Eq.(\ref{depp}) under the OBC:
\begin{align}
    \sum_j (s_j)^2
    \to 2L(t^2+g^2).\label{q5A}
\end{align}
Therefore, from Eqs.(\ref{42A}) and (\ref{q5A}), ${\rm dep}_{\rm F}(H)$ under the OBC is given by,
\begin{align}
    \mathrm{dep}_{\mathrm{OBC}}(H)&=\sqrt{2L(t^2+g^2)-2L(t^2-g^2)}\notag\\
    &=2g\sqrt{L}.\label{q5.5A}
\end{align}

Combing Eq.(\ref{q3A}) with Eq.(\ref{q5.5A}), we have 
\begin{align}
{\rm dep}_{\rm D}=2g\sqrt{L}.       
\end{align}
Thus, the enhancement of non-normality occurs only when the non-Hermitian skin effect presents ($g\neq 0$).

\subsection{Time-reversal invariant Hatano-Nelson model without disorder}
\label{dep-clean}
In this subsection, we consider the model in Eq.(\ref{41-1}). We assume that $t>g\ge 0$ and $t>\Delta\ge 0$ for simplicity.
As discussed in Sec.\ref{sec:spse}, this model shows the symmetry-protected skin effect when $g>\Delta$.

\subsubsection{condition number}
We assume that $g\neq\Delta$, because the Bloch Hamiltonian \eqref{41-1} is not diagonalizable when $g=\Delta$.

We introduce an explicit expression of the one-particle Hamiltonian under the OBC or the PBC using the corresponding Bloch Hamiltonian in a generic way.
For $j=-L,-L+1,\dots,0,\dots,L-1,L$, let $\EOBC{j}$ be an $L\times L$ singular matrix whose elements are
\begin{align}
    (\EOBC{j})_{m,n}
    = \delta_{m,n+j}.
\end{align}
For a given Bloch Hamiltonian $H(k)$, the corresponding one-particle Hamiltonian under the OBC is of the form
\begin{align}
    H_{\mathrm{OBC}}
    = \sum_{j=-L}^L \frac{1}{2\pi}\int_{-\pi}^\pi e^{-ikj} H(k)\otimes \EOBC{j} dk.
\end{align}
Note that
\begin{align}\label{eq:ad_of_EOBC}
    \EOBC{-j} = (\EOBC{j})^T = (\EOBC{j})^\dag
\end{align}
for $-L\le j\le L$, and
\begin{align}
    \EOBC{\pm j} = (\EOBC{\pm1})^j
\end{align}
for $j>0$.
Introducing an $L\times L$ unitary matrix
\begin{align}
    \EPBC
    \coloneqq \begin{pmatrix}
        0 & 0 & 0 & \cdots & 0 & 1 \\
        1 & 0 & 0 & \cdots & 0 & 0 \\
        0 & 1 & 0 & \ddots & 0 & 0 \\
        \vdots & \vdots & \vdots & \ddots & \ddots & \vdots \\
        0 & 0 & 0 & \cdots & 0 & 0 \\
        0 & 0 & 0 & \cdots & 1 & 0
    \end{pmatrix},
\end{align}
on the other hand, the one-particle Hamiltonian corresponding to $H(k)$ under the PBC is of the form
\begin{align}
    H_{\mathrm{PBC}}
    = \sum_{j=-L}^L\frac{1}{2\pi}\int_{-\pi}^\pi e^{-ikj} H(k)\otimes (\EPBC)^j dk.
\end{align}
Note that if $H(k)$ is Hermitian then $H_{\mathrm{OBC}}$ and $H_{\mathrm{PBC}}$ are also Hermitian, because of Eq.\eqref{eq:ad_of_EOBC} and the unitarity of $\EPBC$, respectively.

Here we take Eq.\eqref{41-1} as $H(k)$, and let $\AIIdag{OBC}$ and $\AIIdag{PBC}$ bes the corresponding one-particle Hamiltonian under the OBC and the PBC, respectively, both of which are $2L\times 2L$ matrices.
Since $H(k)$ has the transpose-version of time-reversal symmetry \eqref{eq:TRSdag}, $\AIIdag{OBC}$ and $\AIIdag{PBC}$ also respect the same type of time symmetry:
\begin{gather}
    T' (\AIIdag{OBC})^T {T'}^{-1}
    = \AIIdag{OBC}, \\
    T' (\AIIdag{PBC})^T {T'}^{-1}
    = \AIIdag{PBC},
\end{gather}
with $T'\coloneqq i\sigma_y \otimes {\bm 1}_{L\times L}$, which lead to the presence of the biorthogonal Kramers pairs.
We have to orthogonalize the Kramers pairs in terms of the right eigenstates when we calculate the condition number with respecting the constraint \eqref{eq:normalization}.

In preparation for the estimation of the condition numbers associated by $\AIIdag{OBC}$ and $\AIIdag{PBC}$, we diagonalize the Bloch Hamiltonian \eqref{41-1}:
\begin{align}\label{eq:diagonalization_of_AIIdag}
    &S^{-1}H(k)S = \notag\\ & \begin{pmatrix}
        2t\cos k+2i\sqrt{g^2-\Delta^2}\sin k & 0 \\
        0 & 2t\cos k-2i\sqrt{g^2-\Delta^2}\sin k
    \end{pmatrix},
\end{align}
where $S$ is a $2\times 2$ regular matrix such that
\begin{align}\label{eq:def_S}
    S = \Delta + (\sqrt{g^2-\Delta^2} - g)\sigma_y.
\end{align}
Note that $S$ is generally not unitary.
We extend $S$ onto $\mathbb{C}^{2L}$ by $\tilde{S}\coloneqq S\otimes {\bm 1}_{L\times L}$ to investigate
\begin{align}
    \tilde{S}^{-1}\AIIdag{OBC}\tilde{S}
    = \sum_{j=-L}^L\frac{1}{2\pi}\int_{-\pi}^\pi e^{-ikj} (S^{-1}H(k)S)\otimes \EOBC{j} dk
\end{align}
and
\begin{align}
    \tilde{S}^{-1}\AIIdag{PBC}\tilde{S}
    = \sum_{j=-L}^L\frac{1}{2\pi}\int_{-\pi}^\pi e^{-ikj} (S^{-1}H(k)S)\otimes (\EPBC)^j dk.
\end{align}
Using the following $L\times L$ 
diagonalizable matrices
\begin{align}\label{eq:HOBCpm}
    H_{\mathrm{OBC}}^\pm
    \coloneqq (t\pm\sqrt{g^2 - \Delta^2})\EOBC{1} + (t\mp\sqrt{g^2 - \Delta^2})\EOBC{-1},
\end{align}
and
\begin{align}\label{eq:HPBCpm}
    H_{\mathrm{PBC}}^\pm
    \coloneqq (t\pm\sqrt{g^2 - \Delta^2})\EPBC + (t\mp\sqrt{g^2 - \Delta^2})(\EPBC)^{-1},
\end{align}
we can express $\tilde{S}^{-1}\AIIdag{OBC/PBC}\tilde{S}$ as
\begin{align}\label{eq:SHOBCAIIdagGinv}
    \tilde{S}^{-1}\AIIdag{OBC}\tilde{S}
    &= \begin{pmatrix}
        1 & 0 \\
        0 & 0
    \end{pmatrix}
    \otimes H_{\mathrm{OBC}}^+
    + \begin{pmatrix}
        0 & 0 \\
        0 & 1
    \end{pmatrix}
    \otimes H_{\mathrm{OBC}}^-, \\
    \label{eq:SHPBCAIIdagGinv}
    \tilde{S}^{-1}\AIIdag{PBC}\tilde{S}
    &= \begin{pmatrix}
        1 & 0 \\
        0 & 0
    \end{pmatrix}
    \otimes H_{\mathrm{PBC}}^+
    + \begin{pmatrix}
        0 & 0 \\
        0 & 1
    \end{pmatrix}
    \otimes H_{\mathrm{PBC}}^-.
\end{align}
Note that $H_{\mathrm{OBC}}^\pm$ and $H_{\mathrm{PBC}}^\pm$ have no degenerate eigenvalue.
When $g>\Delta$, $H_{\mathrm{OBC}}^+$ and $H_{\mathrm{PBC}}^+$ are nothing but the Hatano-Nelson model with the non-Hermitian asymmetric hopping $\sqrt{g^2 - \Delta^2}\in\mathbb{R}$.
When $g<\Delta$, on the other hand, $H_{\mathrm{OBC}}^\pm$ and $H_{\mathrm{PBC}}^\pm$ are Hermitian.

Let $\ket{k}_{\mathrm{OBC}}^\pm$ (resp. $\ket{k}_{\mathrm{PBC}}^\pm$) ($k=1,2,\dots,L$) be the linearly independent right eigenstates of $H_{\mathrm{OBC}}^\pm$ (resp. $H_{\mathrm{PBC}}^\pm$), and $\kket{k}_{\mathrm{OBC}}^\pm$ (resp. $\kket{k}_{\mathrm{PBC}}^\pm$) be the left eigenstates corresponding to $\ket{k}_{\mathrm{OBC}}^\pm$ (resp. $\ket{k}_{\mathrm{PBC}}^\pm$).
In the following, we omit $\mathrm{OBC}$ or $\mathrm{PBC}$ in equations when not necessary, such as $\AIIdag{}$, $H^\pm$, $\ket{k}^\pm$, and $\kket{k}^\pm$.
We impose the biorthogonal condition
\begin{align}\label{eq:biorthogonality_AIIdag}
    ^\pm\langle k\kket{k'}^\pm
    = \delta_{k,k'}
\end{align}
for the right and left eigenstates $\ket{k}^\pm$, $\kket{k}^\pm$.
It is clear that
\begin{align}
    (H^\pm)^T
    = H^\mp,
\end{align}
which follows that
\begin{align}
    \tilde{S}^{-1}\AIIdag{}\tilde{S}
    = \begin{pmatrix}
        1 & 0 \\ 
        0 & 0
    \end{pmatrix}\otimes (H^-)^T
    + \begin{pmatrix}
        0 & 0 \\
        0 & 1
    \end{pmatrix}\otimes H^-
\end{align}
from Eqs.\eqref{eq:SHOBCAIIdagGinv} and \eqref{eq:SHPBCAIIdagGinv}.

Let $\mathcal{P}$ be an $L\times L$ regular matrix representing the parity transformation:
\begin{align}
    \mathcal{P} 
    \coloneqq \begin{pmatrix}
        0 & 0 & \cdots & 0 & 1 \\
        0 & 0 & \cdots & 1 & 0 \\
        \vdots & \vdots & \iddots & 0 & 0 \\
        0 & 1 & \cdots & 0 & 0 \\
        1 & 0 & \cdots & 0 & 0
    \end{pmatrix}.
\end{align}
Note that $\mathcal{P}$ satisfies that
\begin{align}
    \mathcal{P}^\dag = \mathcal{P} = \mathcal{P}^{-1}.
\end{align}
The transpose of $H^\pm$ is connected to $H^\pm$ by $\mathcal{P}$ as follows:
\begin{align}
    (H^\pm)^T
    = \mathcal{P}H^\pm\mathcal{P},
\end{align}
because it holds that
\begin{align}
    \EOBC{\mp 1} = \mathcal{P}\EOBC{\pm 1}\mathcal{P},
    \quad
    (\EPBC)^{\mp 1} = \mathcal{P}(\EPBC)^{\pm 1}\mathcal{P}.
\end{align}
Then the relation between the right eigenstate $\ket{k}^-$ of $H^-$ and the eigenstate $(\kket{k}^-)^*$ of $(H^-)^T$, which are associated by a common eigenvalue, is given by
\begin{align}
    \left(\frac{\kket{k}^-}{\|\kket{k}^-\|_2}\right)^*
    = \mathcal{P}\frac{\ket{k}^-}{\|\ket{k}^-\|_2}.
\end{align}
Thus $\AIIdag{}$ has two right eigenstates 
\begin{align}\label{eq:tildekplus}
    \tilde{\ket{k}}^+
    &\coloneqq \tilde{S}\left(\begin{pmatrix}
        1 \\ 0
    \end{pmatrix}\otimes\mathcal{P}\frac{\ket{k}^-}{\|\ket{k}^-\|_2}\right) \\
    \label{eq:tildekplus2}
    &= \begin{pmatrix}
        \Delta \\ i(\sqrt{g^2-\Delta^2}-g)
    \end{pmatrix}\otimes\mathcal{P}\frac{\ket{k}^-}{\|\ket{k}^-\|_2}
\end{align}
and
\begin{align}\label{eq:tildekminus}
    \tilde{\ket{k}}^-
    &\coloneqq \tilde{S}\left(\begin{pmatrix}
        0 \\ 1
    \end{pmatrix}\otimes\frac{\ket{k}^-}{\|\ket{k}^-\|_2}\right) \\
    \label{eq:tildekminus2}
    &= \begin{pmatrix}
        -i(\sqrt{g^2-\Delta^2}-g) \\ \Delta
    \end{pmatrix}\otimes\frac{\ket{k}^-}{\|\ket{k}^-\|_2}
\end{align}
associated with a same eigenvalue.

In order to orthogonalize the right eigenstates, we calculate a $2\times 2$ Gram matrix
\begin{align}\label{eq:Gammak}
    \Gamma_k
    \coloneqq \begin{pmatrix}
        ^+\tilde{\bra{k}} \\ ^-\tilde{\bra{k}}
    \end{pmatrix}
    \begin{pmatrix}
        \tilde{\ket{k}}^+ & \tilde{\ket{k}}^-
    \end{pmatrix}.
\end{align}
Note that the Gram matrix $\Gamma_k$, which consists of the linearly independent eigenstates $\tilde{\ket{k}}^\pm$, is positive-definite and Hermitian.
Let $\eta_k$ be a $2\times 2$ unitary matrix which diagonalizes $\Gamma_k$:
\begin{align}\label{eq:diagonalize_Gammak}
    \eta_k^\dag \Gamma_k \eta_k
    = \begin{pmatrix}
        \gamma_k^{(1)} & 0 \\
        0 & \gamma_k^{(2)}
    \end{pmatrix},
\end{align}
where $\gamma_k^{(1)},\gamma_k^{(2)}\in\mathbb{R}$ are the eigenvalues of $\Gamma_k$ such that $0<\gamma_k^{(1)}\le\gamma_k^{(2)}$.
Then $\tilde{\ket{k}}^1$ and $\tilde{\ket{k}}^2$ defined by
\begin{align}\label{eq:tildekalpha}
    \tilde{\ket{k}}^\alpha
    \coloneqq \frac{(\eta_k)_{1,\alpha}\tilde{\ket{k}}^+ + (\eta_k)_{2,\alpha}\tilde{\ket{k}}^-}{\sqrt{\gamma_k^{(\alpha)}}}
    \quad
    (\alpha=1,2)
\end{align}
are the eigenstates of $\AIIdag{}$ associated with the same eigenvalue, and are orthonormal:
\begin{align}
\label{orthonormal}
    \begin{pmatrix}
        ^1\tilde{\bra{k}} \\ ^2\tilde{\bra{k}}
    \end{pmatrix}
    \begin{pmatrix}
        \tilde{\ket{k}}^1 & \tilde{\ket{k}}^2
    \end{pmatrix}
    = \begin{pmatrix}
        1 & 0 \\
        0 & 1 
    \end{pmatrix}.
\end{align}

We arrange the eigenstates $\tilde{\ket{k}}^1$ and $\tilde{\ket{k}}^2$ to construct a $2L\times 2L$ regular matrix $\tilde{V}$ which diagonalizes $\AIIdag{}$ as follows:
\begin{align}\label{eq:tildeV}
    \tilde{V}
    \coloneqq \begin{pmatrix}
        \tilde{\ket{1}}^1 & \tilde{\ket{1}}^2 & \tilde{\ket{2}}^1 & \tilde{\ket{2}}^2 & \cdots & \tilde{\ket{L}}^1 & \tilde{\ket{L}}^2
    \end{pmatrix}.
\end{align}
This choice of $\tilde{V}$ obeys the normalization condition \eqref{eq:normalization} due to Eq.(\ref{orthonormal}). 

As proven in Appendix \ref{sec:estimate_AIIdag}, the condition number $\kappa(\tilde{V})$ has the following bound:
\begin{align}\label{eq:estimate_AIIdag}
    \begin{dcases*}
        \begin{aligned}
            &\sqrt{\frac{g-\Delta}{g+\Delta}}\left(\frac{2}{L+1}\right)^{3/2}(r^-)^{L-1} \\
            &< \kappa(\tilde{V})
            \le \frac{g+\Delta}{g-\Delta}\sqrt{L}(r^-)^{L-1}
        \end{aligned}
         & if $g>\Delta$ and OBC\\
        \sqrt{\frac{g-\Delta}{g+\Delta}}
        \le \kappa(\tilde{V})
        \le \frac{g+\Delta}{g-\Delta} & otherwise
    \end{dcases*}
\end{align}
with $r^-=\sqrt{\dfrac{t+\sqrt{g^2-\Delta^2}}{t-\sqrt{g^2-\Delta^2}}}$, which indicates that if $g>\Delta$ then the ratio $\kappa_\mathrm{R}=\kappa_{\mathrm{OBC}}/\kappa_{\mathrm{PBC}}$ is of the order $O(e^{L\ln r^-})$, and that if $g<\Delta$ then $\kappa_{\mathrm{R}}$ is the order $O(1)$, with respect to $L$.
Such an enhancement transition of $    \kappa_\mathrm{R}$ at $g=\Delta$ is compatible with the presence/absence-transition of the symmetry-protected skin effect in the model \eqref{41-1}; the skin effect occurs and $\kappa_\mathrm{R}\sim O(e^{L\ln r^-})$ when $g>\Delta$, while it does not occur and $\kappa_\mathrm{R}\sim O(1)$ when $g<\Delta$.

\subsubsection{departure from normality}
The energy eigenvalues of $H(k)$ in Eq.(\ref{41-1}) are given by 
\begin{equation}
    E_{\pm}(k)=2t\cos k\pm 2i\sqrt{g^2-\Delta^2}\sin k,
\end{equation}
where $k$ is the corresponding crystal momentum. Thus, under the PBC, we can evaluate the second term of Eq.(\ref{depp}) as 
\begin{align}
    \sum_{j}|E_j|^2
    &\to \frac{L}{2\pi}\int_0^{2\pi}dk\left[|2t\cos k+2i\sqrt{g^2-\Delta^2}\sin k|^2\right.\notag\\
    &\qquad\qquad\qquad\left.+|2t\cos k-2i\sqrt{g^2-\Delta^2}\sin k|^2\right]\notag\\
    &=\left\{
\begin{array}{ll}
    4L(t^2+g^2-\Delta^2)     &  \mbox{for $g>\Delta$}\\
 4L(t^2+\Delta^2-g^2)   & \mbox{for $g\le\Delta$}
\end{array}
\right.
.
\label{q1}
\end{align}
We also have the square of the singular value $s(k)$ of $H(k)$ as the eigenvalues of
$H(k)H(k)^{\dagger}$
\begin{align}
    H(k)H(k)^{\dagger}&=4t^2\cos^2k+4\Delta^2\sin^2 k+4g^2\sin^2 k\notag\\
    &+8\Delta t\cos k\sin k\sigma_x -8g\Delta \sin^2 k\sigma_y,
\end{align}
which leads to
\begin{align}
    (s(k))^2&=4t^2\cos^2 k+4\Delta^2\sin^2 k+4g^2\sin^2 k\notag\\
    &\pm\sqrt{64\Delta^2 t^2\cos^2 k\sin^2 k+64g^2\Delta^2\sin^4 k}.
\end{align}
Therefore, we can evaluate the first term of Eq.~(\ref{depp}) under the PBC as
\begin{align}
    &\sum_j(s_j)^2\notag\\
    &\to\frac{L}{2\pi}\int_0^{2\pi}dk (8t^2\cos^2 k+8\Delta^2 \sin^2 k +8g^2\sin^2 k)\notag\\
    &=4L(t^2+\Delta^2+g^2).\label{q2}
\end{align}
From Eqs.(\ref{q1}) and (\ref{q2}), we obtain the departure from normality (\ref{depp}) of the model (\ref{41-1}) under the PBC:
\begin{align}
\mathrm{dep}_{\mathrm{PBC}}(H)
=\left\{
\begin{array}{ll}
 2\sqrt{2}\Delta\sqrt{L}    & \mbox{for $g>\Delta$} \\
 2\sqrt{2}g\sqrt{L}    & \mbox{for $g\le \Delta$}     
\end{array}
\right.
    \label{q3}.
\end{align}

Now consider the OBC.
 To calculate the departure from normality, we use the result of the non-Bloch band theory for symplectic class \cite{r9.1-kos}.
As shown in Ref.\cite{r9.1-kos}, the energy spectrum of the model in Eq.(\ref{41-1}) under the OBC is doubly degenerate and given by
\begin{align}
    E(\beta)=
    t(\beta+\beta^{-1})+\sqrt{g^2-\Delta^2}(\beta-\beta^{-1}),
 \end{align}
where $\beta$ is
\begin{align}
    \beta=\sqrt{\left|\frac{t-\sqrt{g^2-\Delta^2}}{t+\sqrt{g^2-\Delta^2}}\right|}e^{i\theta},
\quad
\theta\in[0,2\pi].
\end{align}

Therefore, the second term of Eq.~(\ref{depp}) is evaluated as
\begin{align}
    &\sum_j|E_j|^2\notag\\
    &\to\frac{L}{2\pi}\int_0^{2\pi}d\theta\left[2\left|t(\beta+\beta^{-1})
    +\sqrt{g^2-\Delta^2}(\beta-\beta^{-1})\right|^2\right]\notag\\
&=4L(t^2+\Delta^2-g^2),
%    &=\left\{
%    \begin{array}{ll}
%    4L(t^2-g^2+\Delta^2) & \mbox{for $g>\Delta$}\\
%     4L(t^2+\Delta^2-g^2) & \mbox{for $g\le \Delta$}  
%    \end{array}
%\right.
%,
    \label{q4}
\end{align}
where we have used the fact that the integrations of $\beta(\beta^{-1})^*$ and $\beta^*\beta^{-1}$ over $\theta$ become zero since they are periodic in $\theta$.  
As the singular value
of the Hamiltonian is independent of the boundary conditions in the
infinite volume limit, Eq.(\ref{q2}) also gives the
first term of Eq.(\ref{depp})  under the OBC,
\begin{align}
    \sum_j(s_j)^2
    \to L(4t^2+4\Delta^2+4g^2).\label{q5}
\end{align}
Therefore, the departure from normality (\ref{depp}) of the model (\ref{41-1}) under the OBC is given by
\begin{align}
    \mathrm{dep}_{\mathrm{OBC}}(H)=
    2\sqrt{2}g\sqrt{L}.
    \label{q5.5}
\end{align}

From Eqs.(\ref{q3}) and (\ref{q5.5}), we have 
\begin{align}
{\rm dep}_{\rm D}=
\left\{
\begin{array}{cl}
2\sqrt{2}(g-\Delta)\sqrt{L}& \mbox{for $g>\Delta$}\\
0 & \mbox{for $g\le \Delta$}
\end{array}
\right.
.
\end{align}
Thus, the enhancement of  non-normality under OBC occurs only in the presence of the skin effect.

\section{Disorder-induced topological phase transition}
\label{sec:ditpt}
As shown in Sec.\ref{subsec:Anderson}, disorders induce the Anderson localization that shrinks the loop part of the complex spectrum and makes wings outside of the loop under the PBC. When the strength of the disorder exceeds the critical value, the loop under the PBC completely collapses, and skin modes disappear under the OBC.  
In this section, we show that vanishing enhanced non-normality correctly characterizes the disorder-induced topological phase transition.

\subsection{Exact result}
\label{sec:exact}

Before examining the disorder-induced topological phase transition in terms of enhanced non-normality, we derive here the exact result of the critical disorder strength for the transition.
We use the exact result to check the validity of the description for the transition by non-normality.

The matrix Hamiltonians we consider in this paper have the following common form
\begin{align}
H=H_0+W,
\end{align}
where $H_0$ is a disorder-independent part and $W$ is a disorder potential.
We also assume that the disorder potential $W$ is onsite, $W_{ij}=w_{i}\delta_{ij}$, where $w_i$ follows the Cauchy distribution in Eq.(\ref{cauchy}).
As shown in Appendix~\ref{sec:exa}, for the Cauchy distribution, 
we can take the disorder average exactly and obtain the exact effective Hamiltonian~\cite{r98-fz-1997,r99-bz-1998,r100-zee-world-1998,r101-fz-1999},  
\begin{align}
H_{\rm eff}=H_0-i\gamma{\rm sgn}[{\rm Im}E],
\end{align}
which gives a pole of the Green function in the complex energy plane $E$. 
The effective Hamiltonian provides a loop shape spectrum in the complex energy plane under the PBC, thus responsible for the non-Hermitian skin effect.
Hence, we can determine the critical strength of the disorder at which the loop spectrum shrinks to a point. 
  
\begin{figure}[tbp]
 \begin{center}
  \includegraphics[scale=0.4]{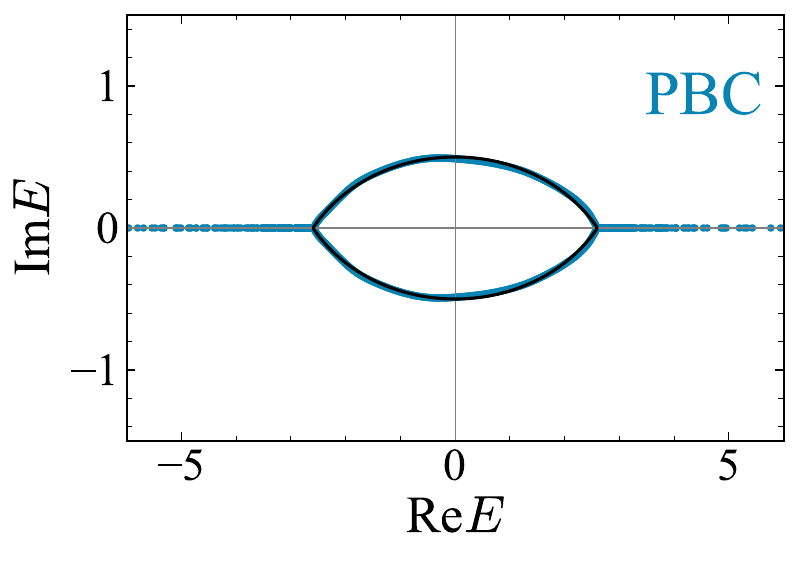}
  \caption{The blue dots represent the numerical PBC spectrum of Eq.(\ref{61-1}) with $t=1.5, g=0.5, L=1000$, and $\gamma=0.5$. The black curve represents the analytical form of the loop structure.
  \label{FIG5b}}
  \end{center}
\end{figure}

In the case of the Hatano-Nelson model in Eq.(\ref{61-1}), $H_0$ gives the PBC spectrum in Eq.(\ref{2-2-4-1-}), which is an ellipse in the complex plane:
\begin{equation}
\label{a7A}
    \frac{(\Re E)^2}{4t^2}+\frac{(\Im E)^{2}}{4g^2}=1.
\end{equation}
Therefore, $H_{\rm eff}$ gives
\begin{align}
\label{a10A}
    \begin{split}
        \frac{(\Re E)^2}{4t^2}+\frac{(\Im E-\gamma)^{2}}{4g^2}=1,\quad \mbox{for ${\rm Im}E>0$}\\
        \frac{(\Re E)^2}{4t^2}+\frac{(\Im E+\gamma)^{2}}{4g^2}=1,\quad \mbox{for ${\rm Im}E<0$}.
    \end{split}
\end{align}
Actually, the above analytical expression for the loop spectrum well reproduces the numerically obtained spectrum in Fig.~\ref{FIG5b}.
Then, when increasing the disorder strength $\gamma$, the loop of the PBC spectrum shrinks and eventually disappears if the two ellipses (\ref{a10A}) pass through the origin, namely if $\gamma=2g>0$. Therefore, the critical disorder strength $\gamma_c$ for the disorder-induced phase transition is
\begin{align}
\gamma_c=2g.\label{a11A}
\end{align}

We also obtain the same result from consideration of the winding number in Eq.(\ref{21-1}). 
To evaluate this number in the present case, we first note that Eq.(\ref{21-1}) can be rewritten as
\begin{align}
W(E)= \int_{-\pi}^{\pi}\frac{dk}{2\pi i}
\frac{\partial}{\partial k}\log \det [(G_k(E))^{-1}],  
\end{align}
where $G_k(E)$ is the Green function in the momentum space, 
\begin{align}
G_k(E)=\frac{1}{E-H(k)}.    
\end{align}
Then, in the presence of disorder, we define the winding number as
\begin{align}
W(E)= \int_{-\pi}^{\pi}\frac{dk}{2\pi i}
\frac{\partial}{\partial k}\log \det (\left<G_k(E)\right>^{-1}),  
\end{align}
where $\left<G_k(E)\right>$ is the disorder averaged Green function.
As shown in Appendix~\ref{sec:exa}, for the Cauchy distribution, we have
\begin{align}
\left<G(E)\right>=\frac{1}{E+i\gamma{\rm sgn}[{\rm Im}E]-H_0}, 
\end{align}
and thus, for the Hatano-Nelson model, we have
\begin{align}
\left<G_k(E)\right>=\frac{1}{E+i\gamma{\rm sgn}[{\rm Im}E]-H_0(k)}, 
\end{align}
where $H_0(k)=(t+g)e^{-ik}+(t-g)e^{ik}$. 
Then, it is easy to see that 
$W(E)=1$ for $E$ inside the loop spectrum in the complex energy plane if $\gamma<2g$, while $W(E)=0$ for any $E$ if $\gamma>2g$. 

\begin{figure}[tbp]
 \begin{center}
  \includegraphics[scale=0.35]{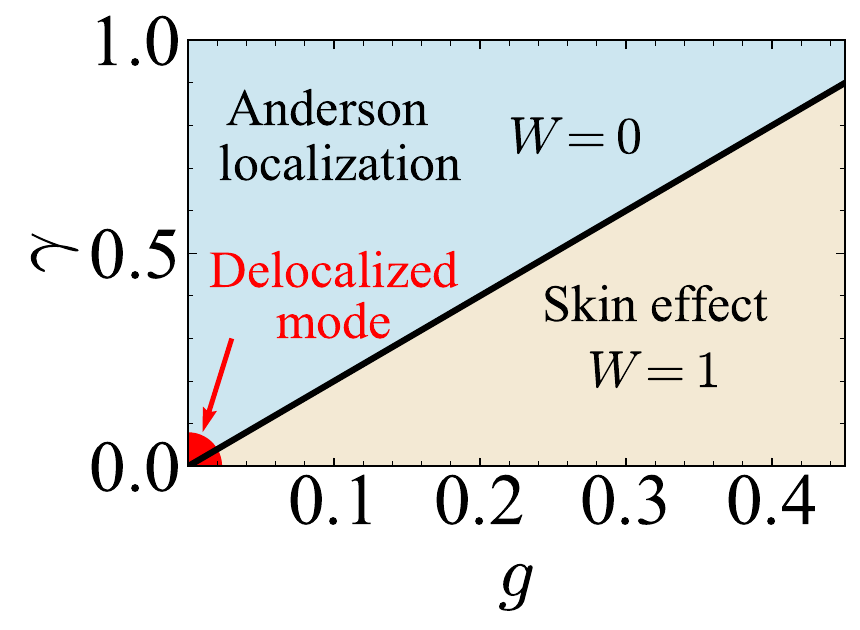}
  \caption{The phase diagram of the OBC Hatano-Nelson model. The blue and yellow regions correspond to the phase where all eigenstates are Anderson modes ($W=0$) and the phase where skin modes exist ($W=1$), respectively. In addition, the red point represents only the origin. At this point, both Anderson localization and skin effect do not occur.  \label{FIG5}}
  \end{center}
\end{figure}

We summarize the phase diagram of the Hatano-Nelson model with the Cauchy distribution disorder in Fig.~\ref{FIG5}: 
All modes show the Anderson localization in the region $\gamma>2g$, while the non-Hermitian skin effect occurs in the region $\gamma<2g$. In addition, if $\gamma=g=0$, both Anderson localization and skin effect do not occur.

In a manner similar to the above, we can also derive the critical strength of disorder in the time-reversal invariant Hatano-Nelson model in Eq.(\ref{a4}). 
Generalizing the $\mathbb{Z}_2$ number in Eq.(\ref{10-13-0-}) as

\begin{align}
  %\label{10-13-0-}
  &(-1)^{\nu(E)}\notag\\
  &=\mathrm{sgn}\left[\frac{\mathrm{Pf}[\left<G_{k=\pi}(E)\right>^{-1}T]}{\mathrm{Pf}[\left<G_{k=0}(E)\right>^{-1}T]}\right.\notag\\
  &\times\exp\left.\left[-\frac{1}{2}\int_{k=0}^{k=\pi}dk\frac{\partial}{\partial k}\log \det[\left<G_k(E)\right>^{-1}T]\right] \right],
\end{align}
we find that the critical strength $\gamma_c$ of this model is 
\begin{align}
\gamma_c=2\sqrt{g^2-\Delta^2}.    
\end{align}
Actually, if $\gamma<2\sqrt{g^2-\Delta^2}$, we have $(-1)^{\nu(E)}=-1$ for $E$ inside the loop spectrum and the symmetry-protected non-Hermitian skin effect occurs, while for $\gamma>2\sqrt{g^2-\Delta^2}$, $(-1)^{\nu(E)}=1$ for any $E$ and the skin effect does not occur.

%\color{red}

\subsection{Enhanced non-normality}
\label{sec:enn}

Here we numerically examine the disorder-induced phase transitions of the Hatano-Neloson model (\ref{61-1}) and its time-reversal invariant variant (\ref{a4}) in terms of enhanced non-normality.

Since the disorder-induced phase transitions are topological, they should be examined by the relevant bulk topological numbers. 
However, the topological numbers are not suitable to numerically examine the phase transition, whereas one can define the bulk topological numbers even in the presence of disorder, as shown in Appendix \ref{sec:invtwiA}.
This is because the topological numbers depend on the reference energy $E$, and the phase transition requires the vanishing of the topological numbers for any $E$. 
In contrast, the scalar measures of non-normality introduced in Sec.\ref{enhance} are independent of the reference energy $E$, and thus no such problem exists.

\subsubsection{Hatano-Nelson model}

\begin{figure*}[tbp]
 \begin{center}
  \includegraphics[scale=0.12]{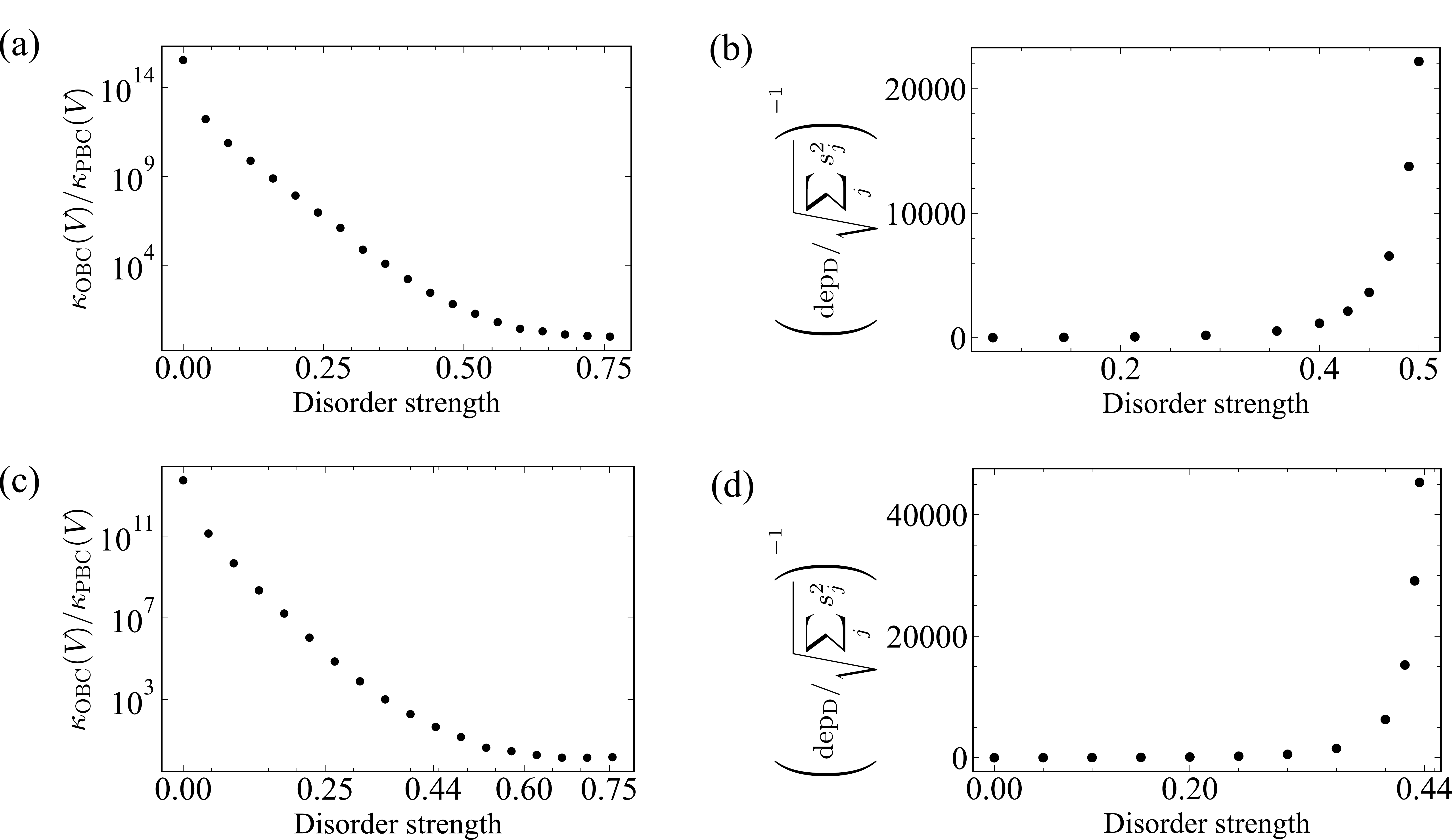}
  \caption{\mbox{(a)(c)} Disorder dependence of $\kappa_{\mathrm{R}}=\kappa_{\mathrm{OBC}}(V)/\kappa_{\mathrm{PBC}}(V)$. We numerically calculate $\kappa_{\mathrm{OBC}}(V), \kappa_{\mathrm{PBC}}(V)$, and their ratio $\kappa_{\mathrm{OBC}}(V)/\kappa_{\mathrm{PBC}}(V)$. We perform these calculations using one hundred different configurations of random potential and calculate the geometric mean of these one hundred obtained ratios. We plot these geometric means in these figures. \mbox{(b)(d)} Disorder dependence of $\left(\mathrm{dep_{D}}/\sqrt{\sum_j s_j^2}\right)^{-1}$. We numerically calculate $\left(\mathrm{dep_{D}}/\sqrt{\sum_j s_j^2}\right)^{-1}$. We perform this calculation using five different configurations of random potential and calculate the arithmetic mean of these five obtained differences. We plot these arithmetic means in these figures. 
  \mbox{(a)(b)} Hatano-Nelson model (\ref{61-1}) with $t=1.75$ $g=0.25$. (a) $L=250$. (b) $L=1000$. \mbox{(c)(d)} Time-reversal invariant Hatano-Nelson model (\ref{a4}) with $t=1, g=0.3, \Delta=0.2$. (c) $L=140$. (d) $L=1000$.
   \label{FIG11}}
  \end{center}
\end{figure*}

First, we show how the ratio $\kappa_{\rm R}$ of the condition numbers under the OBC and the PBC behaves when changing the disorder strength $\gamma$ of the Hatano-Nelson model.
%In the numerical calculation, we take $t=1.5$, $g=0.25$, and $L=1000$. 
We use the model in Eq.(\ref{61-1}) with $t=1.75$ $g=0.25$, and $L=250$, 
and adapt the disorder following the Cauchy distribution. 
The obtained $\gamma$-dependence of the condition number is shown in Fig.~\ref{FIG11}(a), where the ratio $\kappa_{\rm R}$ is exponentially enhanced when $\gamma<0.6$ while it becomes $O(1)$ for $\gamma\ge 0.6$. 
As shown in Fig.~\ref{figa13-1} (a), for $\gamma<0.6$, we also confirm the exponential growth against the system size $L$.
This behavior is consistent with Eq.(\ref{eq:krscale}) and the fact that the skin effect occurs in the smaller value region of $\gamma$.
The numerically obtained critical strength $\gamma_c\sim 0.6$ is close to the exact value $\gamma_c=2g=0.5$. 
We also confirmed that the numerically obtained critical value approaches to the exact value when increasing the system size $L$, and thus the deviation comes from the finite size effect.

In Fig.~\ref{FIG11}(b), we also illustrate how the departure from normality behaves when changing the disorder strength.
The data shows that ${\rm dep}_{\rm D}$ goes to zero when $\gamma$ goes to the critical value $\gamma_c=2g=0.5$.
Moreover, the departure from normality behaves as the square root of the system size if $\gamma_c>\gamma$, while it goes to $O(1)$ if $\gamma_c<\gamma$, as shown in Fig.~\ref{figa13-1}(d).
This behavior is also consistent with Eq.(\ref{eq:ddscale}). 
Therefore, the enhanced non-normality correctly describes the disorder-induced topological phase transition.

\begin{figure}[tbp]
 \begin{center}
  \includegraphics[scale=0.415]{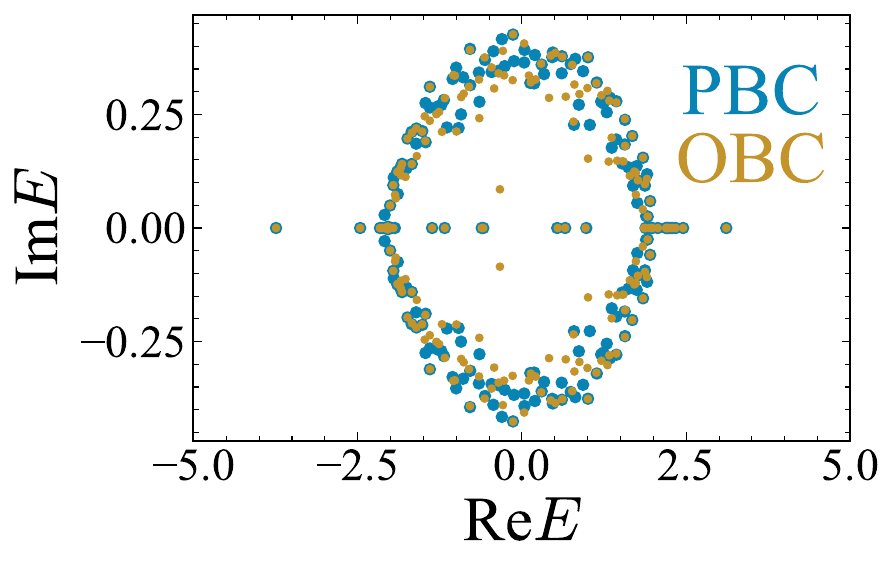}
  \caption{The blue and yellow dots represent the PBC and the OBC spectra of the model (\ref{a2bb}) with time-reversal symmetry broken disorder, respectively ($t=1, g=0.3, \Delta=0.2,\gamma=0.1$, and $L=100$). The disorder follows the Cauchy distribution. \label{figDDD-0.1}}
  \end{center}
\end{figure}

\begin{figure*}[tbp]
 \begin{center}
  \includegraphics[scale=0.15]{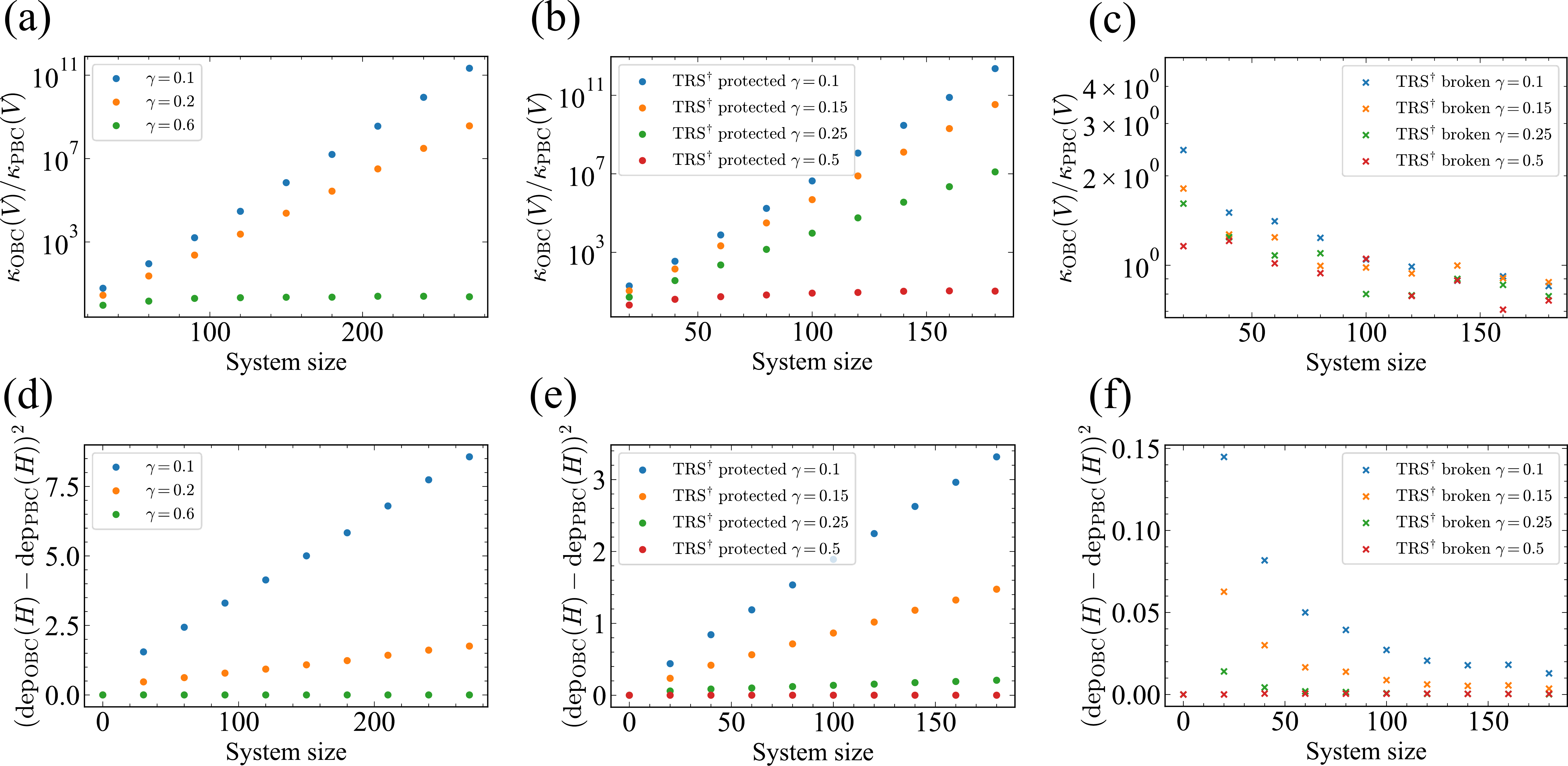}
  \caption{System size dependence of non-normality. (a) (d) Non-normality of the Hatano-Nelson model (\ref{61-1}) with $t=1.75$ and $g=0.25$. (b) (e) Non-normality of the time-reversal invariant Hatano-Nelson model (\ref{a4}) with $t=1, g=0.3$, and $\Delta=0.2$. (c) (f) Non-normality of the time-reversal symmetry broken Hatano-Nelson model (\ref{a2bb}) with $t=1, g=0.3$, and $\Delta=0.2$.
    \label{figa13-1}}
  \end{center}
\end{figure*}

\subsubsection{Time-reversal invariant Hatano-Nelson model}

We also numerically examine the time-reversal invariant Hatano-Nelson model in Eq.(\ref{a4}).
As shown in Sec.\ref{sec:exact}, the exact result tells us that the system exhibits the disorder-induced topological phase transition at $\gamma_c=2\sqrt{g^2-\Delta^2}$.
Then, for $\gamma<2\sqrt{g^2-\Delta^2}$ ($\gamma>2\sqrt{g^2-\Delta^2}$), the system is topologically non-trivial (trivial) and the symmetry-protected non-Hermitian skin effect occurs (does not occur).

In Fig.~\ref{FIG11}(c), 
we illustrate the disorder dependence of $\kappa_{\rm R}$, where we take the model parameter in Eq.(\ref{a4}) as $t=1$, $g=0.3$, $\Delta=0.2$, and $L=140$ and consider the disorder with the Cauchy distribution. 
Our numerical calculation shows that $\kappa_{\rm R}$ is enhanced for $\gamma<0.55$ while it remains $O(1)$ for $\gamma>0.55$. 
The obtained critical strength $\gamma_c\sim 0.55$ is consistent with the exact value $\gamma_c=2\sqrt{g^2-\Delta^2}=0.44$ derived in Sec.\ref{sec:exact}, where
the deviation originates from the finite size effect.
In Fig.~\ref{figa13-1}(b), we also show the size dependence of $\kappa_{\rm R}$, which is consistent with our prediction in Eq.(\ref{eq:krscale}).

Fig.~\ref{FIG11}(d) shows disorder dependence of the departure from normality.
We find that the inverse of ${\rm dep}_{\rm D}$ is very tiny when $\gamma<0.44$, which implies the enhancement of non-normality under OBC in the presence of the symmetry-protected non-Hermitian skin effect.
Then, it diverges as $\gamma$ goes to the critical value $\gamma_c=0.44$, 
meaning that the enhancement of non-normality disappears at the phase transition, where the system becomes topologically trivial. 
In accordance with our general result in Eq.(\ref{eq:ddscale}), 
the departure from normality behaves as the square root of the system size for $\gamma<0.44$, while it goes to $O(1)$ for $\gamma>0.44$, as shown in Fig.~\ref{figa13-1}(c).

\subsubsection{Absence of average symmetry-protected skin effect}
\label{absence}

It has been known that symmetry-protected topological modes may survive even when the relevant symmetry is locally broken as long as the breaking vanishes globally on average \cite{FK12,RBM12,RKYS12,FvHEA14}.
As an application of enhanced non-normality, we also examine this problem for the symmetry-protected non-Hermitian skin effect.
Interestingly, contrary to the above wisdom, 
our numerical calculations reveal that average symmetry protection does not work at all for the symmetry-protected non-Hermitian skin effects. 

%Here we examine the effect of disorders with average time-reversal symmetry in Eq.(\ref{trsd}).
For this purpose, we consider the model in Eq.(\ref{a2d}) with the following disorders
\begin{align}
  \hat{H}&=\sum_{j=1}^{L}\left[
  (t+g)\hat{c}_{j+1,\uparrow}^{\dagger}\hat{c}_{j,\uparrow}+(t-g)\hat{c}_{j,\uparrow}^{\dagger}\hat{c}_{j+1,\uparrow}
  +w_j^{\uparrow}\hat{c}_{j,\uparrow}^{\dagger}\hat{c}_{j,\uparrow}\right]
  \notag\\
  &+\sum_{j=1}^{L}\left[
  (t+g)\hat{c}_{j,\downarrow}^{\dagger}\hat{c}_{j+1,\downarrow}
  +(t-g)\hat{c}_{j+1,\downarrow}^{\dagger}\hat{c}_{j,\downarrow}
  +w_j^{\downarrow}\hat{c}_{j,\downarrow}^{\dagger}\hat{c}_{j,\downarrow}\right]\notag\\
  &-i\Delta\sum_{j=1}^{L}(\hat{c}_{j+1,\uparrow}^{\dagger}\hat{c}_{j,\downarrow}-\hat{c}_{j\uparrow}^{\dagger}\hat{c}_{j+1\downarrow})\notag\\
  &-i\Delta\sum_{j=1}^{L}(\hat{c}_{j+1,\downarrow}^{\dagger}\hat{c}_{j,\uparrow}-\hat{c}_{j,\downarrow}^{\dagger}\hat{c}_{j+1,\uparrow}),
\label{a2bb}
\end{align}
where $w_j^{\uparrow}$ and $w_j^{\downarrow}$ are the onsite disorder with the Cauchy distribution,
\begin{align}
    \label{dis1}
    P(w_j^{\uparrow})=\frac{\gamma}{\pi}\frac{1}{(w_j^{\uparrow})^2+\gamma^2},
\end{align}
and
\begin{align}
    \label{dis2}
    P(w_j^{\downarrow})=\frac{\gamma}{\pi}\frac{1}{(w_j^{\downarrow})^2+\gamma^2}.
\end{align}
Since the disorder potentials $w_j^\uparrow$ and $w_j^\downarrow$ are independent, they locally break time-reversal symmetry in Eq.(\ref{trsd}).
However, their average values vanish, and thus they keep time-reversal symmetry on average. 

In Fig.~\ref{figDDD-0.1}, we show typical complex spectra of the above system under the PBC and the OBC. 
Whereas this result suggests that the symmetry-protected skin effect immediately disappears in the presence of such disorders, we need a more qualitative characterization to establish the disappearance. 

Using the enhanced non-normality, we can establish the absence of average symmetry protection of the non-Hermitian skin effect.  
%Correspondingly, the non-normality is not strongly enhanced as in the cases of symmetry-protected non-Hermitian skin effects,as shown in Fig.~\ref{figa13-1
In Figs.~\ref{figa13-1}(c) and (f), we show the size dependence of the enhanced non-normality for the system with average time-reversal symmetry.
In contrast to Figs. \ref{figa13-1} (b) and (e), the enhancement of non-normality 
quickly disappears when increasing the system size $L$.
Therefore, we conclude that average time-reversal symmetry is not sufficient, and exact time-reversal symmetry is indispensable for the symmetry-protected skin effect.

Now let us discuss why the average symmetry fails to protect the non-Hermitian skin modes.
As discussed in Sec.\ref{sec:spse}, using the doubled Hamiltonian in Eq. (\ref{eq:doubled_Hamiltonian}), we can map the skin modes to topological Majorana end states of a Hermitian system.
For the topological end states of the Hermitian system, the average symmetry protection works as expected:
Because the disorder term fluctuates along the system so that the average value becomes zero,  
the topological end states forming a Kramers pair rarely mix with each other. Therefore, the topological end states remain to be at least approximately zero energy states.
However, the correspondence between the topological end states and the skin modes requires exactly zero energy on the Hermitian side, and thus the skin modes are not protected by the average symmetry.
We also know that the time-reversal invariant superconductor described by the doubled Hamiltonian may have additional symmetry, from which they can keep exact zero-energy Majorana end states under particular time-reversal breaking terms \cite{KYYS21}.
However, the time-reversal breaking terms in superconductors are essentially different from those in non-Hermitian systems: The former keep particle-hole symmetry, while the latter respect chiral symmetry in Eq.(\ref{eq:chiral}). 
Thus, the response to time-reversal breaking terms can be different between them.
Therefore, any time-reversal breaking terms may destabilize the symmetry-protected skin modes in Eq.(\ref{a2d}).

\section{Implications}\label{sec:Implications}

Finally, we would like to discuss the physical meaning of enhanced non-normality.
Remarkably, the enhanced non-normality directly affects the so-called pseudo-spectrum of the system, 
governing the dynamical properties of the system.

Let $\sigma(H)$ be the spectrum of $H$. 
Then, the pseudo-spectrum $\sigma_\epsilon(H)$ is defined as the set of the complex energy $E\in\mathbb{C}$ in the spectrum $\sigma(H+\Delta H)$  for some $\Delta H$ with $\|\Delta H\|_2<\epsilon$. 
%where $\sigma(H+\Delta H)$ is the conventional spectrum of $H+\Delta H$.
In other words, it describes the change of the spectrum of $H$ under a perturbation $\Delta H$ with $\|\Delta H\|_2<\epsilon$.
%The pseudo-spectrum $\sigma_\epsilon(H)$ 
%is the union of the spectra of perturbed Hamiltonians in the form of  $H+E$ with any perturbation $H$ satisfying $\|E\|<\epsilon$.  
Notably, the pseudo-spectrum is equivalently defined as the set of approximate energy eigenvalues $E$ of $H$
with the accuracy 
\begin{align}
\|(H-E)|u\rangle\|_2<\epsilon,
\quad \| |u\rangle\|_2=1.
\end{align}
Therefore, at the same time, it provides approximate energy eigenstates of $H$, and governs the dynamics of the system with accuracy $\epsilon$. 

For a normal $H$, the pseudo-spectrum is just the $\epsilon$-neighborhood of the original spectrum \cite{r67-trefethen-book-2005}:
\begin{align}
    \sigma_{\epsilon}(H)=\sigma(H)+\Delta_\epsilon,
\end{align}
where $\Delta_{\epsilon}$ is the complex energy region $E\in\mathbb{C}$ with the radius less than $\epsilon$:
\begin{align}
    \Delta_\epsilon=\{E\in\mathbb{C}:|E|<\epsilon\}.
\end{align}
This means that a small perturbation only has a small effect on the normal system.
However, for a non-normal $H$, very different behavior may appear:
The pseudo-spectrum can go beyond the $\epsilon$-neighborhood of the unperturbed spectrum:
\begin{align}
    \sigma_{\epsilon}(H)\supseteq\sigma(H)+\Delta_\epsilon
\end{align}
(see Appendix \ref{app:pss}).
Therefore, a small perturbation may change the spectrum drastically, and the dynamical properties may differ extensively from those expected from the original spectrum.

A crucial question here is how the pseudo-spectrum can deviate from the conventional one. 
The answer is given by the Bauer-Fike theorem.  
As shown in Appendix \ref{app:pss}, 
the upper bound for the deviation is given by the scalar measures of non-normality, 
\begin{align}
\sigma_{\epsilon}(H)\subseteq \sigma(H)+\Delta_{\epsilon\kappa(V)},    
\label{eq:BF}
\end{align}
or
\begin{align}
    \sigma_{\epsilon}(H)\subseteq\sigma(H)+\Delta_{\epsilon+\mathrm{dep}_{\mathrm{F}}(H)}.
\label{eq:BF2}
\end{align}
Hence, the enhanced non-normality due to the non-Hermitian skin effects allows much more deviation of the pseudo-spectrum from the original.

%For $\epsilon\ll 1$, Eq.(\ref{app:pss}) gives a more severe upper bound than Eq.(\ref{eq:BF2}). 
For $\epsilon\ll 1$, Eq.(\ref{eq:BF}) gives a more severe upper bound than Eq.(\ref{eq:BF2}). 
Since the skin effects exhibits $\kappa(V)\sim e^{cL}$ with a positive constant $c$, 
a very tiny perturbation $\Delta H$ with
\begin{align}
\|\Delta H\|_2\sim 1/\kappa(V)\sim e^{-cL}    
\end{align}
may give an $O(1)$ correction of the spectrum. 
Such an extraordinary sensitivity against a perturbation applies to highly sensitive sensors, as discussed in Ref.\cite{budich-Bergholtz2020}.

The enhanced non-normality also affects the time-dependent dynamics of the system.
For the diagonalizable $H$ in Eq.(\ref{eq:vhv}), the time-evolution operator $e^{-iHt}$ obeys
\begin{align}
\|e^{-iHt}\|_2&=\|Ve^{-i\Lambda t}V^{-1}\|_2
\nonumber\\
&\le \|V\|_2 \|e^{-i\Lambda t}\|_2\|V^{-1}\|_2
\nonumber\\
&= \kappa(V) e^{-\Delta(H)t},
\label{eq:gap}
\end{align}
where $\Lambda$ is the diagonal matrix in the right hand side of Eq.(\ref{eq:vhv}), and $\Delta(H)=-{\rm max}_n[{\rm Im}E_n^a]$.
Thus, if $\Delta(H)>0$, the enhanced non-normality $\kappa(V)\sim e^{cL}$ implies that 
the upper bound in Eq.(\ref{eq:gap}) 
goes to $O(e^{-1})$ at the relaxation time $\tau$ 
\begin{align}
\tau\sim \frac{1}{\Delta(H)}+\frac{cL}{\Delta(H)}.
\end{align}
Thus, the relaxation time $\tau$ becomes much longer than the usually expected value $1/\Delta(H)$ from the spectrum because of the enhanced non-normality. Such a length-dependent relaxation time caused by the skin effect was discussed in Ref.\cite{r22-haga-liouvilian-2021}.

Whereas Eq.(\ref{eq:gap}) gives a good reference for the dynamics as discussed above, it has been known that the pseudo-spectrum provides a much sharper bound \cite{r67-trefethen-book-2005}, 
\begin{align}
\|e^{-iHt}\|_2\le  \frac{L_\epsilon e^{\alpha_{\epsilon}(H)t}}{2\pi\epsilon},   
\end{align}
for any $t\ge 0$ and any $\epsilon>0$, where $L_\epsilon$ is the arc length of the boundary of $\sigma_{\epsilon}(H)$ and $\alpha_\epsilon(H)={\rm sup}[{\rm Im}\sigma_\epsilon(H)]$.
See Appendix \ref{app:pss} for the derivation.
Thus, the enhanced non-normality also affects the system's dynamics via the enlarged pseudo-spectrum. 

\section{CONCLUSION}
\label{conclusion}
In this paper, we reveal that the non-Hermitian skin effects enhance the non-normality under the OBC for a topological reason.
Correspondingly to state and spectrum changes of the non-Hermitian skin effects, we introduce two different scalar measures and prove that the non-Hermitian skin effects enhance the former for any case and the latter for relatively simple cases. 
Using concrete models, we also confirm both analytically and numerically the validity of the topologically enhanced non-normality as an order parameter of the topological phase transition for the non-Hermitian skin effects.
Furthermore, no average symmetry protection for the symmetry-protected skin effect is established in terms of enhanced non-normality.
The enhanced non-normality results in extraordinary spectrum sensitivity against perturbations and anomalous time-dependent evolution through the pseudo-spectrum and the Bauer-Fike theorem, and thus provides fundamentals in non-Hermitian physics.

\section*{Acknowledgment}
We thank H. Oka, T. Ando, and U. Miura for the helpful discussions.
This work was supported by JST CREST Grant No. JPMJCR19T2 and KAKENHI Grant No. JP20H00131.
Y.O.N. was supported by JST SPRING, Grant Number JPMJSP2110. D.N. was supported by JST, the establishment of university fellowships towards the creation of science technology innovation, Grant Number JPMJFS2123.
N.O. was supported by JSPS KAKENHI Grant No.~JP20K14373.

\appendix

\section{Enhanced non-normality as an order parameter of non-Hermitian skin effects}
\label{app:kappa}
In this section, we present an argument establishing the relationship between the scalars $\kappa_{\rm R}$ and ${\rm dep}_{\rm D}$ for enhanced non-normality and the non-Hermitian skin effects.
Below, we assume that the Hamiltonian has the translation invariance under the periodic boundary condition and the range of the hopping terms is finite, and thus, we can use the non-Bloch theory developed in Refs.\cite{r6.5-yao-wang-edge-2018,r9-yokomizo-murakami-nonbloch-2019}.

First,
we show that if no non-Hermitian skin effect occurs, $\kappa_{\rm R}$ behaves as 
\begin{equation}
\kappa_{\rm R}=o(e^{cL})        
\end{equation}
with a positive constant $c$ and a sufficiently large system size $L$. It should be noted that we use the small $o$ on the right-hand side.

To show this relation, we first notice that 
in the absence of a non-Hermitian skin effect, the non-Bloch theory tells us that the bulk energy eigenstates under the open boundary condition 
are identical to those under the periodic boundary condition in the large $L$ limit \cite{r9-yokomizo-murakami-nonbloch-2019}.
Thus, the bulk states give $\kappa_{\rm R}=O(1)$, and only the finite number of boundary states could provide a more significant contribution to $\kappa_{\rm R}$. Note that this conclusion is generally true even when the bulk spectrum contains exceptional points in the Brillouin zone.
The Hamiltonian becomes non-diagonalizable when the bulk spectrum contains exceptional points in the Brillouin zone, so $\kappa(V)=\infty$. 
However, for the finite-size lattice representation of the Hamiltonian, the momentum is discretized and can avoid the exceptional points by introducing a small perturbation of the Hamiltonian if necessary.   
Under this regularization, $\kappa(V)$ becomes finite under both periodic and open boundary conditions, and thus, the bulk contribution to $\kappa_{\rm R}$ becomes $O(1)$.

Now evaluate the contribution from the boundary states. 
From Eqs. (\ref{eq:bound}) and (\ref{eq:xi}) in the main text, to have a significant contribution to $\kappa_{\rm R}$, the norm of the left eigenstate $|E\rangle\!\rangle$ of a boundary state must be large under the normalizations in Eqs. (\ref{eq:normalization}) and (\ref{eq:biorthogonal}).
This happens if the left eigenstate $|E\rangle\!\rangle$ rarely overlaps with the right eigenstate $|E\rangle$. 
In this situation, $\langle\!\langle E|E\rangle\!\rangle$ must be large 
to satisfy the normalizations $\langle\!\langle E|E\rangle=1$ and $\langle E|E\rangle=1$.

There are two possibilities where the left and right eigenstates of a boundary state rarely overlap.
The first possibility comes when the left and right eigenstates $|E\rangle$ and $|E\rangle\!\rangle$ of a boundary mode are spatially separated. In this case,  they are localized at opposite boundaries.
However, as we see immediately, this situation contradicts our assumption that no non-Hermitian skin effect occurs:
If the left and right eigenstates $|E\rangle$ and $|E\rangle\!\rangle$ are localized at opposite boundaries, so are the following states,
\begin{equation}
\left(
\begin{array}{c}
0\\
|E\rangle 
\end{array}
\right),
\quad
\left(
\begin{array}{c}
|E\rangle\!\rangle\\
0
\end{array}
\right).
\label{eq: zero}
\end{equation}
Therefore, they give boundary zero modes of the Hermitian Hamiltonian
\begin{equation}
\tilde{H}=\left(
\begin{array}{cc}
 0    & H-E \\
H^\dagger-E^*     & 0
\end{array}
\right),
\label{eq: tilde_Hamiltonian}
\end{equation}
where $H$ is the non-Hermitian Hamiltonian considered and $E$ is the complex energy of the boundary mode. 
Since each of the zero modes in Eq.(\ref{eq: zero}) has a definite chirality for the chiral operator $\Sigma=\sigma_z$, it gives a nonzero chiral topological number for each boundary. 
Then, from the bulk-boundary correspondence for the Hermitian Hamiltonian $\tilde{H}$, the nonzero chiral topological number leads to a nonzero winding number of $\tilde{H}$ under the periodic boundary condition.
This result contradicts our assumption since the nonzero winding number implies the non-Hermitian skin effect, as shown in Ref.\cite{r11-okss}.

One might think that another boundary state can cancel the chiral topological number so that no contradiction appears. 
However, this is unlikely to occur.
First, we note that the other state should have exactly the same energy as the original boundary state in the large $L$ limit.
Otherwise, the second boundary state is not a zero mode of $\tilde{H}$ in Eq.(\ref{eq: tilde_Hamiltonian}), and thus it can not have a chiral topological number required for the cancellation. 
Therefore, we assume an exact degeneracy.
Then, let $|E^a\rangle$ and $|E^a\rangle\!\rangle$ $(a=1,2)$ denote the right and left eigenstates of these boundary states. 
They satisfy the normalization conditions,
\begin{equation}
\langle E^a|E^b \rangle=\delta_{ab},
\quad
\langle\!\langle E^a|E^b\rangle=\delta_{ab}.
\end{equation}
The cancellation of the chiral topological number also requires that $|E^1\rangle$ and $|E^2\rangle\!\rangle$ must be localized at the same boundary since $|E^a\rangle$ and $|E^a\rangle\!\rangle$ have the definite chiral topological number $\Sigma=1$ and $\Sigma=-1$, respectively. (See Eq.(\ref{eq: zero}).) 
Therefore, $|E^1\rangle$ and $|E^2\rangle\!\rangle$ must be orthogonal $\langle\!\langle E^2|E^1\rangle=0$, while they substantially overlap in space.
A physically reasonable reason for such orthogonality is symmetry: If these states have distinct values of a quantum number of a unitary symmetry or form a Kramers pair, they cannot mix.
However, both cases result in a symmetry-protected skin effect, which contradicts our assumption:
In the former case, using the unitary symmetry, we can block-diagonalize $H$, where each block has a definite quantum number of the symmetry. 
Since $|E^1\rangle$ and $|E^2\rangle\!\rangle$ have different quantum numbers, they belong to different blocks, so they give nonzero chiral topological numbers to each of the blocks they belong.
Therefore, from the bulk-boundary correspondence, these blocks have nonzero winding numbers for non-Hermitian skin effects, contradicting our assumption. 
In the latter case, $|E^2\rangle\!\rangle$ and $|E^1\rangle$ form a generalized Kramers pair in the form of
\begin{equation}
|E^2\rangle\!\rangle=\lambda U_T(|E^1\rangle)^*,    
\end{equation}
where $U_T$ is an antisymmetric unitary operator and $\lambda$ is a constant.
(Here $\lambda$ is necessary since we do not require $\langle\!\langle E^2|E^2\rangle\!\rangle= 1$.)
Whereas the transpose version of time-reversal symmetry $U_T H^TU_T^{-1}=H$ with $U_TU_T^*=-1$ ensures the generalized Kramers degeneracy, it also enables to relate the boundary states with the symmetry-protected non-Hermitian skin effects in class AII$^\dagger$ in Ref.\cite{r11-okss}. 
Therefore, the latter case also contradicts our assumption.

%According to the general theory for topological boundary zero modes in Hermitian systems, 
%each representation of these symmetries defines an emergent Altland-Zirnbauer (AZ) symmetry class, and if the emergent AZ symmetry class has a %one-dimensional topological number, then we have topological protection of the boundary zero modes. 

\begin{figure}[tbp]
 \begin{center}
  \includegraphics[scale=0.17]{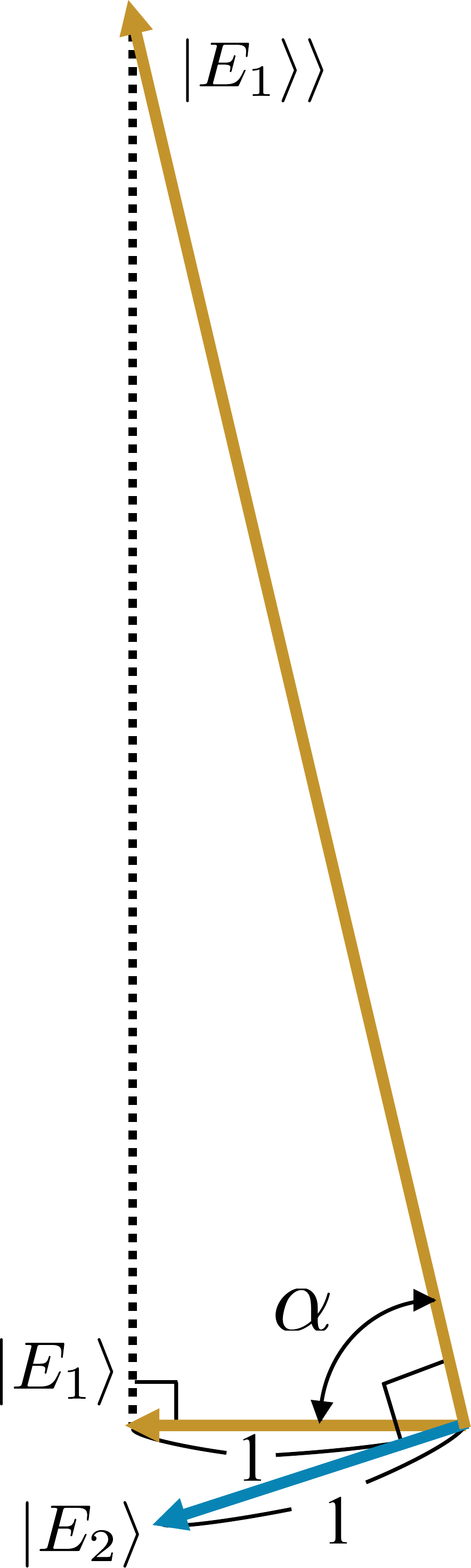}
  \caption{Three eigenstates ($|E^1\rangle$, $|E^1\rangle\!\rangle$, $|E^2\rangle$) under the normalization conditions ($\langle E^1|E^1\rangle=1$, $\langle E^2|E^2\rangle=1$) and the biorthogonal condition ($\langle\!\langle E^1|E^2\rangle=0$, $\langle E^1|E^1\rangle\!\rangle=1$). For almost parallel $|E^1\rangle$ and $|E^2\rangle$, the angle $\alpha$ is close to $\pi/2$, resulting in significantly large $\langle\!\langle E^1|E^1\rangle\!\rangle$.
  \label{Replyy}}
  \end{center}
\end{figure}

Having established that the first possibility contradicts our assumption, we now discuss the situation where the left and right eigenstates of any boundary state are localized at the same boundary.
A possible large contribution to $\kappa_{\rm R}$ in this situation
comes when two boundary states $|E^1\rangle$ and $|E^2\rangle$ are (almost) parallel because $\langle\!\langle E^1|E^1\rangle\!\rangle$ and $\langle\!\langle E^2|E^2\rangle\!\rangle$ in Eq.(38) must diverge to satisfy the biorthogonality
\begin{equation}
\langle\!\langle E^1|E^2\rangle=0, 
\quad        
\langle\!\langle E^2|E^1\rangle=0.
\end{equation}
See Fig.\ref{Replyy}.
Whereas this situation typically occurs when boundary states are about to form an exceptional point, the divergence should rarely depend on $L$ because $|E^1\rangle$ and $|E^2\rangle$ are localized on the same boundary so they do not have the information of the system size $L$.
Therefore, for a sufficiently large $L$, the contribution is bounded by $o(e^{cL})$ unless they exactly form an exceptional point.
When the boundary states exactly form an exceptional point, it gives an infinite contribution to $\kappa_{\rm R}$. However, this situation needs a fine-tuning of Hamiltonian parameters and thus can be avoided by a small perturbation.   

In summary, if no non-Hermitian skin effect occurs, $\kappa_{\rm R}$ behave like $o(e^{cL})$, not like $O(e^{cL})$. 
Therefore, we conclude that $\kappa_{\rm R}=O(e^{cL})$ gives an order parameter of a non-Hermitian skin effect. 

For the departure from non-normality, ${\rm dep}_{\rm D}$, the relation to the non-Hermitian skin effect is much more straightforward: From the definition of the departure from non-normality in the main text, we find immediately that
when ${\rm dep}_{\rm D}$ behaves as $O(L^{1/2})$, $O(L)$ modes must change their energies under the open boundary condition in comparison to those under the periodic boundary condition, and the converse is also true. Therefore, ${\rm dep}_{\rm D}=O(L^{1/2})$ is an order parameter of a non-Hermitian skin effect.

\section{The general estimation of the condition numbers}\label{app:uniquely-defined-condition-number}
In this appendix, we show that the choice of the condition number in the main text has an upper bound of the possible values suppressed by the minimum value.

As mentioned in Sec. \ref{sec:condition-number}, the condition number $\kappa(V)$ is not unique for a given diagonalizable Hamiltonian $H$ due to the redundancy of how the eigenstates are taken.
In order to avoid such an ambiguity, we have imposed the normalization condition \eqref{eq:normalization} on the right eigenstates $\ket{E_\alpha^a}$ ($a=1,2,\ldots,d_\alpha$) corresponding to the eigenvalues $E_\alpha$ ($\alpha=1,2,\ldots,p$) of $H$, which allows the value of the condition number to be determined uniquely. 
This method is useful for estimating the condition number for a given concrete Hamiltonian analytically or numerically. 

We can also employ another choice of the condition number to fix the value, i.e., the minimization.
Recall that the eigenstates $\ket{E_\alpha^a}$ corresponding to $E_\alpha$ have the degree of freedom of a gauge transformation:
\begin{align}
    \ket{E_\alpha^a}
    \to \sum_{b=1}^{d_\alpha}\ket{E_\alpha^b}G_{ba}^\alpha
\end{align}
with $G^\alpha\in GL(d_\alpha)$, where $d_\alpha$ is the degree of degeneracy of $E_\alpha$.
The whole eigenstates of $H$ have the degree of freedom of $GL(d_1)\times \cdots \times GL(d_p)$:
\begin{align}
    &V = (\ldots,\ket{E_\alpha^1},\ldots,\ket{E_\alpha^{d_\alpha}},\ldots) \notag\\
    &\to VG 
    = \left(\ldots,\sum_a\ket{E_\alpha^a}G_{a1}^\alpha,\ldots,\sum_a\ket{E_1^{d_1}}G_{ad_\alpha}^\alpha,\ldots\right),
\end{align}
    where $G$ is an $N\times N$ regular matrix such that
\begin{align}
    G = G^1\oplus \cdots \oplus G^p
    \in GL(d_1)\times\cdots\times GL(d_p). 
\end{align}
Then we can introduce the minimized condition number
\begin{align}
    \kappa^\star
    \coloneqq \min_{G\in GL(d_1)\times\cdots\times GL(d_p)}\kappa(VG)
\end{align}
without any ambiguity of definition.
This method is suitable for the analysis of perturbation sensitivity by the Bauer-Fike theorem referred to in Sec.\ref{sec:Implications}, because $\kappa^\star$ provides the strictest bound for the extent of pseudo-spectra in Eq.\eqref{eq:BF}.

In the following in this appendix, let us suppose that the right eigenstates $\ket{E_\alpha^a}$ corresponding to the same eigenvalue $E_\alpha$ of $H$ are orthonormalized as in \eqref{eq:normalization}, and that the regular matrix $V$ is fixed to the form \eqref{eq:def-of-V} with such eigenstates.
Such a choice of the condition number $\kappa(V)$ is the method used in the main text.
We shall prove an inequality:
\begin{align}\label{eq:ineq-cond-num}
    \kappa^\star
    \le \kappa(V)
    \le \sqrt{p}\kappa^\star
    \le \sqrt{L}\kappa^\star,
\end{align}
which indicates that $\kappa(V)$ is of the same order of $L$ with the minimum value $\kappa^\star$ if $\kappa(V)\sim e^{\alpha L}$ for a constant $\alpha>0$. 

$\kappa^\star\le\kappa(V)$ follows from the definition of $\kappa^\star$, and $\sqrt{p}\kappa^\star\le\sqrt{L}\kappa^\star$ is also trivial. 

In order to prove that $\kappa(V)\le\sqrt{p}\kappa^\star$, we introduce the function $\psi$ such that
\begin{align}
    \psi(M)
    \coloneqq \max_{\alpha=1,2,\ldots,p}\norm{(\ket{M_\alpha^1},\ket{M_\alpha^2},\ldots,\ket{M_\alpha^{d_\alpha}})}_2
\end{align}
for any $N\times N$ matrices $M=(\ket{M_1^1},$ $\ldots,$ $ \ket{M_1^{d_1}},$ $\ket{M_2^1},$ $\ldots,$ $\ket{M_2^{d_2}},$ $\ldots,$ $\ket{M_p^1},$ $\ldots,$ $\ket{M_p^{d_p}})$ with any $N$ vectors $\ket{M_\alpha^a}\in\mathbb{C}^N$. 
For any $G=G^1\oplus\cdots\oplus G^p\in GL(d_1)\times\cdots\times GL(d_p)$, it holds that
\begin{align}\label{eq:equality-for-psi}
    \psi(VG)
    &= \max_\alpha\norm{(\ket{E_\alpha^1},\ldots,\ket{E_\alpha^{d_\alpha}})G^\alpha}_2 \notag\\
    &= \max_\alpha\norm{G^\alpha}
    = \psi(G),
\end{align}
because $(\ket{E_\alpha^1},\ldots,\ket{E_\alpha^{d_\alpha}})$ is 
an $N\times d_\alpha$ unitary matrix thanks to the normalization condition \eqref{eq:normalization}. 

For any $N\times N$ matrix $M$, the function $\psi$ is related to the 2-norm of matrices by the inequality:
\begin{align}\label{eq:psi-and-2norm}
    \norm{M}_2
    \ge \psi(M).
\end{align}
In order to check Eq.\eqref{eq:psi-and-2norm}, we prepare orthogonal projectors $Q^\alpha$ on $\mathbb{C}^N$ such that
\begin{align}
    Q^\alpha
    = \bigoplus_{\beta=1}^p \delta_{\alpha,\beta}\boldsymbol{1}_{d_\beta\times d_\beta}.
\end{align}
Since $MQ=(\ldots,0,\ket{M_\alpha^1},\ldots,\ket{M_\alpha^{d_\alpha}},0,\ldots)$, we have
\begin{align}
    \norm{MQ^\alpha}_2
    = \norm{(\ket{M_\alpha^1},\ldots,\ket{M_\alpha^{d_\alpha}})}_2.
\end{align}
Since $Q^\alpha$ is an orthogonal projection and thus $\norm{Q^\alpha}_2=1$, it holds that
\begin{align}
    \norm{M}_2
    \ge\norm{(\ket{M_\alpha^1},\ldots,\ket{M_\alpha^{d_\alpha}})}_2
\end{align}
for any $\alpha=1,2,\ldots,p$,
which becomes \eqref{eq:psi-and-2norm}
by maximizing the right-hand side with respect to $\alpha$.

Particularly if $M=G$ then the reverse inequality also holds:
\begin{align}\label{eq:psi-and-2norm-G}
    \norm{G}_2
    \le \psi(G).
\end{align}
We can straightforwardly verify \eqref{eq:psi-and-2norm-G} by using the following inequality for any vector 
$\boldsymbol{c}=\boldsymbol{c}_1\oplus\boldsymbol{c}_2\oplus\cdots\oplus\boldsymbol{c}_p\in\mathbb{C}^N$ with $\boldsymbol{c}_\alpha\in\mathbb{C}^{d_\alpha}$:
\begin{align}
    \norm{G\boldsymbol{c}}_2^2
    &= \sum_\alpha\norm{G^\alpha\boldsymbol{c}_\alpha}_2^2 \\
    &\le \sum_\alpha\norm{G^\alpha}_2^2\norm{\boldsymbol{c}_\alpha}_2^2 \\
    &\le 
    \left( \max_\alpha\norm{G^\alpha}_2\right)^2 \sum_\alpha \norm{\boldsymbol{c}_\alpha}_2^2 \\
    &= \psi(G)^2\norm{\boldsymbol{c}}_2^2.
\end{align}

Using \eqref{eq:equality-for-psi}, \eqref{eq:psi-and-2norm}, and \eqref{eq:psi-and-2norm-G}, we obtain
\begin{align}
    \kappa(VG)
    &= \norm{VG}_2\norm{(VG)^{-1}}_2 \\
    &\ge\psi(VG)\norm{(VG)^{-1}}_2
    = \psi(G)\norm{G^{-1}V^{-1}} \\
    &\ge \norm{G}_2\norm{G^{-1}V^{-1}}
    \ge \norm{V^{-1}},
\end{align}
which follows that
\begin{align}\label{eq:V-VVG}
    \kappa(V)
    = \norm{V}_2\norm{V^{-1}}_2
    \le \norm{V}_2\kappa(VG).
\end{align}
Minimizing \eqref{eq:V-VVG} with respect to $G\in GL(d_1)\times\cdots\times GL(d_p)$, we have
\begin{align}
    \kappa(V)
    \le \norm{V}_2\kappa^\star.
\end{align}

At last we prove that $\norm{V}_2^2\le p$ to conclude that $\kappa(V)\le\sqrt{p}\kappa^\star$. 
For any $\boldsymbol{c}=\boldsymbol{c}_1\oplus\cdots\oplus\boldsymbol{c}_p\in\mathbb{C}^N$, we can derive
\begin{align}
    \norm{V\boldsymbol{c}}_2
    &= \norm{\sum_\alpha (\ket{E_\alpha^1},\ldots,\ket{E_\alpha^{d_\alpha}})\boldsymbol{c}_\alpha} \\ 
    &\le \sum_\alpha\norm{(\ket{E_\alpha^1},\ldots,\ket{E_\alpha^{d_\alpha}})\boldsymbol{c}_\alpha}_2
    = \sum_\alpha\norm{\boldsymbol{c}_\alpha}_2,
\end{align}
and applying the Cauchy-Schwarz inequality we have
\begin{align}
    \norm{V\boldsymbol{c}}_2^2
    \le \left( \sum_{\alpha=1}^p \norm{\boldsymbol{c}_\alpha}_2 \right)^2
    \le p\sum_\alpha\norm{\boldsymbol{c}_\alpha}^2
    = p\norm{\boldsymbol{c}}_2^2.
\end{align}
Therefore $\norm{V}_2\le\sqrt{p}$ holds.

Note that one can see the proof of \eqref{eq:ineq-cond-num} for the case of $d_1=d_2=\cdots=d_p=1$, i.e., no degeneracy for any eigenvalues in Ref. \cite{vanderSluis1969}.
The above proof includes and extends the proof in Ref.\cite{vanderSluis1969}.

\section{Proof of Eq.(\ref{48-inequality})}
\label{48-inequality-d}
In this appendix, we prove the inequality (\ref{48-inequality}).
\subsection{Preparation}
In this subsection, we calculate the norm of the diagonal matrix $R$ in Eq.\eqref{r} and the vector $\ket{n}_{\mathrm{OBC}}$ in Eq.\eqref{eq:skinmode}. 
% The matrix $R$ has only diagonal elements, which are given by Eq.(\ref{r}). Thus
Since $r\geq1$, we have
\begin{align}
\label{R}
    \|R\|_2=\max_{j=1,2,\cdots, L}r^{j}=r^L,
\end{align}
\begin{align}
\label{R-1}
    \|R^{-1}\|_2=\max_{k=1,2,\cdots, L}r^{-k}=r^{-1}.
\end{align}
Therefore we obtain the condition number of the matrix $R$:
\begin{align}
\label{Rcon}
    \kappa(R)=r^{L-1}.
\end{align}
In the following, we calculate the crucial inequality for the norm of $\ket{n}_{\mathrm{OBC}}$ $(n=1,2,\cdots, L)$:
\begin{align}
    \|\ket{n}_{\mathrm{OBC}} \|^2_{2}&=\sum_{m=1}^{L}\left(\sqrt{\frac{2}{L+1}}r^m\sin\frac{\pi mn}{L+1}\right)^2\notag\\
    &=\frac{r^2}{L+1}\frac{(r^{2L+2}-1)(r^2+1)(1-\cos\frac{2\pi n}{L+1})}{(r^2-1)(r^4-2r^2\cos\frac{2\pi n}{L+1}+1)}\notag\\
    &>\frac{r^2}{L+1}\frac{(r^{2L+2}-1)(r^2+1)(1-\cos\frac{2\pi n}{L+1})}{(r^2-1)(r^4+2r^2+1)}\notag\\
    &=\frac{2r^2}{L+1}\frac{(r^{2L+2}-1)}{(r^4-1)}\sin^2\frac{\pi n}{L+1}\notag\\
    &\geq\frac{2r^2}{L+1}\frac{(r^{2L+2}-1)}{(r^4-1)}\sin^2\frac{\pi}{L+1}\notag\\
    &\ (\because n=1,2,\cdots, L)\notag\\
    &\geq \frac{2r^2}{L+1}\frac{(r^{2L+2}-1)}{(r^4-1)}\left(\frac{2}{\pi}\frac{\pi}{L+1}\right)^2\notag\\
    &=\left(\frac{2}{L+1}\right)^3\frac{1-r^{-2(L+1)}}{1-r^{-4}}r^{2L}\notag\\
    &\geq\left(\frac{2}{L+1}\right)^3\frac{1-r^{-4}}{1-r^{-4}}r^{2L}\notag\\
    &=\left(\frac{2}{L+1}\right)^3r^{2L}.    
\end{align}
Thus we obtain the inequality for the 2-norm of $\ket{n}$,
\begin{align}
\label{15}
    \|\ket{n}_{\mathrm{OBC}}\|_2>
    \left(\frac{2}{L+1}\right)^{3/2}r^L.
\end{align}

% \subsection{The estimation of Eq.(\ref{48-inequality}) from the above}
\subsection{The upper bound in Eq.(\ref{48-inequality})}
Using Eq.(\ref{eq:ineq-cond-num}), we obtain the inequality for the condition number of the matrix $V_{\mathrm{OBC}}$:
\begin{align}
    \kappa_{\mathrm{OBC}}(V)\leq \sqrt{L}\kappa^{\star}\coloneqq\sqrt{L}\kappa(V_{\mathrm{OBC}}G),
\end{align}
where $G$ denotes the regular diagonal matrix, which minimizes the condition number about the matrix $V_{\mathrm{OBC}}$ as mentioned in Appendix \ref{app:uniquely-defined-condition-number}.
Since $\kappa^{\star}$ denotes the minimum value of the condition number of the matrix which diagonalizes Eq.~(\ref{a-1}) under the OBC, we obtain the following inequality:
\begin{align}
\kappa^{\star}&\leq\kappa(RU)
=\kappa(R)\quad(\because UU^{\dagger}=1).
\end{align}
Thus we obtain the inequality for the upper bound in Eq.(\ref{48-inequality}):
\begin{align}
    \label{17}
    \kappa_{\mathrm{OBC}}(V)\leq \sqrt{L}\kappa(R)=\sqrt{L}r^{L-1}\quad(\because \text{(\ref{Rcon})}).
\end{align}
% \subsection{The estimation of Eq.(\ref{48-inequality}) from the bellow}
\subsection{The lower bound in Eq.(\ref{48-inequality})}
The Frobenius norm of the matrix $V_{\mathrm{OBC}}$ is $\sqrt{L}$:
\begin{align}
    \label{18}
    \|V_{\mathrm{OBC}}\|_{\mathrm{F}}^{2}&=\operatorname{tr}(V_{\mathrm{OBC}}^{\dag}V_{\mathrm{OBC}})\notag\\
    &=\sum_{n=1}^{L}\left(\frac{\ket{n}_{\mathrm{OBC}}}{\|\ket{n}_{\mathrm{OBC}}\|_2}\right)^{\dag}\frac{\ket{n}_{\mathrm{OBC}}}{\|\ket{n}_{\mathrm{OBC}}\|_2}\notag\\
    &=L.
\end{align}
Substituting Eq.(\ref{18}) to the inequality between the 2-norm and Frobenius norm (i.e.,~$(1/\sqrt{L})\|V_{\mathrm{OBC}}\|_{\mathrm{F}}\leq\|V_{\mathrm{OBC}}\|_2\leq\|V_{\mathrm{OBC}}\|_{\mathrm{F}}$), we obtain the following inequality for the 2-norm of the matrix $V_{\mathrm{OBC}}$:
\begin{align}
\label{20}
    1\leq\|V_{\mathrm{OBC}}\|_2\leq \sqrt{L}.
\end{align}

Next, we derive the inequality for the 2-norm of the matrix $V^{-1}_{\mathrm{OBC}}$. Using the Cauchy–Schwarz inequality, we obtain the following inequality:
\begin{align}
    \label{21}
    \|NN^{-1}U^{\dag}R^{-1}\|_2&\leq \|N\|_2\|N^{-1}U^{\dag}R^{-1}\|_2\notag\\
    \therefore \|N^{-1}U^{\dag}R^{-1}\|_2& \geq \|N\|_2^{-1}\|U^{\dagger}R^{-1}\|_{2}.
\end{align}
From Eq.(\ref{21}), we obtain the inequality for 2-norm of the matrix $V_{\rm OBC}^{-1}$:
\begin{align}\label{eq:lower_VOBCinv}
    \|V^{-1}_{\mathrm{OBC}}\|&=\|N^{-1}U^{\dag}R^{-1}\|_2\notag\\
    &\geq\|N\|_2^{-1}\|U^{\dagger}R^{-1}\|_{2}\notag\\
    &=\|N\|_2^{-1}\|R^{-1}\|_2\notag\\
    &=\left(\max_{n=1,2,\cdots,L}\|\ket{n}_{\mathrm{OBC}}\|_2^{-1}\right)^{-1}\|R^{-1}\|_2\notag\\
    &=\left(\min_{n=1,2,\cdots,L}\|\ket{n}_{\mathrm{OBC}}   \|_2\right)\|R^{-1}\|_2\notag\\
    &> \left(\frac{2}{L+1}\right)^{3/2}r^{L}r^{-1}\quad(\because\text{(\ref{R-1}),(\ref{15})})\notag\\
    &=\left(\frac{2}{L+1}\right)^{3/2}r^{L-1}.
\end{align}
Therefore we obtain the inequality for the lower bound in Eq.(\ref{48-inequality}):
\begin{align}
\label{26}
    \kappa_{\mathrm{OBC}}(V)&=\|V_{\mathrm{OBC}}\|_2\|V^{-1}_{\mathrm{OBC}}\|_2\notag\\
    &>\left(\frac{2}{L+1}\right)^{3/2}r^{L-1}.
\end{align}
\subsection{Result}
In conclusion, from Eqs.(\ref{17}) and (\ref{26}), we obtain the inequality (\ref{48-inequality}):
\begin{align}
    \left(\frac{2}{L+1}\right)^{3/2}r^{L-1}<\kappa_{\mathrm{OBC}}(V)\leq \sqrt{L}r^{L-1}.
\end{align}

\section{Proof of Eq.(\ref{eq:estimate_AIIdag})}\label{sec:estimate_AIIdag}
In this appendix, we prove the inequality \eqref{eq:estimate_AIIdag} about the condition number $\kappa(\tilde{V})$ corresponding to the Hamiltonian \eqref{41-1}.
Suppose that $0<g<t$, $0<\Delta<t$, and $g\neq \Delta$ hold.

\subsection{Rewriting of Eq.(\ref{eq:tildeV})}
We shall rewrite the matrix $\tilde{V}$ defined in \eqref{eq:tildeV} into an explicit form using $S$, $\ket{k}^-$, $\eta_k$, and $\gamma_k^{(\alpha)}$. 

From the definition \eqref{eq:tildekalpha} of $\tilde{\ket{k}}^{\alpha}$, we obtain the relation
\begin{align}
    \begin{pmatrix}
        \tilde{\ket{k}}^1 & \tilde{\ket{k}}^2
    \end{pmatrix}
    = \begin{pmatrix}
        \tilde{\ket{k}}^+ & \tilde{\ket{k}}^-
    \end{pmatrix}
    \eta_k
    \begin{pmatrix}
        \frac{1}{\sqrt{\gamma_k^{(1)}}} & 0 \\
        0 & \frac{1}{\sqrt{\gamma_k^{(2)}}}
    \end{pmatrix}
\end{align}
between $\tilde{\ket{k}}^\alpha$ and $\tilde{\ket{k}}^\pm$, which provides an expression of $\tilde{V}$ with $\tilde{\ket{k}}^\pm$:
\begin{align}
    \tilde{V}
    &\coloneqq \begin{pmatrix}
        \tilde{\ket{1}}^1 & \tilde{\ket{1}}^2 & \cdots & \tilde{\ket{L}}^1 & \tilde{\ket{L}}^2
    \end{pmatrix} \\
    &= \begin{pmatrix}
        \tilde{\ket{1}}^+ & \tilde{\ket{1}}^- & \cdots & \tilde{\ket{L}}^+ & \tilde{\ket{L}}^-
    \end{pmatrix} \bigoplus_{k=1}^L \eta_k\begin{pmatrix}
        \frac{1}{\sqrt{\gamma_k^{(1)}}} & 0 \\
        0 & \frac{1}{\sqrt{\gamma_k^{(2)}}}
    \end{pmatrix}.
\end{align}
From the definition \eqref{eq:tildekplus} and \eqref{eq:tildekminus} of $\tilde{\ket{k}}^\pm$, we have
\begin{align}
    &\begin{pmatrix}
        \tilde{\ket{1}}^+ & \tilde{\ket{1}}^- & \cdots & \tilde{\ket{L}}^+ & \tilde{\ket{L}}^-
    \end{pmatrix} \notag \\
    &= \begin{pmatrix}
        \tilde{\ket{1}}^+ & \cdots & \tilde{\ket{L}}^+
    \end{pmatrix}
    \left( \begin{pmatrix}
        1 & 0
    \end{pmatrix}\otimes{\bm 1}_{L\times L} \right) \notag \\
    &\quad + \begin{pmatrix}
        \tilde{\ket{1}}^- & \cdots & \tilde{\ket{L}}^-
    \end{pmatrix}
    \left( \begin{pmatrix}
        0 & 1
    \end{pmatrix}\otimes{\bm 1}_{L\times L} \right) \\
    &= \tilde{S}\left( \begin{pmatrix}
        1 \\ 0
    \end{pmatrix}\otimes\begin{pmatrix}
        \mathcal{P}\frac{\ket{1}^-}{\|\ket{1}^-\|_2} & \cdots & \mathcal{P}\frac{\ket{L}^-}{\|\ket{L}^-\|_2}
    \end{pmatrix} \right) \left( \begin{pmatrix}
        1 & 0
    \end{pmatrix}\otimes{\bm 1}_{L\times L} \right) \notag \\
    &\quad + \tilde{S}\left( \begin{pmatrix}
        0 \\ 1
    \end{pmatrix}\otimes\begin{pmatrix}
        \frac{\ket{1}^-}{\|\ket{1}^-\|_2} & \cdots & \frac{\ket{L}^-}{\|\ket{L}^-\|_2}
    \end{pmatrix} \right) \left( \begin{pmatrix}
        0 & 1
    \end{pmatrix}\otimes{\bm 1}_{L\times L} \right) \\
    &= \tilde{S}\left(\begin{pmatrix}
        1 & 0 \\
        0 & 0 
    \end{pmatrix}\otimes\mathcal{P}\begin{pmatrix}
        \frac{\ket{1}^-}{\|\ket{1}^-\|_2} & \cdots & \frac{\ket{L}^-}{\|\ket{L}^-\|_2}
    \end{pmatrix}\right) \notag\\
    &\quad + \tilde{S}\left(\begin{pmatrix}
        0 & 0 \\
        0 & 1 
    \end{pmatrix}\otimes\begin{pmatrix}
        \frac{\ket{1}^-}{\|\ket{1}^-\|_2} & \cdots & \frac{\ket{L}^-}{\|\ket{L}^-\|_2}
    \end{pmatrix}\right).
\end{align}
Therefore we obtain 
\begin{align}\label{eq:tildeVexplicit}
    \tilde{V}
    &= \tilde{S}\Bigg[ \begin{pmatrix}
        1 & 0 \\
        0 & 0 
    \end{pmatrix}\otimes\mathcal{P}\begin{pmatrix}
        \frac{\ket{1}^-}{\|\ket{1}^-\|_2} & \cdots & \frac{\ket{L}^-}{\|\ket{L}^-\|_2}
    \end{pmatrix} \notag \\
    &\qquad +\begin{pmatrix}
        0 & 0 \\
        0 & 1 
    \end{pmatrix}\otimes\begin{pmatrix}
        \frac{\ket{1}^-}{\|\ket{1}^-\|_2} & \cdots & \frac{\ket{L}^-}{\|\ket{L}^-\|_2}
    \end{pmatrix}\Bigg] \notag \\
    &\qquad \times \bigoplus_{k=1}^L \eta_k\begin{pmatrix}
        \frac{1}{\sqrt{\gamma_k^{(1)}}} & 0 \\
        0 & \frac{1}{\sqrt{\gamma_k^{(2)}}}
    \end{pmatrix},
\end{align}
whose condition number $\kappa(\tilde{V})$ is to be evaluated.
The inverse of Eq.\eqref{eq:tildeVexplicit} is
\begin{align}\label{eq:tildeVinvexplicit}
    \tilde{V}^{-1}
    &= \left(\bigoplus_{k=1}^L \begin{pmatrix}
        \sqrt{\gamma_k^{(1)}} & 0 \\
        0 & \sqrt{\gamma_k^{(2)}}
    \end{pmatrix}\eta_k^\dag\right) \notag \\
    &\quad \times \Bigg[ \begin{pmatrix}
        1 & 0 \\
        0 & 0 
    \end{pmatrix} \otimes \left(\mathcal{P}\begin{pmatrix}
        \frac{\ket{1}^-}{\|\ket{1}^-\|_2} & \cdots & \frac{\ket{L}^-}{\|\ket{L}^-\|_2}
    \end{pmatrix}\right)^{-1} \notag \\
    &\qquad + \begin{pmatrix}
        0 & 0 \\
        0 & 1 
    \end{pmatrix} \otimes \begin{pmatrix}
        \frac{\ket{1}^-}{\|\ket{1}^-\|_2} & \cdots & \frac{\ket{L}^-}{\|\ket{L}^-\|_2}
    \end{pmatrix}^{-1} \Bigg] \tilde{S}^{-1}.
\end{align}

\subsection{The common estimate of the condition number}
In this subsection, we employ the submultiplicativity of the 2-norm to find a bound of $\|\tilde{V}\|_2$ and $\|\tilde{V}^{-1}\|_2$ without considering a certain boundary condition (i.e., the OBC or the PBC) or a certain relation between $g$ and $\Delta$. 
As a result, we shall obtain an upper and lower estimate of the condition number
\begin{align}
    \kappa(\tilde{V})
    \coloneqq \|\tilde{V}\|_2 \|\tilde{V}^{-1}\|_2.
\end{align}

\subsubsection{The upper estimate}
From the explicit form \eqref{eq:tildeVexplicit} of $\tilde{V}$, we have
\begin{align}
    \|\tilde{V}\|_2
    &\le \|\tilde{S}\|_2 \Bigg\| \begin{pmatrix}
        1 & 0 \\
        0 & 0 
    \end{pmatrix}\otimes\mathcal{P}\begin{pmatrix}
        \frac{\ket{1}^-}{\|\ket{1}^-\|_2} & \cdots & \frac{\ket{L}^-}{\|\ket{L}^-\|_2}
    \end{pmatrix} \notag \\
    &\qquad\qquad +\begin{pmatrix}
        0 & 0 \\
        0 & 1 
    \end{pmatrix}\otimes\begin{pmatrix}
        \frac{\ket{1}^-}{\|\ket{1}^-\|_2} & \cdots & \frac{\ket{L}^-}{\|\ket{L}^-\|_2}
    \end{pmatrix}\Bigg\|_2 \notag \\
    &\quad \times \norm{\bigoplus_{k=1}^L \eta_k\begin{pmatrix}
        \frac{1}{\sqrt{\gamma_k^{(1)}}}  & 0 \\
        0 & \frac{1}{\sqrt{\gamma_k^{(2)}}}
    \end{pmatrix}}_2 \\
    &= \norm{S}_2\max_{k=1,2,\ldots,L}\norm{\eta_k\begin{pmatrix}
        \frac{1}{\sqrt{\gamma_k^{(1)}}}  & 0 \\
        0 & \frac{1}{\sqrt{\gamma_k^{(2)}}}
    \end{pmatrix}}_2 \notag \\
    &\quad \times \max\Bigl\{ \norm{\mathcal{P}\begin{pmatrix}
        \frac{\ket{1}^-}{\|\ket{1}^-\|_2} & \cdots & \frac{\ket{L}^-}{\|\ket{L}^-\|_2}
    \end{pmatrix}}_2, \notag \\
    &\qquad\qquad\qquad \norm{\begin{pmatrix}
        \frac{\ket{1}^-}{\|\ket{1}^-\|_2} & \cdots & \frac{\ket{L}^-}{\|\ket{L}^-\|_2}
    \end{pmatrix}}_2 \Bigl\}.
    \label{eq:upper_tildeV1}
\end{align}
Recalling that $\eta_k$ and $\mathcal{P}$ are unitary, and that $\gamma_k^{(1)}\le\gamma_k^{(2)}$ holds for each $k$, we can rearrange Eq.\eqref{eq:upper_tildeV1} as follows:
\begin{align}
    \|\tilde{V}\|_2
    &\le \norm{S}_2\left(\max_{k=1,2,\ldots,L}\max_{\alpha=1,2}\frac{1}{\sqrt{\gamma_k^{(\alpha)}}}\right) \notag \\
    &\qquad \times \norm{\begin{pmatrix}
        \frac{\ket{1}^-}{\|\ket{1}^-\|_2} & \cdots & \frac{\ket{L}^-}{\|\ket{L}^-\|_2}
    \end{pmatrix}}_2 \\
    &= \norm{S}_2\left(\max_{k=1,2,\ldots,L}\frac{1}{\sqrt{\gamma_k^{(1)}}}\right) \notag \\
    &\qquad \times \norm{\begin{pmatrix}
        \frac{\ket{1}^-}{\|\ket{1}^-\|_2} & \cdots & \frac{\ket{L}^-}{\|\ket{L}^-\|_2}
    \end{pmatrix}}_2.
\end{align}
In the same way as $\tilde{V}$, the norm of \eqref{eq:tildeVinvexplicit} is bounded from the above by 
\begin{align}
    \|\tilde{V}^{-1}\|_2
    &\le \norm{S^{-1}}_2\left(\max_{k=1,2,\ldots,L}\max_{\alpha=1,2}\sqrt{\gamma_k^{(\alpha)}}\right) \notag \\
    &\quad \times \max\Bigg\{ \norm{\begin{pmatrix}
        \frac{\ket{1}^-}{\|\ket{1}^-\|_2} & \cdots & \frac{\ket{L}^-}{\|\ket{L}^-\|_2}
    \end{pmatrix}^{-1}\mathcal{P}}_2, \notag \\
    &\qquad\qquad\qquad \norm{\begin{pmatrix}
        \frac{\ket{1}^-}{\|\ket{1}^-\|_2} & \cdots & \frac{\ket{L}^-}{\|\ket{L}^-\|_2}
    \end{pmatrix}^{-1}}_2 \Bigg\} \\
    &= \norm{S^{-1}}_2\left(\max_{k=1,2,\ldots,L}\sqrt{\gamma_k^{(2)}}\right) \notag \\
    &\qquad \times \norm{\begin{pmatrix}
        \frac{\ket{1}^-}{\|\ket{1}^-\|_2} & \cdots & \frac{\ket{L}^-}{\|\ket{L}^-\|_2}
    \end{pmatrix}^{-1}}_2.
\end{align}

Then we have
\begin{align}\label{eq:upper_cond}
    \kappa(\tilde{V})
    &\le \kappa(S)\kappa\left(\begin{pmatrix}
        \frac{\ket{1}^-}{\|\ket{1}^-\|_2} & \cdots & \frac{\ket{L}^-}{\|\ket{L}^-\|_2}
    \end{pmatrix}\right) \notag \\
    &\quad \times \left(\max_{k=1,2,\ldots,L}\frac{1}{\sqrt{\gamma_k^{(1)}}}\right) \left(\max_{k=1,2,\ldots,L}\sqrt{\gamma_k^{(2)}}\right).
\end{align}

\subsubsection{The lower estimate}
The orthonormality \eqref{orthonormal} of $\tilde{\ket{k}}^1$ and $\tilde{\ket{k}}^2$ leads to a lower bound of $\|\tilde{V}\|_2$:
\begin{align}
    \|\tilde{V}\|_2^2
    &\ge \frac{1}{2L}\|\tilde{V}\|_{\mathrm{F}}^2
    = \frac{1}{2L}\operatorname{tr}(\tilde{V}^\dag\tilde{V}) \\
    &= \frac{1}{2L}\sum_{k=1}^L \sum_{\alpha=1,2} {}^\alpha\tilde{\langle k}\tilde{\ket{k}}{}^\alpha
    = 1.
    \label{eq:lower_tildeV}
\end{align}
On the other hand, the submultiplicativity of the 2-norm provides a lower bound of $\|\tilde{V}^{-1}\|_2$:
\begin{align}
    \|\tilde{V}^{-1}\|_2
    &\ge \norm{\bigoplus_{k=1}^L \eta_k\begin{pmatrix}
        \frac{1}{\sqrt{\gamma_k^{(1)}}} & 0 \\
        0 & \frac{1}{\sqrt{\gamma_k^{(2)}}}
    \end{pmatrix}}_2^{-1} \notag \\
    &\quad \times \Bigg\| \begin{pmatrix}
        1 & 0 \\
        0 & 0 
    \end{pmatrix} \otimes \left(\mathcal{P}\begin{pmatrix}
        \frac{\ket{1}^-}{\|\ket{1}^-\|_2} & \cdots & \frac{\ket{L}^-}{\|\ket{L}^-\|_2}
    \end{pmatrix}\right)^{-1} \notag \\
    &\qquad + \begin{pmatrix}
        0 & 0 \\
        0 & 1 
    \end{pmatrix} \otimes \begin{pmatrix}
        \frac{\ket{1}^-}{\|\ket{1}^-\|_2} & \cdots & \frac{\ket{L}^-}{\|\ket{L}^-\|_2}
    \end{pmatrix}^{-1} \Bigg\|_2 \|\tilde{S}\|_2^{-1} \\
    &= \norm{S}_2^{-1}\left(\max_{k=1,2,\ldots,L}\max_{\alpha=1,2}\frac{1}{\sqrt{\gamma_k^{(\alpha)}}}\right)^{-1} \notag \\
    &\quad \times \max\Bigg\{ \norm{\begin{pmatrix}
        \frac{\ket{1}^-}{\|\ket{1}^-\|_2} & \cdots & \frac{\ket{L}^-}{\|\ket{L}^-\|_2}
    \end{pmatrix}^{-1}\mathcal{P}}_2, \notag \\
    &\qquad\qquad\qquad \norm{\begin{pmatrix}
        \frac{\ket{1}^-}{\|\ket{1}^-\|_2} & \cdots & \frac{\ket{L}^-}{\|\ket{L}^-\|_2}
    \end{pmatrix}^{-1}}_2 \Bigg\} \\
    &= \norm{S}_2^{-1}\left(\min_{k=1,2,\ldots,L}\sqrt{\gamma_k^{(1)}}\right) \notag \\
    &\qquad \times \norm{\begin{pmatrix}
        \frac{\ket{1}^-}{\|\ket{1}^-\|_2} & \cdots & \frac{\ket{L}^-}{\|\ket{L}^-\|_2}
    \end{pmatrix}^{-1}}_2.
    \label{eq:lower_tildeVinv}
\end{align}

The product of Eqs.\eqref{eq:lower_tildeV} and \eqref{eq:lower_tildeVinv} is just a lower estimate of the condition number:
\begin{align}\label{eq:lower_cond}
    \kappa(\tilde{V})
    &\ge \norm{S}_2^{-1}\left(\min_{k=1,2,\ldots,L}\sqrt{\gamma_k^{(1)}}\right) \notag \\
    &\qquad \times \norm{\begin{pmatrix}
        \frac{\ket{1}^-}{\|\ket{1}^-\|_2} & \cdots & \frac{\ket{L}^-}{\|\ket{L}^-\|_2}
    \end{pmatrix}^{-1}}_2.
\end{align}

\subsection{Case analysis}
In the previous subsection, we had the upper and lower bound
\begin{align}\label{eq:preestimate_cond}
    &\norm{S}_2^{-1}\left(\min_{k=1,2,\ldots,L}\sqrt{\gamma_k^{(1)}}\right)  \norm{\begin{pmatrix}
        \frac{\ket{1}^-}{\|\ket{1}^-\|_2} & \cdots & \frac{\ket{L}^-}{\|\ket{L}^-\|_2}
    \end{pmatrix}^{-1}}_2 \notag \\
    &\le \kappa(\tilde{V}) \notag \\
    &\le \kappa(S)\kappa\left(\begin{pmatrix}
        \frac{\ket{1}^-}{\|\ket{1}^-\|_2} & \cdots & \frac{\ket{L}^-}{\|\ket{L}^-\|_2}
    \end{pmatrix}\right) \notag \\
    &\quad \times \left(\max_{k=1,2,\ldots,L}\frac{1}{\sqrt{\gamma_k^{(1)}}}\right) \left(\max_{k=1,2,\ldots,L}\sqrt{\gamma_k^{(2)}}\right)
\end{align}
for the condition number $\kappa(\tilde{V})$.
Toward the further estimate to derive Eq.\eqref{eq:estimate_AIIdag}, it is necessary to consider separately the cases of (a) $g<\Delta$ or (b) $g>\Delta$, and (i) the PBC or (ii) the OBC.
In this subsection, we calculate or estimate some quantities in Eq.\eqref{eq:preestimate_cond} for each of the above cases.

\subsubsection{The case of \texorpdfstring{$g<\Delta$}{TEXT}}\label{sec:g<Delta}
In this case, $H^-$ in Eqs.\eqref{eq:HOBCpm} and \eqref{eq:HPBCpm} is Hermitian, and thus is diagonalized by the unitary matrix
\begin{align}
    \begin{pmatrix}
        \frac{\ket{1}^-}{\|\ket{1}^-\|_2} & \cdots & \frac{\ket{L}^-}{\|\ket{L}^-\|_2}
    \end{pmatrix}.
\end{align}
Then we have
\begin{align}
    \norm{\begin{pmatrix}
        \frac{\ket{1}^-}{\|\ket{1}^-\|_2} & \cdots & \frac{\ket{L}^-}{\|\ket{L}^-\|_2}
    \end{pmatrix}^{-1}}_2
    = 1
\end{align}
and
\begin{align}
    \kappa\left(\begin{pmatrix}
        \frac{\ket{1}^-}{\|\ket{1}^-\|_2} & \cdots & \frac{\ket{L}^-}{\|\ket{L}^-\|_2}
    \end{pmatrix}\right)
    = 1.
\end{align}

In addition, the right eigenstate $\ket{k}^-$ of $H^-$ agrees with the left one $\kket{k}^-$, by which the biorthogonality \eqref{eq:biorthogonality_AIIdag} results in the normalization condition
\begin{align}
    \|\ket{k}^-\|_2
    = 1
    = \|\kket{k}^-\|_2.
\end{align}
Recalling Eqs.\eqref{eq:tildekplus2} and \eqref{eq:tildekminus2}, i.e.,
\begin{align}
    \tilde{\ket{k}}^+
    = \begin{pmatrix}
        \Delta \\
        -\sqrt{\Delta^2-g^2} -ig
    \end{pmatrix}\otimes \mathcal{P}\ket{k}^-
\end{align}
and
\begin{align}
    \tilde{\ket{k}}^-
    = \begin{pmatrix}
        \sqrt{\Delta^2-g^2} +ig \\
        \Delta
    \end{pmatrix}\otimes\ket{k}^-,
\end{align}
we obtain a concrete form of the Gram matrix $\Gamma_k$ in Eq.\eqref{eq:Gammak}:
\begin{align}
    \Gamma_k
    &\coloneqq \begin{pmatrix}
        ^+\tilde{\langle k}\tilde{\ket{k}}{}^+ & ^+\tilde{\langle k}\tilde{\ket{k}}{}^- \\
        ^-\tilde{\langle k}\tilde{\ket{k}}{}^+ & ^-\tilde{\langle k}\tilde{\ket{k}}{}^-
    \end{pmatrix} \\
    &= \begin{pmatrix}
        2\Delta^2 & 2i\Delta g {}^-\bra{k}\mathcal{P}\ket{k}{}^- \\
        -2i\Delta g {}^-\bra{k}\mathcal{P}\ket{k}{}^- & 2\Delta^2
    \end{pmatrix} \\
    &= 2\Delta\left(\Delta - g {}^-\bra{k}\mathcal{P}\ket{k}{}^-\sigma_y\right).
\end{align}
Note that ${}^-\tilde{\bra{k}}\mathcal{P}\tilde{\ket{k}}{}^-$ is real due to the Hermiticity of $\mathcal{P}$.
Since the eigenvalues of $\sigma_y$ are $\pm 1$, we have
\begin{align}
    \gamma_k^{(1)}
    = 2\Delta\left(\Delta - g \abs{{}^-\bra{k}\mathcal{P}\ket{k}{}^-}\right)
\end{align}
and
\begin{align}    
    \gamma_k^{(2)}
    = 2\Delta\left(\Delta + g \abs{{}^-\bra{k}\mathcal{P}\ket{k}{}^-}\right),
\end{align}
which are bounded uniformly with respect to $k$ as
\begin{align}
    \gamma_k^{(1)}
    \ge 2\Delta\left(\Delta-g\right)
\end{align}
and
\begin{align}
    \gamma_k^{(2)}
    \le 2\Delta\left(\Delta+g\right)
\end{align}
by the Cauchy-Schwarz inequality.

In order to calculate $\norm{S}_2$ and $\kappa(S)$, we find the singular values of $S$ in Eq.\eqref{eq:def_S}.
Since
\begin{align}
    SS^\dag
    &= \left(\Delta+(i\sqrt{\Delta^2-g^2}-g)\sigma_y\right) \notag \\
    &\qquad \times\left(\Delta+(-i\sqrt{\Delta^2-g^2}-g)\sigma_y\right) \\
    &= 2\Delta(\Delta-g\sigma_y),
\end{align}
the singular values of $S$ are $\sqrt{2\Delta(\Delta\pm g)}$.
Thus 
\begin{align}
    \norm{S}_2
    = \sqrt{2\Delta(\Delta+g)}
\end{align}
and
\begin{align}
    \kappa(S)
    = \sqrt{\frac{2\Delta(\Delta+g)}{2\Delta(\Delta-g)}}
    = \sqrt{\frac{\Delta+g}{\Delta-g}}
\end{align}
hold.

Therefore the upper and lower estimate \eqref{eq:preestimate_cond} is reduced to
\begin{align}\label{eq:estimate_g<Delta}
    \sqrt{\frac{\Delta-g}{\Delta+g}}
    \le \kappa(\tilde{V})
    \le \frac{\Delta+g}{\Delta-g}.
\end{align}
Note that the estimate \eqref{eq:estimate_g<Delta} is valid whether we consider the OBC or the PBC.

\subsubsection{The case of \texorpdfstring{$g>\Delta$}{TEXT}}
In this case, since
\begin{align}
    SS^\dag
    &= \left(\Delta+(\sqrt{g^2-\Delta^2}-g)\sigma_y\right)^2 \\
    &= 2(g-\sqrt{g^2-\Delta^2})(g-\Delta\sigma_y),
\end{align}
the singular values of $S$ are 
\begin{align}
    \sqrt{2(g-\sqrt{g^2-\Delta^2})(g\pm\Delta)},
\end{align}
which lead to 
\begin{align}
    \norm{S}_2
    = \sqrt{2(g-\sqrt{g^2-\Delta^2})(g+\Delta)}
\end{align}
and
\begin{align}
    \kappa(S)
    = \sqrt{\frac{g+\Delta}{g-\Delta}}.
\end{align}

Recalling Eqs.\eqref{eq:tildekplus2} and \eqref{eq:tildekminus2}, i.e.,
\begin{align}
    \tilde{\ket{k}}^+
    = \begin{pmatrix}
        \Delta \\
        i(\sqrt{g^2-\Delta^2} -g)
    \end{pmatrix}\otimes \mathcal{P}\frac{\ket{k}^-}{\|\ket{k}^-\|_2}
\end{align}
and
\begin{align}
    \tilde{\ket{k}}^-
    = \begin{pmatrix}
        -i(\sqrt{g^2-\Delta^2} -g) \\
        \Delta
    \end{pmatrix}\otimes\frac{\ket{k}^-}{\|\ket{k}^-\|_2},
\end{align}
we obtain a concrete form of the Gram matrix $\Gamma_k$ in Eq.\eqref{eq:Gammak}:
\begin{align}
    \Gamma_k
    = 2(g-\sqrt{g^2-\Delta^2})\left(g - \Delta \frac{{}^-\bra{k}\mathcal{P}\ket{k}{}^-}{\|\ket{k}^-\|_2^2}\sigma_y\right).
\end{align}
Note that ${}^-\tilde{\bra{k}}\mathcal{P}\tilde{\ket{k}}{}^-$ is real due to the Hermiticity of $\mathcal{P}$.
Since the eigenvalues of $\sigma_y$ are $\pm 1$, we have
\begin{align}
    \gamma_k^{(1)}
    = 2(g-\sqrt{g^2-\Delta^2})\left(g - \Delta\abs{\frac{{}^-\bra{k}\mathcal{P}\ket{k}{}^-}{\|\ket{k}^-\|_2^2}}\right)
\end{align}
and
\begin{align}    
    \gamma_k^{(2)}
    = 2(g-\sqrt{g^2-\Delta^2})\left(g + \Delta\abs{\frac{{}^-\bra{k}\mathcal{P}\ket{k}{}^-}{\|\ket{k}^-\|_2^2}}\right),
\end{align}
which are bounded uniformly with respect to $k$ as
\begin{align}
    \gamma_k^{(1)}
    \ge 2(g-\sqrt{g^2-\Delta^2})(g - \Delta)
\end{align}
and
\begin{align}
    \gamma_k^{(2)}
    \le 2(g-\sqrt{g^2-\Delta^2})(g + \Delta)
\end{align}
by the Cauchy-Schwarz inequality.

In order to calculate or estimate the norm or the condition number of $\begin{pmatrix} \frac{\ket{1}^-}{\|\ket{1}^-\|_2} & \cdots & \frac{\ket{L}^-}{\|\ket{L}^-\|_2} \end{pmatrix}$, it is necessary to examine the properties of $H^-$, which depend on the boundary conditions.
In the following, we consider the cases of the PBC and the OBC, separately.

(i) \textit{The case of the PBC}\quad In this case, $H^-$ in Eq.\eqref{eq:HPBCpm} is normal, and thus is diagonalized by the unitary matrix
\begin{align}
    \begin{pmatrix}
        \frac{\ket{1}_{\mathrm{PBC}}^-}{\|\ket{1}_{\mathrm{PBC}}^-\|_2} & \cdots & \frac{\ket{L}_{\mathrm{PBC}}^-}{\|\ket{L}_{\mathrm{PBC}}^-\|_2}
    \end{pmatrix}.
\end{align}
Then, in the same way as the case of $g<\Delta$ in Appendix \ref{sec:g<Delta}, we have
\begin{align}
    \norm{\begin{pmatrix}
        \frac{\ket{1}_{\mathrm{PBC}}^-}{\|\ket{1}_{\mathrm{PBC}}^-\|_2} & \cdots & \frac{\ket{L}_{\mathrm{PBC}}^-}{\|\ket{L}_{\mathrm{PBC}}^-\|_2}
    \end{pmatrix}^{-1}}_2
    = 1
\end{align}
and
\begin{align}
    \kappa\left(\begin{pmatrix}
        \frac{\ket{1}_{\mathrm{PBC}}^-}{\|\ket{1}_{\mathrm{PBC}}^-\|_2} & \cdots & \frac{\ket{L}_{\mathrm{PBC}}^-}{\|\ket{L}_{\mathrm{PBC}}^-\|_2}
    \end{pmatrix}\right)
    = 1.
\end{align}

From Eq.\eqref{eq:preestimate_cond}, therefore, we have
\begin{align}
    \sqrt{\frac{g-\Delta}{g+\Delta}}
    \le \kappa(\tilde{V}_{\mathrm{PBC}})
    \le \frac{g+\Delta}{g-\Delta}.
\end{align}

(ii) \textit{The case of the OBC}\quad In this case, $H^+$ in Eq.\eqref{eq:HOBCpm} is nothing but the Hatano-Nelson model under the OBC with the non-Hermitian asymmetric hopping $\sqrt{g^2-\Delta^2}$; replacing $g$ with $\sqrt{g^2-\Delta^2}$ in Eq.\eqref{a-1} gives $H_{\mathrm{OBC}}^+$ in Eq.\eqref{eq:HOBCpm}.
Note that the matrix
\begin{align}\label{eq:PVOBC}
    \mathcal{P}\begin{pmatrix}
        \frac{\ket{1}_{\mathrm{OBC}}^-}{\|\ket{1}_{\mathrm{OBC}}^-\|_2} & \cdots & \frac{\ket{L}_{\mathrm{OBC}}^-}{\|\ket{L}_{\mathrm{OBC}}^-\|_2}
    \end{pmatrix}
\end{align}
diagonalizes $H_{\mathrm{OBC}}^+$ because it holds that
\begin{align}
    H_{\mathrm{OBC}}^+
    = \mathcal{P}H_{\mathrm{OBC}}^-\mathcal{P}.
\end{align}
Then introducing a parameter 
\begin{align}
    r^-\coloneqq \sqrt{\frac{t+\sqrt{g^2-\Delta^2}}{t-\sqrt{g^2-\Delta^2}}}(>1),
\end{align}
the matrix \eqref{eq:PVOBC} corresponds to the value at $r=r^-$ of $V_{\mathrm{OBC}}$ in Eq.\eqref{1-1} in Section \ref{sec:cond_HN} up to the order of columns. 
Using the inequalities \eqref{17} and \eqref{eq:lower_VOBCinv}, and recalling that $\mathcal{P}$ is unitary, we have 
\begin{align}
    \kappa\left(\begin{pmatrix}
        \frac{\ket{1}_{\mathrm{OBC}}^-}{\|\ket{1}_{\mathrm{OBC}}^-\|_2} & \cdots & \frac{\ket{L}_{\mathrm{OBC}}^-}{\|\ket{L}_{\mathrm{OBC}}^-\|_2}
    \end{pmatrix}\right)
    \le \sqrt{L}(r^-)^{L-1}
\end{align}
and
\begin{align}
    \norm{\begin{pmatrix}
        \frac{\ket{1}_{\mathrm{OBC}}^-}{\|\ket{1}_{\mathrm{OBC}}^-\|_2} & \cdots & \frac{\ket{L}_{\mathrm{OBC}}^-}{\|\ket{L}_{\mathrm{OBC}}^-\|_2}
    \end{pmatrix}^{-1}}_2
    > \left(\frac{2}{L+1}\right)^{3/2}(r^-)^{L-1}.
\end{align}

From Eq.\eqref{eq:preestimate_cond}, therefore, we obtain
\begin{align}
    &\sqrt{\frac{g-\Delta}{g+\Delta}}\left(\frac{2}{L+1}\right)^{3/2}(r^-)^{L-1} \notag \\
    &< \kappa(\tilde{V}_{\mathrm{OBC}})
    \le \frac{g+\Delta}{g-\Delta}\sqrt{L}(r^-)^{L-1}.
\end{align}

\subsection{Conclusion}
From the above, it holds that
\begin{align}
    &\sqrt{\frac{g-\Delta}{g+\Delta}}\left(\frac{2}{L+1}\right)^{3/2}(r^-)^{L-1} \notag \\
    &< \kappa(\tilde{V})
    \le \frac{g+\Delta}{g-\Delta}\sqrt{L}(r^-)^{L-1}
\end{align}
for $g>\Delta$ and under the OBC, otherwise
\begin{align}
    \sqrt{\frac{g-\Delta}{g+\Delta}}
    \le \kappa(\tilde{V})
    \le \frac{g+\Delta}{g-\Delta},
\end{align}
which are simply Eq.\eqref{eq:estimate_AIIdag}.
It is clear that the former is of the order $O(e^{L\ln r^-})$, while the latter is of the order $O(1)$.

\section{Disorder-averaged effective Hamiltonian for Cauchy distribution disorder}
\label{sec:exa}

Let us consider the (matrix) Hamiltonian $H$ consisting of a term $H_0$ without disorder and a term $W$ with onsite disorder,
\begin{align}
H=H_0+W.
\end{align}
Assuming that the onsite disorder potential $w_j$, which is given by $W_{ij}=w_i\delta_{ij}$, follows the Cauchy distribution,
\begin{align}
P(w_j)=\frac{\gamma}{\pi}\frac{1}{w_j^2+\gamma^2},
\end{align}
we derive here the disorder-averaged effective Hamiltonian \footnote{For Hermitian $H_0$, this result was first given by Lloyd \cite{Lloyd1969-wc}, and the generalization to the non-Hermitian case was done in Ref.\cite{r99-bz-1998}. Here we follow the derivation in Ref.\cite{Kozlov_2014}. } 

To derive the effective Hamiltonian, we consider the Green function $G(z)$ of $H$
\begin{equation}
\label{48-0}
    G(z)=\frac{1}{z-H},
\end{equation}
where $z$ is a complex parameter.
A remarkable property of the Cauchy distribution is that the disorder-average of the Green function can be done exactly:
Following Ref.\cite{Kozlov_2014}, we first expand the Green function as
\begin{align}
G(z)&=\frac{1}{z-H_0-W}
\nonumber\\
&=\frac{1}{(z-W)(1-(z-W)^{-1}H_0)}
\nonumber\\
&=\frac{1}{z-W}+\frac{1}{z-W}H_0\frac{1}{z-W}+\cdots
,
\end{align}
then perform the disorder-average. We typically have the contribution in the form of 
\begin{align}
\left<\left(\frac{1}{z-w_i}\right)^{n_i}\left(\frac{1}{z-w_j}\right)^{n_j}\cdots\right>,
\label{eq:z-w}
\end{align}
and from the residue theorem, $w_j$s' in Eq.(\ref{eq:z-w}) are replaced by $-i\gamma$ ($i\gamma$) after the disorder-average if ${\rm Im}z>0$ (${\rm Im}z<0$).
As a result, the disorder-averaged Green function is given by
\begin{align}
\left<G(z)\right>&=\frac{1}{z+i\gamma{\rm sgn}[{\rm Im}z]}
\nonumber\\
&+\frac{1}{z+i\gamma{\rm sgn}[{\rm Im}z]}H_0\frac{1}{z+i\gamma{\rm sgn}[{\rm Im} z]}+\cdots
\nonumber\\
&=\frac{1}{z+i\gamma{\rm sgn}[{\rm Im}z]-H_0}.
\label{eq:<g>}
\end{align}
Then, defining the effective Hamiltonian $H_{\rm eff}$ such that it gives a pole of $\left<G(z)\right>$ in the complex plane of $z$, we have
\begin{align}
H_{\rm eff}=H_0-i\gamma{\rm sgn}[{\rm Im}z].
\end{align}

From the disorder-averaged Green function, we can also calculate the density of state \cite{r99-bz-1998}.
The density of states at the complex energy $z=x+iy$ is defined as
\begin{equation}
    \rho(x,y)= \left<\frac{1}{L}\sum_{i=1}^{L}\delta(x-\Re E_i)\delta(y-\Im E_i)\right>,
\end{equation}
where $E_i$ are eigenvalues of the matrix Hamiltonian $H$ and $L$ is the number of eigenvalues. 
%and $\left<~\cdot~\right>$ represents the impurity-average. 
Then, we have
\begin{equation}
\label{49-1}
    \rho(x,y)=\frac{1}{\pi}\frac{\partial}{\partial z^{\ast}}\overline{G(z)},
\end{equation}
where $\overline{G(z)}$ is defined as
\begin{align}
\overline{G(z)}=\left<\frac{1}{L}{\rm tr}G(z)\right>.
\end{align}
To show this, using the Jordan decomposition of $H$, we first rewrite the Green function in terms of the eigenvalues of $H$,
\begin{align}
   \overline{G(z)}&=\left<\frac{1}{L}\sum_{i=1}^{L}\frac{1}{z-E_i}\right>, \nonumber\\
   &=\left<\frac{1}{L}\sum_{i=1}^{L}\frac{1}{(x-\Re E_i)+i(y-\Im E_i)}\right>,
   \label{AA99}
\end{align}
where $z=x+iy$. 
%If the complex parameter $z$ is equal to the complex energy $x+iy$ where we consider the density of states $\rho(x,y)$: $z=x+iy$~($x\in \mathbb{R}, y\in \mathbb{R}$), then
%\begin{equation}
%    \overline{G(z)}=\left<\frac{1}{N}\sum_{i=1}^{N}\frac{1}{(x-\Re E_i)+i(y-\Im E_i)}\right>.
%\end{equation}
Then, differentiating Eq.~(\ref{AA99}) with respect to $z^{\ast}$, we have 
\begin{align}
    &\frac{1}{\pi}\frac{\partial}{\partial z^{\ast}}\overline{G(z)}\notag\\
    &=\frac{1}{\pi}\frac{\partial}{\partial z^{\ast}}\left<\frac{1}{L}\sum_{i=1}^{L}\frac{1}{(x-\Re E_i)+i(y-\Im E_i)}\right>\notag\\
    &=\left<\frac{1}{L}\sum_{i=1}^{L}\delta(x-\Re E_i)\delta(y-\Im E_i)\right>\notag\\
    &=\rho(x,y),
\end{align}
where we have used the identity
\begin{equation}
    \frac{\partial}{\partial z^{\ast}}\left(\frac{1}{z}\right)=\pi \delta(\Re z)\delta(\Im z),
\end{equation}
in the second equality.
The Green function in Eq.~(\ref{eq:<g>}) is rewritten as 
\begin{equation}
\label{60-1}
    \overline{G(z)}=\overline{G_0(z+i\gamma)}\theta(\Im z)+\overline{G_0(z-i\gamma)}\theta(-\Im z),
\end{equation}
where $\theta(x)$ is the Heaviside step function and $\overline{G_0(z)}$ is the trace of the Green function without disorder
\begin{align}
\overline{G_0(z)}
=\frac{1}{L}{\rm tr}
\frac{1}{z-H_0}.
\end{align}
Thus, differentiating Eq.~(\ref{60-1}) with respect to $z^{\ast}$, we obtain
\begin{align}
    \rho(x,y)
&=\rho_{0}(x,y+\gamma)\theta(y)+\rho_0(x,y-\gamma)\theta(-y)\notag\\
&+\frac{i}{2\pi}\delta(y)\left[\overline{G_0(x+i\gamma)}-\overline{G_0(x-i\gamma)} \right], \label{62-1}
\end{align}
where $\rho_0(x,y)$ is the density of states without disorder:
\begin{equation}
    \rho_0(x,y)=\frac{1}{\pi}\frac{\partial}{\partial z^{\ast}}\overline{G_0(z)}.
\end{equation}

\begin{figure*}[tbp]
 \begin{center}
  \includegraphics[scale=0.5]{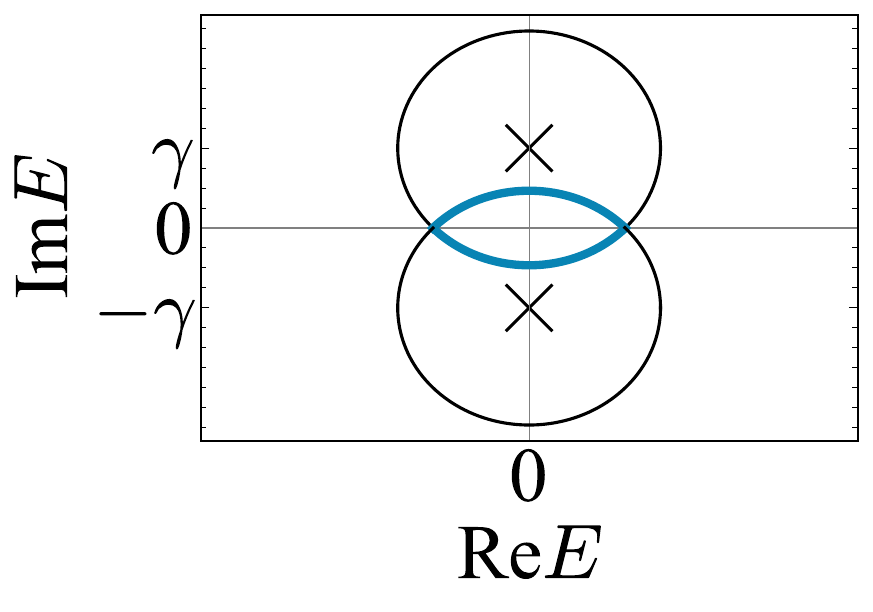}
  \caption{The blue line represents the spectrum of $H=H_{0}+W$, where $W$ follows the Cauchy distribution. The black line represents the energy spectrum that shifts the energy spectrum of $H_{0}$ parallel to the imaginary axis by $\pm \gamma$.\label{Fig63-1}}
  \end{center}
\end{figure*}
Note that we can apply this exact result to the PBC and the OBC.

Equation~(\ref{62-1}) nicely explains the disorder-average spectrum of the Hatano-Nelson model.
The first and second terms give the spectrum in the lower ($\Im z<0$) and the upper ($\Im z>0$) half parts of the complex energy plane, respectively.
In the case of the Hatano-Nelson model, $\rho_0(x,y)$ forms a loop in the complex energy plane, and thus the first and second terms give a blue curve in Fig.~\ref{Fig63-1}.
Furthermore, the third term in Eq.~(\ref{62-1}) provides the energy spectrum on the real axis of the complex energy plane. This part explains the spectrum of modes with Anderson localization.

\section{Topological invariants in the presence of disorders}
\label{sec:invtwiA}

In the main text, we introduce the topological numbers for the non-Hermitian skin effects in Eqs. (\ref{21-1}) and (\ref{10-13-0-}). 
However, when disorders exist, one cannot employ them  
because the crystal momentum $k$ is no longer a good quantum number.
In disordered Hermitian systems, we can avoid this difficulty by using topological invariants in the phase space of the twisted boundary condition\cite{r106-niu-thouless-wu-1985,supercell2007,supercell2010,supercell2012}.
Here we generalize this method to non-Hermitian systems.

\begin{figure*}[tbp]
  \begin{center}
  \includegraphics[scale=0.13]{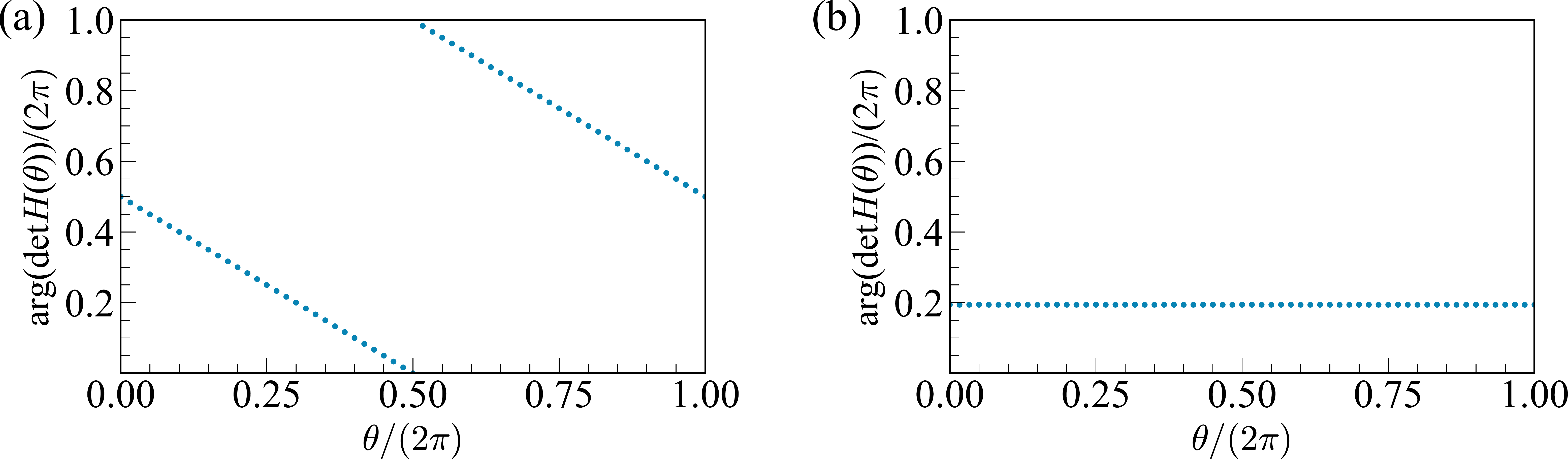}
  \caption{The flow of $\mathrm{arg}(\det H(\theta))$ with respect to $\theta$ for the Hatano-Nelson model (\ref{61-1}) with $t=1.5$, $g=0.5$, $L=100$, $\gamma=0.5$. (a) The reference energy is inside the PBC spectrum ($E=0$). (b) The reference energy is outside the PBC spectrum ($E=10+10i$). \label{FIG6}}
  \end{center}
\end{figure*}

Following Ref.\cite{r1-Gongashida}, we first define the phase space version of Eq.(\ref{21-1}).
For this purpose, we introduce the twisted boundary condition by attaching a phase $e^{i\theta}$ to bonds at the boundary. For the model in Eq.(\ref{61-1}), the twisted boundary condition is given by the following form:
\begin{widetext}
\begin{equation}
  H(\theta)=
  \begin{pmatrix}
    w_1&t-g &0&\cdots&0&(t+g)e^{-i\theta}&\\
    t+g&w_{2}&t-g&\cdots&0&0&\\
    0 &t+g &w_3 &\cdots &0&0&\\
    \vdots&\vdots &\vdots & \ddots&\vdots &\vdots&\\
    0&0&0& \cdots&w_{L-1} &t-g&\\
    (t-g)e^{i\theta}&0&0 &\cdots &t+g&w_{L}
  \end{pmatrix}.\\
\end{equation}
\end{widetext}
Then, the phase space version of Eq.(\ref{21-1}) is given by
\begin{equation}
\label{71-1}
  W(E)\coloneqq\int_{0}^{2\pi}\frac{d\theta}{2\pi i}\frac{\partial}{\partial \theta}\log \det (H(\theta)-E),
\end{equation}
which coincides with Eq.(\ref{21-1}) in the absence of disorders \cite{r1-Gongashida}). 
The quantization of $W(E)$ follows from 
\begin{align}
  W(E)&=\int_{0}^{2 \pi}\frac{d\theta}{2\pi i}\frac{\partial}{\partial \theta}
  \left[\log \left|\det \left(H(\theta)-E\right)\right|\right.\notag\\
  &+\left.i \mathrm{arg}\left(\det\left(H(\theta)-E\right)\right)\right]\notag\\
  &=\int_{0}^{2 \pi}\frac{d\theta}{2\pi }\frac{\partial}{\partial \theta}\mathrm{arg}\left(\det \left(H(\theta)-E\right)\right). 
  \label{www}
\end{align}
Therefore, if the reference energy $E$ is inside (outside) the PBC spectrum, we have $W(E)\neq0$ ($W(E)=0$).
In Figs. \ref{FIG6} (a) and (b), we illustrate how the argument of $\det H(\theta)$ behaves inside and outside the PBC spectrum, respectively, for 
the model in Eq.(\ref{61-1}) with $t=1.5$, $g=0.5$, $L=100$ and $\gamma=0.5$, where 
the corresponding winding numbers are $-1$ and $0$.

In a similar manner, we can define the phase space version of Eq.(\ref{10-13-0-}).
We introduce the twisted boundary condition by attaching a phase $e^{i\theta}$ to bonds at
the boundary so that 
the resultant Hamiltonian $H(\theta)$ keeps the transpose version of time-reversal symmetry,  $TH^T(\theta)T^{-1}=H(-\theta)$ with $TT^*=-1$. 
Then, the following equation defines the $\mathbb{Z}_2$ invariant in the phase space,
\begin{align}
(-1)^{\nu(E)}=
  &\frac{\mathrm{Pf}[(H(\pi)-E) T]}{\mathrm{Pf}[( H(0)-E) T]}\notag\\
  &\times\exp\left[-\frac{1}{2}\int_{\theta=0}^{\theta=\pi}d\log \det[(H(\theta)-E) T]\right],
\label{a8}
\end{align}
since the square of the right-hand side becomes an identity. 

\begin{figure*}[tbp]
  \begin{center}
  \includegraphics[scale=.13]{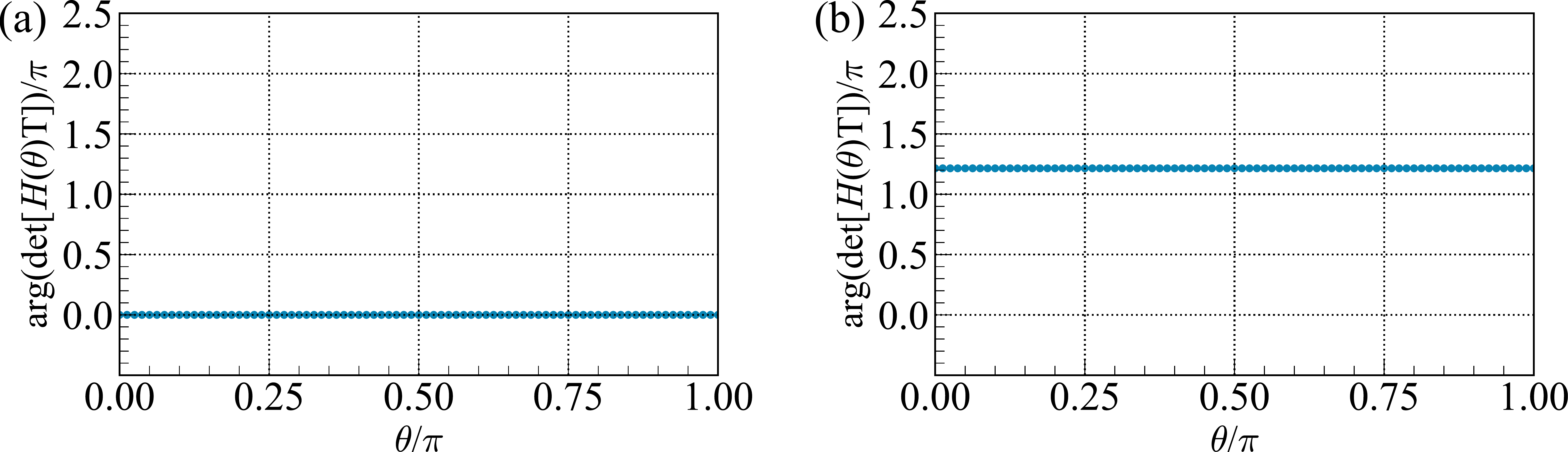}
  \caption{The flow of $\arg(\det [(H(\theta)-E])T)$ with respect to $\theta$ for the model (\ref{a4}) with $t=1$, $g=0.3$, $\Delta=0.2$, $L=100$ and $\gamma=0.2$. (a) The reference energy is inside the PBC spectrum ($E=0$). (b) The reference energy is outside the PBC spectrum ($E=10+10i$).\label{FIG10}}
  \end{center}
\end{figure*}

%\section{CALUCULATION OF $\mathbb{Z}_2$ TOPOLOGICAL NUMBER}
%\label{sec:check}
We calculate the $\mathbb{Z}_2$ topological number in Eq.(\ref{a8}) for  the model (\ref{a4}) with $t=1$, $g=0.3$, $\Delta=0.2$, $L=100$, and $\gamma=0.2$.
For this purpose, we rewrite Eq.(\ref{a8}) as
\begin{align}
(-1)^{\nu(E)}
  &=\frac{\mathrm{Pf}[( H(\pi)-E) T]}{\mathrm{Pf}[( H(0)-E) T]}\times\left[\frac{|\det(H(\pi)-E) T|}{|\det( H(0)-E) T|}\right]^{-\frac{1}{2}}
  \notag\\
  &\exp\left[-\frac{i}{2}\int_0^\pi d\theta \frac{\partial}{\partial \theta}{\rm arg}\det[ (H(\theta)-E)T]
  \right],
\end{align}
When $E$ is inside the PBC spectrum, we have
\begin{equation}
  \left[\frac{|\det[(H(\pi)-E)T]|}{|\det[(H(0)-E)T)]|}\right]^{-\frac{1}{2}}=1,
\end{equation}
and
\begin{equation}
  \frac{\mathrm{Pf}[(H(\pi)-E)T]}{\mathrm{Pf}[(H(0)-E)T]}=-1.
\end{equation} 
Our numerical calculation also finds that $\arg(\det[(H(\theta)-E)T])$ is independent of $\theta$, as illustrated in Fig.~\ref{FIG10}(a). 
Thus, we have the non-trivial $\mathbb{Z}_2$ invariant $(-1)^{\nu(E)}=-1$.
On the other hand, if  $E$ is outside the PBC spectrum, we have
\begin{equation}
  \left[\frac{|\det[(H(\pi)-E)T]|}{|\det[(H(0)-E)T]|}\right]^{-\frac{1}{2}}=1,
\end{equation}
and
\begin{equation}
  \frac{\mathrm{Pf}[(H(\pi)-E)T]}{\mathrm{Pf}[(H(0)-E)T]}=1, 
\end{equation} 
and the $\theta$ dependence of $\arg(\det[(H(\theta)-E)T])$ in  Fig.~\ref{FIG10}(b). 
Thus, the $\mathbb{Z}_2$ invariant is trivial, $(-1)^{\nu(E)}=1$.

Whereas these topological numbers in the phase space characterize the non-Hermitian skin effects in the presence of disorders, they are unsuitable for numerical detection of the disorder-induced topological phase transition discussed in Sec.\ref{sec:enn}. 
This is because they depend on the reference energy $E$, and the phase transition requires the vanishing of the topological numbers for any $E$. 
In contrast, the scalar measures of non-normality introduced in Sec.\ref{enhance} are independent of the reference energy $E$, and thus no such problem exists.

\section{Numerical algorithm for evaluation of condition number}

We outline here the algorithm for numerical evaluation of the condition number $\kappa(V)$. 
First, note that 
only the right eigenstates of $H$ are necessary to evaluate the condition number:
Once one constructs $V$ consisting of the right eigenstates subject to the normalization condition in Eq.(\ref{eq:normalization}), then one obtains the condition number $\kappa(V)$ as the ratio of the maximum and the minimal singular values of $V$. See Eq.(\ref{eq:kappa_evaluation}).
Therefore, the evaluation process of the condition number is straightforward for most non-Hermitian systems.

For non-Hermitian systems with disorders such as the Hatano-Nelson model, however, we need a large number of samples to perform the disorder average of the condition number.
In particular, under the open boundary condition, the non-Hermitian skin effect fairly enhances the perturbation sensitivity, as indicated in Eq.(\ref{eq:BF}), and thus, the variance of the numerically obtained condition number becomes large, and the convergence of the calculation requires a large number of samples. 
To save the computer resources required by this property, 
we have used the imaginary gauge transformation that maps a non-Hermitian system to a Hermitian one \cite{r50-hatano-1996,r51-hatano-1997,r52-hatano-1998}, as explained below.

For concreteness, let us first consider the Hatano-Nelson model,
\begin{equation}
  \label{61-1}
\hat{H}=\sum_{j=1}^{L}\left[(t+g)\hat{c}_{j+1}^{\dagger}\hat{c}_{j}+(t-g)\hat{c}^{\dagger}_{j}\hat{c}_{j+1}+w_j\hat{c_j}^\dagger\hat{c_j}\right],
\end{equation}
with $t>g$.
The imaginary gauge transformation is given by
\begin{align}
\hat{c}_j\rightarrow e^{\alpha j}\hat{c}_j \quad
\hat{c}_j^{\dagger}\rightarrow e^{-\alpha j}\hat{c}_j^{\dagger},
\quad (e^\alpha=\frac{\sqrt{|t^2-g^2|}}{t-g}),
\end{align}
which maps the non-Hermitian Hamiltonian into 
the following Hermitian one under the OBC,
\begin{equation}
  \label{kyoge-ji}
  \hat{\tilde{H}}=\sum_{j=1}^{L}\left[\sqrt{|t^2-g^2|}\hat{c}_{j+1}^{\dagger}\hat{c}_{j}+\sqrt{|t^2-g^2|}\hat{c}^{\dagger}_{j}\hat{c}_{j+1}+w_j\hat{c_j}^\dagger\hat{c_j}\right].
\end{equation}
Since Eq.(\ref{kyoge-ji}) is Hermitian, the disorder term does not give the problem caused by the enhanced perturbation sensitivity.
After numerically diagonalizing Eq.(\ref{kyoge-ji}), we perform the inverse of the imaginary gauge transformation, then we obtain the matrix diagonalizing Eq.(\ref{61-1}). Finally, by normalizing each column of the diagonalization matrix to satisfy Eq.(\ref{eq:normalization}), we obtain the matrix $V$ required for calculating the condition number.

We also use a similar technique to evaluate the condition number in the time-reversal invariant Hatano-Nelson model, 
\begin{align}
  \hat{H}&=\sum_{j=1}^{L}\left[
  (t+g)\hat{c}_{j+1,\uparrow}^{\dagger}\hat{c}_{j,\uparrow}+(t-g)\hat{c}_{j,\uparrow}^{\dagger}\hat{c}_{j+1,\uparrow}
  +w_j\hat{c}_{j,\uparrow}^{\dagger}\hat{c}_{j,\uparrow}\right]
  \notag\\
  &+\sum_{j=1}^{L}\left[
  (t+g)\hat{c}_{j,\downarrow}^{\dagger}\hat{c}_{j+1,\downarrow}
  +(t-g)\hat{c}_{j+1,\downarrow}^{\dagger}\hat{c}_{j,\downarrow}
  +w_j\hat{c}_{j,\downarrow}^{\dagger}\hat{c}_{j,\downarrow}\right]\notag\\
  &-i\Delta\sum_{j=1}^{L}(\hat{c}_{j+1,\uparrow}^{\dagger}\hat{c}_{j,\downarrow}-\hat{c}_{j\uparrow}^{\dagger}\hat{c}_{j+1\downarrow})\notag\\
  &-i\Delta\sum_{j=1}^{L}(\hat{c}_{j+1,\downarrow}^{\dagger}\hat{c}_{j,\uparrow}-\hat{c}_{j,\downarrow}^{\dagger}\hat{c}_{j+1,\uparrow}).
\label{a2bb}
\end{align}
First, performing the imaginary spin rotation diagonalizing the spin-dependent part of the Hamiltonian,
\begin{align}
&\begin{pmatrix}
c_{i,\uparrow}\\
c_{i,\downarrow}
\end{pmatrix}    
\rightarrow 
\begin{pmatrix}
c_{i,+}\\
c_{i,-}
\end{pmatrix}
=R
\begin{pmatrix}
c_{i,\uparrow}\\
c_{i,\downarrow}
\end{pmatrix},
\nonumber\\
&\begin{pmatrix}
c^\dagger_{i,\uparrow}&
c^\dagger_{i,\downarrow}
\end{pmatrix}
\rightarrow
\begin{pmatrix}
c^\dagger_{i,+}&
c^\dagger_{i,-}
\end{pmatrix}=\begin{pmatrix}
c^\dagger_{i,\uparrow}&
c^\dagger_{i,\downarrow}
\end{pmatrix}R^{-1}
\end{align}
with
\begin{align}
\label{kaiten}
R=
\begin{pmatrix}
    \frac{ig}{\sqrt{\Delta^2-g^2}}&-\frac{\Delta}{\sqrt{\Delta^2-g^2}} &\\
    \frac{\Delta}{\sqrt{\Delta^2-g^2}}&\frac{ig}{\sqrt{\Delta^2-g^2}}&
  \end{pmatrix},
\end{align}
we have
\begin{align}
\hat{\tilde{H}}=
&\sum_{j=1}^{L}\left[
  (t\pm\sqrt{g^2-\Delta^2})\hat{c}_{j+1,\pm}^{\dagger}\hat{c}_{j,\pm}
\right.
\nonumber\\
&\left. +(t\mp\sqrt{g^2-\Delta^2})\hat{c}_{j,\pm}^{\dagger}\hat{c}_{j+1,\pm}
+w_j\hat{c}_{j,\pm}^{\dagger}\hat{c}_{j,\pm}\right],
\label{eq:spin-diag}
\end{align}
where the double sign corresponds and takes the summation.
Then, the imaginary gauge transformation
\begin{align}
\hat{c}_{j,\pm}\rightarrow e^{\alpha_\pm j}\hat{c}_{j,\pm} \quad
\hat{c}_{j,\pm}^{\dagger}\rightarrow e^{-\alpha_\pm j}\hat{c}_{j,\pm}^{\dagger},
\end{align}
with
\begin{align}
e^\alpha_\pm=\frac{\sqrt{|t^2-g^2+\Delta^2|}}
{t\mp\sqrt{g^2-\Delta^2}}
\end{align}
maps Eq.(\ref{eq:spin-diag}) into the Hermitian one under the OBC,
\begin{align}
  \label{1}
\hat{\tilde{\tilde{H}}}=&\sum_{j=1}^{L}\left[\sqrt{|t^2-g^2+\Delta^2|}\hat{c}_{j+1,\pm}^{\dagger}\hat{c}_{j,\pm}
\right.
 \nonumber\\ 
&\left.
  +\sqrt{|t^2-g^2+\Delta^2|}\hat{c}^{\dagger}_{j,\pm}\hat{c}_{j+1,\pm}+w_j\hat{c_{j,\pm}}^\dagger\hat{c_{j,\pm}}\right].
\end{align}
In a manner similar to the Hatano-Nelson model, we numerically diagonalize Eq.(\ref{1}), and then perform the inverse of the imaginary gauge transformation and the imaginary spin rotation. The obtained matrix diagonalizes Eq.(\ref{a2bb}), from which we evaluate the condition number.

\section{Basic properties of pseudo spectrum}
\label{app:pss}

Here we summarize the basic properties of the pseudo spectrum $\sigma_\epsilon(H)$ of a matrix Hamiltonian $H$ used in this paper \cite{r67-trefethen-book-2005}.

Let $\sigma(H)$ be the spectrum of a matrix Hamiltonian $H$,
\begin{align}
\sigma(H)=\{E\in \mathbb{C}|{\rm det}(H-E)=0\},    
\end{align}
then, the pseudo-spectrum $\sigma_\epsilon(H)$ is defined by 
\begin{align}
&\sigma_\epsilon(H)
\nonumber\\
&=\{E\in \mathbb{C}|\exists \Delta H: E\in \sigma(H+\Delta H),\|\Delta H\|_2<\epsilon\},    
\label{eq:pss1}
\end{align}
where the size of $\Delta H$ is the same as that of $H$.
First, we show that the pseudo-spectrum is equivalently defined as 
\begin{align}
\sigma_\epsilon(H)=\{E\in \mathbb{C}|\exists|u\rangle: \|(H-E)|u\rangle\|_2<\epsilon, \| |u\rangle\|_2=1\}.   
\label{eq:pss2}
\end{align}
%First we show Eq.(\ref{eq:pss2} from Eq.(\ref{eq:pss1}).
Actually, 
if $E\in \sigma_\epsilon(H)$ in Eq.(\ref{eq:pss1}), 
then there exist an eigenstate $|u\rangle$ of 
$(H+\Delta H-E)|u\rangle=0$ with $\||u\rangle\|_2=1$, so
we have $\|(H-E)|u\rangle\|_2=\|\Delta H|u\rangle\|_2<\|\Delta H\|_2<\epsilon$, which implies 
$E\in \sigma_\epsilon(H)$ in Eq.(\ref{eq:pss2}).
Conversely, if $E\in \sigma_\epsilon(H)$ in Eq.(\ref{eq:pss2}), we can introduce $s|v\rangle\equiv (E-H)|u\rangle$ with $s<\epsilon$ and $\||v\rangle \|_2=1$. Then, by defining $\Delta H\equiv s|v\rangle\langle u|$, we have 
$(H+\Delta H-E)|u\rangle=0$ with $\|\Delta H\|_2=s<\epsilon$, so $E\in \sigma_\epsilon(H)$ in Eq.(\ref{eq:pss1}).

The pseudo-spectrum can go beyond the $\epsilon$-neighborhood of the unperturbed spectrum:
\begin{align}
    \sigma_{\epsilon}(H)\supseteq\sigma(H)+\Delta_\epsilon,
\end{align}
where $\Delta_\epsilon=\{E\in \mathbb{C}||E|<\epsilon\}$.
The proof of this relation is straightforward: If $E\in \sigma(H)+\Delta_\epsilon$, it holds that $E=E'+\delta$ with $E'\in \sigma(H)$, $|\delta|<\epsilon$, which means that $E\in \sigma(H+\delta {\bm 1})$ with $\|\delta {\bm 1}\|_2=|\delta|<\epsilon$, so $E\in \sigma_\epsilon(H)$.

An important inclusion relation for the pseudo-spectrum is the Bauer-Fike theorem. 
The theorem is given by 
\begin{align}
\sigma_{\epsilon}(H)\subseteq \sigma(H)+\Delta_{\epsilon\kappa(V)},    
\label{eq:apBF}
\end{align}
for a diagonal matrix Hamiltonian $H$, and
\begin{align}
    \sigma_{\epsilon}(H)\subseteq\sigma(H)+\Delta_{\epsilon+\mathrm{dep}_{\mathrm{F}}(H)},
\label{eq:apBF2}
\end{align}
for a general matrix Hamiltonian $H$.

To prove Eq.(\ref{eq:apBF}), let us suppose $E\in \sigma_\epsilon(H)$.
If $E\in \sigma(H)$ at the same time, we trivially have $E\in \sigma(H)+\Delta_{\epsilon\kappa(V)}$, so we only consider the case with $E\notin \sigma(H)$, where $(E-H)^{-1}$ is well-defined. 
Then, for the diagonal matrix Hamiltonian $H=V\Lambda V^{-1}$ in Eq.(\ref{eq:vhv}), we have
\begin{align}
\| (E-H)^{-1}\|_2&=\|(E-V\Lambda V^{-1})^{-1}\|_2   
\nonumber\\
&=\|V(E-\Lambda)^{-1}V^{-1}\|_2
\nonumber\\
&\le \|V\|_2 \|(E-\Lambda)^{-1}\|_2 \|V^{-1}\|_2
\nonumber\\
&=
\kappa(V)/{\rm min}_{E_\alpha\in \sigma(H)}(E-E_\alpha).
\label{eq:bfp}
\end{align}
From Eq.(\ref{eq:pss2}), we can also show that $\|(E-H)^{-1}\|_2>\epsilon^{-1}$:
For $|u\rangle$ in Eq.(\ref{eq:pss2}), if we write $(E-H)|u\rangle\equiv s|v\rangle$ with $\||v\rangle\|_2=1$, we have $s<\epsilon$ and $(E-H)^{-1}|v\rangle=s^{-1}|u\rangle$, which leads to
\begin{align}
\|(E-H)^{-1}\|_2\ge \|(E-H)^{-1}|v\rangle\|_2=s^{-1}>\epsilon^{-1}.    
\label{eq:bfp2}
\end{align}
Combining Eqs.(\ref{eq:bfp}) and (\ref{eq:bfp2}), we have
${\rm min}_{E_\alpha\in\sigma(H)}|E-E_\alpha|<\epsilon\kappa(V)$, 
which implies $E\in\sigma(H)+\Delta_{\epsilon\kappa(V)}$. Therefore, Eq.(\ref{eq:apBF}) holds.

Equation (\ref{eq:apBF2}) is also proved from Eq.(\ref{eq:pss2}). For $E\in\sigma_\epsilon(H)$,
there exists $|u\rangle$ with $\|(H-E)|u\rangle\|_2<\epsilon$ and $\||u\rangle\|_2=1$. Then, from the Schur decomposition minimizing the 2-norm departure from normality, $H=U(\Lambda+R)U^\dagger$, we have
\begin{align}
\epsilon>\|(H-E)|u\rangle\|_2&=\|U(\Lambda+R)U^\dagger|u\rangle-EUU^\dagger|u\rangle\|_2    
\nonumber\\
&=\|(\Lambda+R) U^\dagger|u\rangle-EU^\dagger|u\rangle\|_2,
\end{align}
which leads to
\begin{align}
&{\rm min}_{E_\alpha\in \sigma(H)} |E-E_\alpha|
\nonumber\\
&\le \|\Lambda U^\dagger|u\rangle-EU^\dagger|u\rangle\|_2
\nonumber\\
&=\|(\Lambda+R) U^\dagger|u\rangle-EU^\dagger|u\rangle-RU^\dagger|u\rangle\|_2
\nonumber\\
&\le\|(\Lambda+R) U^\dagger|u\rangle-EU^\dagger|u\rangle\|_2+\|RU^\dagger|u\rangle\|_2
\nonumber\\
&<\epsilon+\|R\|_2
\nonumber\\
&<\epsilon+\|R\|_F=\epsilon+{\rm dep}_{\rm F}(H).
\end{align}
Therefore,  it holds that $E\in \sigma(H)+\Delta_{\epsilon+{\rm dep}_{\rm F}(H)}$, implying Eq.(\ref{eq:apBF2}).

The pseudo-spectrum also governs the dynamics of the system. 
First, using the residue theorem, we have
\begin{align}
e^{-iHt}=\frac{1}{2\pi i}\oint_\Gamma e^{-iEt}(E-H)^{-1}dE,    
\end{align}
where $\Gamma$ is any contour closing $\sigma(H)$ in its interior.
This equation leads to
\begin{align}
\|e^{-iHt}\|_2\le \frac{1}{2\pi}\oint_\Gamma |e^{-i E t}|\|(E-H)^{-1}\|_2 |dE|.   
\label{eq:gamma_int}
\end{align}
To evaluate the right-hand side of the above inequality, we show that if $\|(E-H)^{-1}\|_2>\epsilon^{-1}$, then $E \in \sigma_{\epsilon}(H)$:
Under this assumption, there exists $|v\rangle$ satisfying $\||v\rangle\|_2=1$ and $\|(E-H)^{-1}|v\rangle\|_2>\epsilon^{-1}$. 
Then, introducing $s^{-1}|u\rangle\equiv (E-H)^{-1}|v\rangle$ with $\||u\rangle\|_2=1$, we have $(E-H)|u\rangle=s|v\rangle$ with $s<\epsilon$, implying $\|(H-E)|u\rangle\|_2<\epsilon$, and $E\in \sigma_\epsilon(H)$ in Eq.(\ref{eq:pss2}).
Because of this property, if we choose $\Gamma$ in Eq.(\ref{eq:gamma_int}) so as to enclose $\sigma_\epsilon(H)$, $\|(E-H)^{-1}\|_2$ in Eq.(\ref{eq:gamma_int}) satisfies $\|(E-H)^{-1}\|_2<\epsilon^{-1}$.
Thus, we have
\begin{align}
\|e^{-iHt}\|_2 &\le \frac{1}{2\pi\epsilon}\oint_\Gamma {\rm max}|e^{-i E t}| |dE| 
\nonumber\\
&=\frac{L_\epsilon e^{\alpha_{\epsilon}(H)t}}{2\pi\epsilon},   
\end{align}
where $L_\epsilon$ is the arc length of the boundary of $\sigma_{\epsilon}(H)$ and $\alpha_\epsilon(H)={\rm max}[{\rm Im}\sigma_\epsilon(H)]$.

\section{Characterization of skin effects by condition number of Hamiltonian}
\label{sec:conconcon}
In this section, we show that the condition number of the Hamiltonian itself also detects the disorder-induced topological phase transition of non-Hermitian skin effects. 

\subsection{Condition number and topology}
\label{sec:contopo}
We define the condition number of the Hamiltonian $H$ itself as 
\begin{equation}
\label{a15-0}
    \ka(H)=\|H\|_2\|H^{-1}\|_2=\frac{s_{\max}(H)}{s_{\min}(H)}\geq 1,
\end{equation}
where $\|\cdot\|_2$ is the 2-norm of a matrix
\begin{align}
\|\cdot \|_2={\rm max}_{\bm x}\left[\sqrt{|\cdot{\bm x}|^2}/\sqrt{|{\bm x}|^2}\right].  
\end{align}
Here, $s_{\max}(H)$ and $s_{\min}(H)$ are the largest singular value and the smallest singular value of $H$, respectively.

\begin{figure*}[tbp]
 \begin{center}
  \includegraphics[scale=0.14]{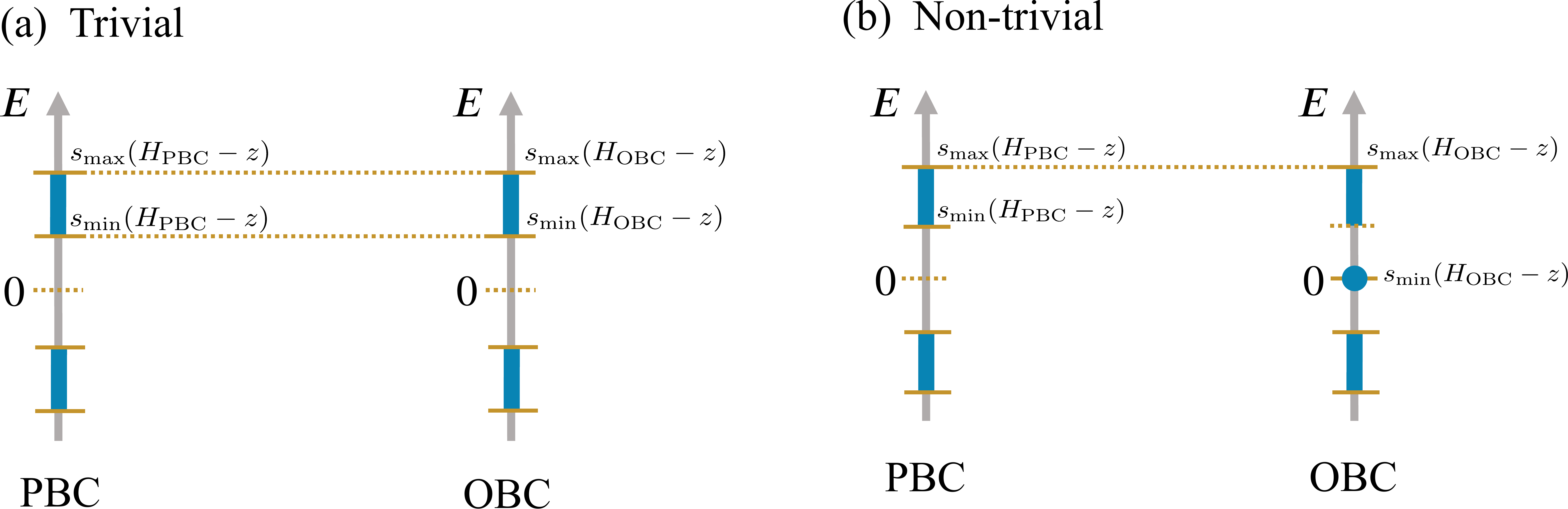}
  \caption{Schematic illustration of energy spectrum for $\tilde{H}$ under the PBC and the OBC. $s_{\min}(H-z)$ and $s_{\max}(H-z)$ represent the smallest and largest singular value of $H-z$, respectively. (a) Trivial case of the Hamiltonian (\ref{a16-2}). (b) Non-trivial case of the Hamiltonian (\ref{a16-2}).
  \label{FIG14}}
  \end{center}
\end{figure*}

Let us consider a non-Hermitian Hamiltonian $H$.
If the reference energy $z$ of $H$ is in the energy region where the topological number calculated from $H_{\rm PBC}$ is non-trivial, then the condition number of $H-z\bm{1}$ ($\bm{1}$ represents the identity matrix. In the following, we omit it.) generally satisfies
\begin{equation}
\label{a15-1}
    \ka(H_{\mathrm{PBC}}-z)\neq\ka(H_{\mathrm{OBC}}-z).
\end{equation}
We check this property in the following. We introduce the following Hermitian Hamiltonian \footnote{$z$ is not the eigenenergy of $H$.}
\begin{equation}
\label{a16-2}
    \tilde{H}=
    \begin{pmatrix}
    0 &H-z\\
    H^{\dag}-z^{\ast}&0
  \end{pmatrix}\\.
\end{equation}
In this parameter region, since non-Hermitian skin effects occur, $\tilde{H}_{\mathrm{OBC}}$ has zero-energy boundary mode as discussed in Sec.\ref{sec:skintopo}.
Because the energy eigenvalue of $\tilde{H}$ equals the singular value of $H-z$ as we check it later, we derive the following relation in this parameter region (Fig.~\ref{FIG14} (a))
\begin{equation}
    s_{\min}(H_{\mathrm{PBC}}-z)\neq s_{\min}(H_{\mathrm{OBC}}-z).
\end{equation}
Noting that $s_{\max}(H-z)$ does not depend on the boundary conditions (Fig.~\ref{FIG14} (a)), we can show that it holds that Eq.(\ref{a15-1}) in this parameter region.

We check the equivalence between the eigenvalue of $\tilde{H}$ and the singular value of $H-z$.
Considering the eigenvalue equation of $\tilde{H}$:
\begin{align}
 &\begin{pmatrix}
    0 &H-z\\
    H^{\dag}-z^{\ast}&0
  \end{pmatrix}
    \begin{pmatrix}
    v\\
    u
  \end{pmatrix}
  =E
  \begin{pmatrix}
    v\\
    u
  \end{pmatrix}\notag\\
  \vspace{10mm}
  &\Leftrightarrow
  \left\{ \,
    \begin{aligned}
    &\ (H-z)u = Ev \\
    &(H^{\dag}-z^{\ast})v=Eu
    \end{aligned}
    \right. ,
\end{align}
we can derive this equivalence:
\begin{align}
    &(H^{\dag}-z^{\ast})(H-z)u\notag\\
    &=E(H^{\dag}-z^{\ast})v\notag\\
    &=E^{2}u.
\end{align}

When the reference energy $z$ is in the energy region where the topological number is trivial, the condition number of $H_{\rm PBC}-z$ equals that of $H_{\rm PBC}-z$ because $\tilde{H}_{\rm OBC}$ has no zero-energy boundary mode:
\begin{equation}
\label{a15-2}
    \ka(H_{\mathrm{PBC}}-z)=\ka(H_{\mathrm{OBC}}-z).
\end{equation}

These above discussions indicate that one can detect the existence of skin modes by calculating the condition numbers of $H_{\mathrm{PBC}}-z$ and $H_{\mathrm{OBC}}-z$, respectively, under the proper reference energy $z$ without directly computing the topological numbers.
In addition, since these arguments do not need to assume translational symmetry, we can apply them to disordered systems.
In the following, we show that the equality of the condition number for $H$ correctly characterizes the disordered phase transition of skin effects. 

\subsection{Application to Hatano-Nelson model}
\label{sec:chaconA}
\begin{figure*}[tbp]
 \begin{center}
  \includegraphics[scale=0.13]{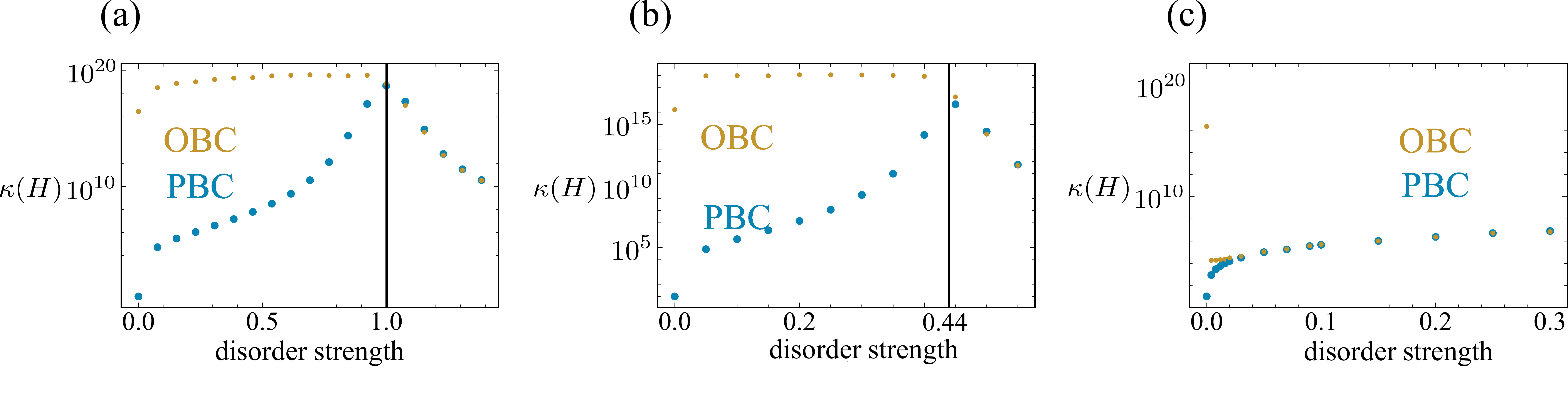}
  \caption{The blue (yellow) dot represents the disorder dependence of $\kappa(H)$ under the PBC (OBC).
  We numerically calculate $\kappa(H)$ under the PBC and the OBC under one hundred different configurations of random potential, and calculate the geometric mean of these one hundred obtained ratios.
  We plot these geometric means in these figures.
  (a) The condition number of the Hatano-Nelson model (\ref{61-1}) with $t=1.5$, $g=0.5$, and $L=3000$. The black line denotes $\gamma=1$.(b) The condition number of the time-reversal invariant Hatano-Nelson model (\ref{a2}) with $t=1, g=0.3, \Delta=0.2$, and $L=2500$. The black line denotes $\gamma=0.44$. (c) The condition number of the time-reversal symmetry broken Hatano-Nelson model (\ref{a2bb}) with $t=1, g=0.3$, $\Delta=0.2$, and $L=1500$.
  \label{FIG15}}
  \end{center}
\end{figure*}

We numerically examine disorder dependence of $\ka(H)$ in the Hatano-Nelson model (\ref{61-1}) with $t=1.5, g=0.5$, and $L=3000$ (Fig.\ref{FIG15} (a)). We adopt the disorder following the Cauchy distribution (\ref{cauchy}).
The reference energy $z$ is set to 0.
Our numerical calculation shows that the condition number of $H$ under the PBC is not equal to that under the OBC when $\gamma<1$ while this equality is not satisfied when $\gamma>1$:
\begin{align}
    \ka(H_{\rm PBC})\neq\ka(H_{\rm OBC})\ \mathrm{for\ \gamma<1},
\end{align}
\begin{align}
    \ka(H_{\rm PBC})=\ka(H_{\rm OBC})\ \mathrm{for\ \gamma>1}.
\end{align}
This behavior is consistent with the fact that the skin effect occurs in $\gamma<1$.
The numerically obtained critical strength $\gamma_c\sim 1$ is close to the exact value $\gamma_c=2g=1$ derived in Sec.\ref{sec:exact}.

\subsection{Application to time-reversal invariant Hatano-Nelson model}
\label{sec:chacon}
We also analyze the behavior of $\ka(H)$ under the time-reversal invariant Hatano-Nelson model (\ref{a4}) when changing the disorder strength $\gamma$. 
In Fig.~\ref{FIG15}(b), 
we illustrate the disorder dependence of $\kappa(H)$, where we take the model parameter in Eq. (\ref{a4}) as $t=1$, $g=0.3$, $\Delta=0.2$, and $L=2500$ and consider the disorder with the Cauchy distribution. 
The reference energy $z$ is set to 0. 
Fig.~\ref{FIG15}(b) shows that $\ka(H)$ under the PBC and the OBC, respectively, are not equal for $\gamma<1$, whereas they are not for $\gamma>1$. 
The obtained critical strength $\gamma_c\sim 0.45$ is consistent with the exact value $\gamma_c=\sqrt{g^2-\Delta^2}=0.44$ derived in Sec.\ref{sec:exact}, where
the deviation originates from the finite size effect.

\subsection{Application to time-reversal broken Hatano-Nelson model}
So far, we have considered the time-reversal symmetry-protected non-Hermitian skin effect, where the disorder locally preserves that symmetry. 
In this subsection, we numerically investigate the disorder dependence of $\ka(H)$ in the time-reversal broken Hatano-Nelson model (\ref{a2bb}) with $t=1, g=0.3, \Delta=0.2$, and $L=1500$. 
The reference energy $z$ is set to 0, and the disorder follows the Cauchy distribution, whose probability density function is given by Eqs.(\ref{dis1}) and (\ref{dis2}). 
In Fig.\ref{FIG15}(c), we exhibit the behavior of $\ka(H)$ when changing the disorder strength $\gamma$. 
This calculation implies that $\ka(H)$ under the PBC is not equal to that under the OBC for $\gamma>0.02$. This behavior is consistent with the fact that the symmetry-protected skin effect immediately disappears in the presence of time-reversal broken disorders.

\newpage

\bibliography{bib}
\end{document}